{}
{}

\documentclass[11pt,a4paper]{article}

\newif\ifpublic\publictrue
\newif\iffancy\fancytrue

\usepackage[a4paper,text={160mm,247mm},centering]{geometry}
\usepackage{setspace}
\usepackage{rotating}

\usepackage{amsmath,amssymb, amsfonts, geometry}

\makeatletter
\providecommand*{\shuffle}{%
  \mathbin{\mathpalette\shuffle@{}}%
}
\newcommand*{\shuffle@}[2]{%
  \sbox0{$#1\vcenter{}$}%
  \kern .15\ht0 
  \rlap{\vrule height .25\ht0 depth 0pt width 2.5\ht0}%
  \raise.1\ht0\hbox to 2.5\ht0{%
    \vrule height 1.75\ht0 depth -.1\ht0 width .17\ht0 %
    \hfill
    \vrule height 1.75\ht0 depth -.1\ht0 width .17\ht0 %
    \hfill
    \vrule height 1.75\ht0 depth -.1\ht0 width .17\ht0 %
  }%
  \kern .15\ht0 
}
\makeatother

\makeatletter
\g@addto@macro\bfseries{\boldmath}
\makeatother

\usepackage{xparse}

\ExplSyntaxOn
\NewDocumentCommand{\omwb}{m m}
{
 \omm\left(\begin{smallmatrix}
 \omwb_print:n {#1} \\
 \omwb_print:n {#2}
 \end{smallmatrix}\right)
}
\seq_new:N \l_omwb_list_seq
\cs_new_protected:Npn \omwb_print:n #1
{
  \seq_set_split:Nnn \l_omwb_list_seq { , } { #1 }
  \seq_use:Nn \l_omwb_list_seq { , & }
}
\ExplSyntaxOff

\usepackage[pdfencoding=auto,bookmarks=true,hyperfigures=true]{hyperref}
\usepackage{float}
\usepackage[nosort]{cite}
\usepackage{lmodern}
\usepackage{amsbsy}
\usepackage[utf8]{inputenc}

\usepackage[font=small,labelfont=bf]{caption}

\usepackage[usenames,dvipsnames]{xcolor}
\definecolor{dgreen}{rgb}{0,0.70,0.30}
\definecolor{gold}{rgb}{0.85,.66,0}
\definecolor{purple}{rgb}{1.0,0.3,0.6}

\usepackage{tikz}
\usetikzlibrary{arrows}

\numberwithin{equation}{section}

\newcommand{\eqn}[1]{eq.~\eqref{#1}}
\newcommand{\Eqn}[1]{Equation~\eqref{#1}}
\newcommand{\eqns}[2]{eqs.~\eqref{#1} and~\eqref{#2}}

\newcommand{\rcite}[1]{ref.~\cite{#1}}
\newcommand{\rcites}[1]{refs.~\cite{#1}}

\providecommand{\href}[2]{#2}

\makeatletter
\def\mr@ignsp#1 {\ifx\:#1\@empty\else #1\expandafter\mr@ignsp\fi}%
\newcommand{\multiref}[1]{\begingroup
\xdef\mr@no@sparg{\expandafter\mr@ignsp#1 \: }%
\def\mr@comma{}%
\@for\mr@refs:=\mr@no@sparg\do{\mr@comma\def\mr@comma{,}\ref{\mr@refs}}%
\endgroup}
\renewcommand{\eqref}[1]{(\multiref{#1})}
\makeatother

\makeatletter
\newcommand{\namedref}[2]{\hyperref[#2]{#1~\ref*{#2}}}
\newcommand{\secref}{\@ifstar{\namedref{Section}}{\namedref{section}}}
\newcommand{\subsecref}{\@ifstar{\namedref{Subsection}}{\namedref{subsection}}}
\newcommand{\appref}{\@ifstar{\namedref{Appendix}}{\namedref{appendix}}}
\newcommand{\tabref}{\@ifstar{\namedref{Table}}{\namedref{table}}}
\newcommand{\figref}{\@ifstar{\namedref{Figure}}{\namedref{figure}}}
\makeatother

\providecommand{\hypersetup}[1]{}

\hypersetup{plainpages=false}
\hypersetup{pdfpagemode=UseNone}
\hypersetup{bookmarksnumbered=true}
\hypersetup{pdfstartview=FitH}
\hypersetup{colorlinks=false}
\hypersetup{citebordercolor={.5 1 .5}}
\hypersetup{urlbordercolor={.5 1 1}}
\hypersetup{linkbordercolor={1 .7 .7}}

\makeatletter
\let\@keywords\@empty
\let\@subject\@empty
\providecommand{\keywords}[1]{\gdef\@keywords{#1}}
\providecommand{\subject}[1]{\gdef\@subject{#1}}
\def\thetitle{\@title}
\def\theauthor{\@author}
\def\thesubject{\@subject}
\def\thedate{\@date}
\def\thekeywords{\@keywords}
\makeatother
\AtBeginDocument{
\hypersetup{pdftitle={\thetitle}}%
\hypersetup{pdfauthor={\theauthor}}%
\hypersetup{pdfsubject={\thesubject}}%
\hypersetup{pdfkeywords={\thekeywords}}%
}

\newif\ifnote 
\notetrue

\allowdisplaybreaks

\newcommand{\SL}{\mathrm{SL}}

\newcommand{\te}{\textrm}

\newcommand{\ap}{\alpha'}

\newcommand{\ZR}{\mathbb R}
\newcommand{\ZC}{\mathbb C}

\newcommand{\ZN}{\mathbb N}
\newcommand{\ZZ}{\mathbb Z}

\def\YM{\textrm{YM}}

\def\open{\textrm{open}}

\DeclareMathOperator{\omm}{\omega}

\DeclareMathOperator{\vol}{vol}

\usepackage{nicefrac}

\DeclareMathOperator{\Res}{Res}

\DeclareMathOperator{\Li}{Li}

\DeclareMathOperator{\id}{id}
\DeclareMathOperator{\im}{im}
\DeclareMathOperator{\PT}{PT}
\DeclareMathOperator{\KN}{KN}

\DeclareMathOperator{\adm}{adm}

\DeclareMathOperator{\bhF}{\boldsymbol{\hat{F}}}

\DeclareMathOperator{\bF}{\boldsymbol{F}}
\DeclareMathOperator{\bL}{\boldsymbol{L}}
\DeclareMathOperator{\bB}{\boldsymbol{B}}

\newcommand{\grEdge}[2]{{\tiny\begin{tikzpicture}
		\tikzset{vertex/.style = {shape=circle,draw,minimum size=0.5em}}
		\tikzset{edge/.style = {->,> = latex'}}
		\tikzset{baseline={(0, -0.4em)}}
		\node[vertex] (1) at (0,0) {$#1$};
		\node[vertex] (2) at (1,0) {$#2$};
		\draw[edge] (1) to (2);
		\end{tikzpicture}}}
\newcommand{\grEdgeLong}[2]{{\tiny\begin{tikzpicture}
		\tikzset{vertex/.style = {shape=circle,draw,minimum size=0.5em}}
		\tikzset{edge/.style = {->,> = latex'}}
		\tikzset{baseline={(0, -0.4em)}}
		\node[vertex] (1) at (0,0) {$#1$};
		\node[vertex] (2) at (1.5,0) {$#2$};
		\draw[edge] (1) to (2);
		\end{tikzpicture}}}
	\newcommand{\grEdgeReverse}[2]{{\tiny\begin{tikzpicture}
			\tikzset{vertex/.style = {shape=circle,draw,minimum size=0.5em}}
			\tikzset{edge/.style = {->,> = latex'}}
			\tikzset{baseline={(0, -0.4em)}}
			\node[vertex] (1) at (0,0) {$#1$};
			\node[vertex] (2) at (1,0) {$#2$};
			\draw[edge] (2) to (1);
			\end{tikzpicture}}}
\newcommand{\grDoubleEdge}[3]{{\tiny\begin{tikzpicture}
		\tikzset{vertex/.style = {shape=circle,draw,minimum size=0.5em}}
		\tikzset{edge/.style = {->,> = latex'}}
		\tikzset{baseline={(0, -0.4em)}}
		\node[vertex] (1) at (0,0) {$#1$};
		\node[vertex] (2) at (1,0) {$#2$};
		\node[vertex] (3) at (2,0) {$#3$};
		\draw[edge] (1) to (2);
		\draw[edge] (2) to (3);
		\end{tikzpicture}}}

\newcommand{\grDoubleEdgeSource}[3]{{\tiny\begin{tikzpicture}
		\tikzset{vertex/.style = {shape=circle,draw,minimum size=0.5em}}
		\tikzset{edge/.style = {->,> = latex'}}
		\tikzset{baseline={(0, -0.4em)}}
		\node[vertex] (1) at (0,0) {$#1$};
		\node[vertex] (2) at (1,0) {$#2$};
		\node[vertex] (3) at (2,0) {$#3$};
		\draw[edge] (2) to (1);
		\draw[edge] (2) to (3);
		\end{tikzpicture}}}
\newcommand{\grTriangleSink}[3]{{\tiny\begin{tikzpicture}
		\tikzset{vertex/.style = {shape=circle,draw,minimum size=0.5em}}
		\tikzset{edge/.style = {->,> = latex'}}
		\tikzset{baseline={(0, -0.4em)}}
		\node[vertex] (1) at (0,0.35) {$#1$};
		\node[vertex] (2) at (1,0.35) {$#2$};
		\node[vertex] (3) at (0.5,-0.35) {$#3$};
		\draw[edge] (1) to (3);
		\draw[edge] (2) to (3);
		\end{tikzpicture}}}
\newcommand{\grTriangleDSink}[3]{{\tiny\begin{tikzpicture}
		\tikzset{vertex/.style = {shape=circle,draw,minimum size=0.5em}}
		\tikzset{edge/.style = {->,> = latex'}}
		\tikzset{baseline={(0, -0.4em)}}
		\node[vertex] (1) at (0,0.35) {$#1$};
		\node[vertex] (2) at (1,0.35) {$#2$};
		\node[vertex] (3) at (0.5,-0.35) {$#3$};
		\draw[double,edge] (1) to (3);
		\draw[double,edge] (2) to (3);
		\draw[edge] (1) to (2);
		\end{tikzpicture}}}
\newcommand{\grTriangleSource}[3]{{\tiny\begin{tikzpicture}
		\tikzset{vertex/.style = {shape=circle,draw,minimum size=0.5em}}
		\tikzset{edge/.style = {->,> = latex'}}
		\tikzset{baseline={(0, -0.4em)}}
		\node[vertex] (1) at (0,0.35) {$#1$};
		\node[vertex] (2) at (1,0.35) {$#2$};
		\node[vertex] (3) at (0.5,-0.35) {$#3$};
		\draw[edge] (1) to (2);
		\draw[edge] (1) to (3);
\end{tikzpicture}}}
\newcommand{\grTriangleClock}[3]{{\tiny\begin{tikzpicture}
		\tikzset{vertex/.style = {shape=circle,draw,minimum size=0.5em}}
		\tikzset{edge/.style = {->,> = latex'}}
		\tikzset{baseline={(0, -0.4em)}}
		\node[vertex] (1) at (0,0.35) {$#1$};
		\node[vertex] (2) at (1,0.35) {$#2$};
		\node[vertex] (3) at (0.5,-0.35) {$#3$};
		\draw[edge] (1) to (2);
		\draw[edge] (2) to (3);
		\end{tikzpicture}}}

\newcommand{\grSquareSink}[4]{{\tiny\begin{tikzpicture}
		\tikzset{vertex/.style = {shape=circle,draw,minimum size=0.5em}}
		\tikzset{edge/.style = {->,> = latex'}}
		\tikzset{baseline={(0, -0.4em)}}
		\node[vertex] (1) at (0,-0.5) {$#1$};
		\node[vertex] (2) at (1,-0.5) {$#2$};
		\node[vertex] (3) at (1,0.5) {$#3$};
		\node[vertex] (4) at (0,0.5) {$#4$};
		\draw[edge] (1) to (2);
		\draw[edge] (3) to (2);
		\draw[edge] (4) to (3);
		\end{tikzpicture}}}
	
\newcommand{\grSquaren}[4]{{\tiny\begin{tikzpicture}
		\tikzset{vertex/.style = {shape=circle,draw,minimum size=0.5em}}
		\tikzset{edge/.style = {->,> = latex'}}
		\tikzset{baseline={(0, -0.4em)}}
		\node[vertex] (1) at (0,-0.5) {$#1$};
		\node[vertex] (2) at (1,-0.5) {$#2$};
		\node[vertex] (3) at (1,0.5) {$#3$};
		\node[vertex] (4) at (0,0.5) {$#4$};
		\draw[edge] (4) to (3);
		\draw[edge] (3) to (2);
		\draw[edge] (4) to (1);
		\end{tikzpicture}}}
\newcommand{\grSquarec}[4]{{\tiny\begin{tikzpicture}
		\tikzset{vertex/.style = {shape=circle,draw,minimum size=0.5em}}
		\tikzset{edge/.style = {->,> = latex'}}
		\tikzset{baseline={(0, -0.4em)}}
		\node[vertex] (1) at (0,-0.5) {$#1$};
		\node[vertex] (2) at (1,-0.5) {$#2$};
		\node[vertex] (3) at (1,0.5) {$#3$};
		\node[vertex] (4) at (0,0.5) {$#4$};
		\draw[edge] (4) to (3);
		\draw[edge] (1) to (2);
		\draw[edge] (4) to (1);
		\end{tikzpicture}}}
\newcommand{\grSquareClock}[4]{{\tiny\begin{tikzpicture}
		\tikzset{vertex/.style = {shape=circle,draw,minimum size=0.5em}}
		\tikzset{edge/.style = {->,> = latex'}}
		\tikzset{baseline={(0, -0.4em)}}
		\node[vertex] (1) at (0,-0.5) {$#1$};
		\node[vertex] (2) at (1,-0.5) {$#2$};
		\node[vertex] (3) at (1,0.5) {$#3$};
		\node[vertex] (4) at (0,0.5) {$#4$};
		\draw[edge] (1) to (3);
		\draw[edge] (3) to (2);
		\draw[edge] (4) to (3);
		\end{tikzpicture}}}
	
\newcommand{\grSquareSource}[4]{{\tiny\begin{tikzpicture}
		\tikzset{vertex/.style = {shape=circle,draw,minimum size=0.5em}}
		\tikzset{edge/.style = {->,> = latex'}}
		\tikzset{baseline={(0, -0.4em)}}
		\node[vertex] (1) at (0,-0.5) {$#1$};
		\node[vertex] (2) at (1,-0.5) {$#2$};
		\node[vertex] (3) at (1,0.5) {$#3$};
		\node[vertex] (4) at (0,0.5) {$#4$};
		\draw[edge] (1) to (3);
		\draw[edge] (1) to (2);
		\draw[edge] (4) to (3);
		\end{tikzpicture}}}
\newcommand{\grSquareDSink}[4]{{\tiny\begin{tikzpicture}
		\tikzset{vertex/.style = {shape=circle,draw,minimum size=0.5em}}
		\tikzset{edge/.style = {->,> = latex'}}
		\tikzset{baseline={(0, -0.4em)}}
		\node[vertex] (1) at (0,-0.5) {$#1$};
		\node[vertex] (2) at (1,-0.5) {$#2$};
		\node[vertex] (3) at (1,0.5) {$#3$};
		\node[vertex] (4) at (0,0.5) {$#4$};
		\draw[double,edge] (1) to (2);
		\draw[double,edge] (3) to (2);
		\draw[edge] (1) to (3);
		\draw[edge] (4) to (3);
		\end{tikzpicture}}}
\newcommand{\grSquareDClock}[4]{{\tiny\begin{tikzpicture}
		\tikzset{vertex/.style = {shape=circle,draw,minimum size=0.5em}}
		\tikzset{edge/.style = {->,> = latex'}}
		\tikzset{baseline={(0, -0.4em)}}
		\node[vertex] (1) at (0,-0.5) {$#1$};
		\node[vertex] (2) at (1,-0.5) {$#2$};
		\node[vertex] (3) at (1,0.5) {$#3$};
		\node[vertex] (4) at (0,0.5) {$#4$};
		\draw[double,edge] (1) to (2);
		\draw[double,edge] (3) to (2);
		\draw[edge] (1) to (3);
		\draw[edge] (4) to (1);
		\end{tikzpicture}}}
\newcommand{\grSquareClockClock}[4]{{\tiny\begin{tikzpicture}
		\tikzset{vertex/.style = {shape=circle,draw,minimum size=0.5em}}
		\tikzset{edge/.style = {->,> = latex'}}
		\tikzset{baseline={(0, -0.4em)}}
		\node[vertex] (1) at (0,-0.5) {$#1$};
		\node[vertex] (2) at (1,-0.5) {$#2$};
		\node[vertex] (3) at (1,0.5) {$#3$};
		\node[vertex] (4) at (0,0.5) {$#4$};
		\draw[edge] (3) to (2);
		\draw[edge] (1) to (3);
		\draw[edge] (4) to (1);
		\end{tikzpicture}}}
\newcommand{\grSquareClockSouce}[4]{{\tiny\begin{tikzpicture}
		\tikzset{vertex/.style = {shape=circle,draw,minimum size=0.5em}}
		\tikzset{edge/.style = {->,> = latex'}}
		\tikzset{baseline={(0, -0.4em)}}
		\node[vertex] (1) at (0,-0.5) {$#1$};
		\node[vertex] (2) at (1,-0.5) {$#2$};
		\node[vertex] (3) at (1,0.5) {$#3$};
		\node[vertex] (4) at (0,0.5) {$#4$};
		\draw[edge] (1) to (2);
		\draw[edge] (1) to (3);
		\draw[edge] (4) to (1);
		\end{tikzpicture}}}
\newcommand{\grSquareDSource}[4]{{\tiny\begin{tikzpicture}
		\tikzset{vertex/.style = {shape=circle,draw,minimum size=0.5em}}
		\tikzset{edge/.style = {->,> = latex'}}
		\tikzset{baseline={(0, -0.4em)}}
		\node[vertex] (1) at (0,-0.5) {$#1$};
		\node[vertex] (2) at (1,-0.5) {$#2$};
		\node[vertex] (3) at (1,0.5) {$#3$};
		\node[vertex] (4) at (0,0.5) {$#4$};
		\draw[double,edge] (1) to (2);
		\draw[double,edge] (3) to (2);
		\draw[edge] (4) to (1);
		\draw[edge] (4) to (3);
		\end{tikzpicture}}}
\newcommand{\grSquareDDSink}[4]{{\tiny\begin{tikzpicture}
		\tikzset{vertex/.style = {shape=circle,draw,minimum size=0.5em}}
		\tikzset{edge/.style = {->,> = latex'}}
		\tikzset{baseline={(0, -0.4em)}}
		\node[vertex] (1) at (0,-0.5) {$#1$};
		\node[vertex] (2) at (1,-0.5) {$#2$};
		\node[vertex] (3) at (1,0.5) {$#3$};
		\node[vertex] (4) at (0,0.5) {$#4$};
		\draw[double,edge] (1) to (2);
		\draw[edge] (4) to (1);
		\draw[double,edge] (3) to (2);
		\draw[double,edge] (1) to (3);
		\draw[double,edge] (4) to (3);
		\end{tikzpicture}}}
\vspace*{0.2cm}
\title{\textbf{
    A note on the Drinfeld associator for genus-zero superstring amplitudes in twisted de Rham theory
    }}
\author{
Andr\'e Kaderli$^{\,\textit{a},\textit{b}}$}
\date{\today}

\begin{document}
\pdfbookmark[1]{Title Page}{title} \thispagestyle{empty}
\begin{flushright}
  \verb!HU-EP-19/40!\\
  \verb!HU-Mathematik-2019-09!
\end{flushright}
\vspace*{0.4cm}
\begin{center}%
  \begingroup\LARGE\bfseries\thetitle\par\endgroup
\vspace{1.0cm}

\begingroup\large\theauthor\par\endgroup
\vspace{9mm}
\begingroup\itshape
$^{\te{a}}$Institut f\"ur Mathematik und Institut f\"ur Physik, Humboldt-Universit\"at zu Berlin\\
IRIS Adlershof, Zum Gro\ss{}en Windkanal 6, 12489 Berlin, Germany
\par\endgroup
\vspace{3mm}
\begingroup\itshape
$^{\te{b}}$Max-Planck-Institut f\"ur Gravitationsphysik, Albert-Einstein-Institut\\
Am M\"uhlenberg 1, 14476 Potsdam, Germany
\par\endgroup
\vspace{3mm}

\vspace{1.0cm}

\begingroup\ttfamily
kaderlia@physik.hu-berlin.de
\par\endgroup

\vspace{1.2cm}

\bigskip

\textbf{Abstract}\vspace{5mm}

\begin{minipage}{13.4cm}
The string corrections of tree-level open-string amplitudes can be described by Selberg integrals satisfying a Knizhnik-Zamolodchikov (KZ) equation. This allows for a recursion of the $\alpha'$-expansion of tree-level string corrections in the number of external states using the Drinfeld associator.\\
While the feasibility of this recursion is well-known, we provide a mathematical description in terms of twisted de Rham theory and intersection numbers of twisted forms. In particular, this leads to purely combinatorial expressions for the matrix representation of the Lie algebra generators appearing in the KZ equation in terms of directed
graphs. This, in turn, admits efficient algorithms for symbolic and numerical computations using adjacency matrices of directed graphs and is a crucial step towards analogous recursions and algorithms at higher genera. 
\end{minipage}

\vspace*{4cm}

\end{center}

\newpage

\setcounter{tocdepth}{2}
\tableofcontents


\section{Introduction}
\label{sec:introduction}

Tree-level amplitudes of superstrings furnish a prime example of the richness of the mathematical structures underlying scattering amplitudes. Recent developments \cite{Brown:2018omk1,Brown:2019wna,Mizera:2017cqs,Mizera:2019gea} revealed the particular importance of twisted de Rham theory, which seems to be a language suitable to express various results for scattering amplitudes in quantum field and string theory in a rigorous mathematical framework. Such fundamental descriptions may reveal new insights, connect known results and promote the understanding of physical phenomena in the context of amplitude calculations.

The calculation of open tree-level superstring amplitudes is an important problem since it might shed some light on the calculation of more complicated scattering amplitudes in physical (quantum field) theories. In particular, recursive methods which generate solutions using linear algebra exclusively instead of direct evaluations of the integrals are of special interest, since matrix multiplications can be readily implemented in computer algebra systems and efficiently evaluated numerically. Examples of such techniques can be found in \rcites{Broedel:2013aza,Mafra:2016mcc}, where tree-level amplitude recursions for the $\alpha'$-expansion of superstring amplitudes are proposed.

The recursion described in \rcite{Broedel:2013aza} is based on the mathematical structure of Selberg integrals \cite{Selberg44,aomoto1987,Terasoma} occurring in tree-level open-superstring amplitudes. However, the relevant matrices necessary for the recursion are not provided and it has not yet been formulated in terms of twisted de Rham theory. In these notes, we state this recursion relation in terms of intersection numbers and add some observations crucial for the understanding of the recursive mechanism. We show in particular that the required matrices are braid matrices and describe a graphical algorithm to calculate them explicitly. Since the relevant properties of the Selberg integrals can be recovered in a certain class of genus-one integrals relevant for loop-level amplitudes, this investigation helps paving the way for amplitude recursions at higher genera. In particular, this work is accompanied by the article \cite{BroedelKaderli}, in which such a genus-one recursion is proposed and an explicit derivation of how to relate the one-loop string corrections to the genus-zero integrals discussed in the present article is given. 

This article is structured as follows: in \secref{sec:preliminaries}, we introduce the mathematical and physical preliminaries by providing a brief introduction to twisted de Rham theory and an overview of tree-level open-superstring amplitudes. Furthermore, we review the Knizhnik-Zamolodchikov (KZ) equation and the Drinfeld associator, which are the fundamental ingredients of the recursion. In \secref{sec:recursion}, we present and reformulate the recursion of \rcite{Broedel:2013aza} in the language of twisted de Rham theory and thereby provide a general formalism delivering the missing matrix representation of the Lie algebra generators which form the alphabet used in the construction of the Drinfeld associator.

\section{Background: string amplitudes in twisted de Rham theory}
\label{sec:preliminaries}
The purpose of this section is to introduce the mathematical and physical preliminaries. However, this introduction remains on the level of a brief overview and we recommend consulting the literature stated below for a more complete and rigorous treatment of the relevant topics. 

\subsection{Twisted de Rham theory}\label{ssec:twistedDeRhamCohomology}
We would like to get started with a brief introduction to twisted de Rham theory, whose main content is the investigation of differential forms with multi-valued coefficients. Such structures are omnipresent in string amplitude calculations, where certain branch choices of the multi-valued coefficients lead to the physical amplitudes. We follow the lines of \rcites{Mizera:2017cqs,Mizera:2019gea,Frellesvig:2019uqt} for the statements about twisted de Rham theory and their connection to superstring amplitudes. The fundamental definitions and their properties are primarily based on \rcite{AomotoKita11}, where the whole theory is constructed rigorously.

The central objects in twisted de Rham theory are integrals of the form
\begin{equation}\label{preliminaries:integral}
\int_{\Delta} u\, \varphi\,,
\end{equation}
where 
\begin{equation}\label{preliminaries:u}
u(z)=\prod_i^k f_i(z)^{\alpha_i}\,,\qquad \alpha_i \in \ZC\setminus \ZZ\,,
\end{equation}
is a multi-valued product of polynomials $f_i(z)=f_i(z_1,z_2,\dots,z_n)$ defined on the $n$-dimensional affine variety
\begin{equation*}
M=\ZC^n\setminus D\,, \qquad D=\bigcup_{i=1}^k D_i\,,\qquad D_i = \{f_i(z)=0\}\,.
\end{equation*}
The $n$-dimensional region of integration $\Delta$ is an $n$-simplex with boundaries on the divisor $D$ and, thus, constitutes a topological cycle. The factor $\varphi$ is a smooth $n$-form on $M$.

Since the function $u$ is multi-valued, instead of working on the covering space of $M$, a certain branch $u_{\Delta}$ of $u$ on $\Delta$ can be specified to render the integral \eqref{preliminaries:integral} well-defined. This specification accounts for the "twist" in twisted de Rham theory and is noted by specifying the integration region via 
\begin{align}\label{preliminaries:integralTwisted}
\int_{\Delta\otimes u_{\Delta}}\varphi&=\int_{\Delta}\left[\text{fixed branch }u_{\Delta}\text{ of }u\text{ on }\Delta\right]\,\varphi\,.
\end{align}
In the above definition, the integration region $\Delta$ is said to be \textit{loaded} with $u_{\Delta}$. Considering a smooth $(n-1)$-form $\varphi$ and defining the single-valued one-form 
\begin{align*}
\omega &= d \log u
\end{align*}
as well as the integrable connection $\nabla_{\omega}$ by the equation
\begin{align}\label{preliminaries:nabla}
\nabla_{\omega}\varphi&= d \varphi+\omega \wedge \varphi\,,
\end{align}
Stokes' theorem implies 
\begin{equation}\label{preliminaries:StokesImplication}
\int_{\partial \Delta}u_{\Delta}\, \varphi=\int_{ \Delta}d\left(u_{\Delta}\, \varphi
\right) =\int_{ \Delta\otimes u_{\Delta}}\nabla_{\omega} \varphi\,.
\end{equation}
Note that \eqn{preliminaries:nabla} indeed defines an integrable connection, since it implies that
\begin{align*}
\nabla_{\omega}\circ \nabla_{\omega}&=0\,.
\end{align*}
Relation \eqref{preliminaries:StokesImplication} can be entirely expressed in terms of loaded integration domains if the boundary operator $\partial_\omega$ for the $n$-simplex $\Delta=\langle 01\cdots n\rangle$ is defined as follows
\begin{align*}
\partial_{\omega}\left(\langle 01\cdots n\rangle\otimes u_{\langle 01\cdots n\rangle}\right)&=\sum_{i=0}^n(-1)^i\langle 01\cdots \hat{i}\cdots n\rangle\otimes u_{\langle 01\cdots \hat{i}\cdots n\rangle}\,,
\end{align*}
where $u_{\langle 01\cdots \hat{i}\cdots n\rangle}$ is the restriction of the branch 
$u_{\langle 01\cdots n\rangle}$ of $u$ to the $i$-th face of $\langle 01\cdots n\rangle$ and $\hat{i}$ denotes that we omit the $i$-th coordinate: $\langle 01\cdots \hat{i}\cdots n\rangle=\langle 01\cdots i-1\, i+1 \cdots n\rangle$. This definition implies in particular that
\begin{align*}
\partial_{\omega}\circ \partial_{\omega}&=0\,.
\end{align*}
Using the above definitions, the twisted version of Stokes' theorem can be expressed as
\begin{align*}
	\int_{\Delta\otimes u_{\Delta}} \nabla_{\omega}\varphi &= \int_{\partial_{\omega}\left(\Delta\otimes u_{\Delta}\right)} \varphi \,.
\end{align*}

Since $u$ vanishes on the boundary $\partial \Delta$ of the $n$-simplex, adding $\nabla_{\omega}\xi$ to $\varphi$, where $\xi$ is an $(n-1)$-form on $M$, does not change the result of the integral~\eqref{preliminaries:integralTwisted}. Therefore, it is convenient to define the quotient vector space 
\begin{align*}
H^n(M,\nabla_{\omega})&=\ker(\nabla_{\omega})/\im(\nabla_{\omega})\,,
\end{align*}
called the $n$-th twisted cohomology. Its elements are referred to as \textit{twisted forms} or \textit{twisted cocycles}, which we denote according to the notation of \rcites{Mizera:2019gea,Frellesvig:2019uqt} by $\langle \varphi |\in H^n(M,\nabla_{\omega})$. Moreover, a dual\footnote{Below, the notion of "duality" among $H^n(M,\nabla_{\omega})$, $H^n(M,\nabla^{\vee}_{\omega})$ and $H_n(M,\mathcal{L}_{\omega})$ as well as $H_n(M,\mathcal{L}^{\vee}_{\omega})$ is discussed by introducing the associated non-degenerate pairings.} vector space $H^n(M,\nabla^{\vee}_{\omega})$ can be defined by replacing the connection $\nabla_{\omega}$ with $\nabla^{\vee}_{\omega}=\nabla_{-\omega}$ and its elements are denoted by $ |\varphi \rangle \in H^n(M,\nabla^{\vee}_{\omega})$ .

Having introduced a twisted version of de Rham cohomology, a twisted analogue of homology can be defined via a brief detour to homology with coefficients. This formalism allows to keep track of the local branches of $u$ in the integration regions $\Delta\otimes u_{\Delta}$ and can be introduced as follows: the differential equation 
\begin{align}\label{preliminaries:degLocalSystVee}
\nabla^{\vee}_{\omega} \xi&=0
\end{align}  
admits the formal solution
\begin{equation*}
\xi=c u\,,\qquad c\in\ZC\,.
\end{equation*}
Therefore, the space of local solutions has the complex dimension one. For a locally finite open cover $X=\bigcup_i U_i$, two local solutions $\xi_i$, $\xi_j$ on $U_i$ and $U_j$, respectively, with $U_{ij}=U_i\cap U_j\neq \varnothing$, satisfy 
\begin{equation*}
\xi_i=\zeta_{ij} \xi_j
\end{equation*} 
for some $\zeta_{ij}\in \ZC$. On the other hand, any local solution $\xi$ on $U_{ij}$ can be expressed as 
\begin{equation*}
\xi=\zeta_i\xi_i=\zeta_j \xi_j
\end{equation*}
for some $\zeta_i,\zeta_j\in\ZC$ such that $\zeta_i=\zeta_{ij}^{-1}\zeta_j$. Therefore, the local solutions of \eqn{preliminaries:degLocalSystVee} define upon gluing together the fibres $\{\zeta_i\}$ by the transition functions $\{\zeta_{ij}^{-1}\}$ a flat line bundle $\mathcal{L}_{\omega}$. Hence, the boundary operator $\partial_{\omega}$ defines a map between chain groups with coefficients in $\mathcal{L}_{\omega}$. This leads to the definition of the $n$-th twisted homology group 
\begin{align*}
H_n(M,\mathcal{L}_{\omega})&=\ker(\partial_{\omega})/\im(\partial_{\omega})\,,
\end{align*}
where the elements are called \textit{twisted cycles} and are denoted by $|\sigma]\in H_n(M,\mathcal{L}_{\omega})$. A dual vector space $H_n(M,\mathcal{L}^{\vee}_{\omega})$ with elements $[\sigma|\in H_n(M,\mathcal{L}^{\vee}_{\omega})$ is analogously defined by the dual line bundle $\mathcal{L}^{\vee}_{\omega}$ of $\mathcal{L}_{\omega}$ which, in turn, is defined by the local solutions of the differential equation 
\begin{align*}
\nabla_{\omega}\xi&=0
\end{align*} 
with generic solutions of the form $c u^{-1}$ for $c\in \ZC$ and hence, with the associated transition functions $\{\zeta_{ij}\}$.

In order to define convergent integrals with twisted cycles and twisted forms for a possibly non-compact manifold $M$, it is convenient to introduce the $n$-th locally finite twisted homology group $H^{\text{lf}}_n(M,\mathcal{L}_{\omega})$, which is constructed in analogy to $H_n(M,\mathcal{L}_{\omega})$ with the simplices required to be locally finite. Similarly, the $n$-th compactly supported twisted cohomology $H^n_{\text{c}}(M,\nabla_{\omega})$ is defined to be the twisted cohomology of differential forms with compact support.

The vector spaces defined above are related by various dualities leading to non-degenerate pairings. Important examples include the following non-degenerate bilinear forms \cite{AomotoKita11}:\begin{itemize}
	\item the pairing of a twisted form and a locally finite cycle
	\begin{equation}
	\begin{matrix}
	H^n(M,\nabla_{\omega})\times H^{\text{lf}}_n(M,\mathcal{L}_{\omega})&\rightarrow & \ZC\\
	(\langle \varphi |,|\sigma ])&\mapsto&\langle \varphi |\sigma ]=\int_{\sigma} u\, \varphi\,,
	\end{matrix}
	\end{equation}
	\item the pairing of a twisted form with compact support and a twisted form
	\begin{equation}\label{intNumb:Forms}
	\begin{matrix}
	H^n(M,\nabla_{\omega}) \times H^n_{\text{c}}(M,\nabla^{\vee}_{\omega})&\rightarrow & \ZC\\
	(\langle \varphi |,|\psi\rangle)&\mapsto&\langle \varphi |\psi\rangle=\int_{M} \varphi\wedge \psi\,,
	\end{matrix}
	\end{equation}
	called intersection number of twisted forms,
	\item and the pairing of a twisted cycle with a locally finite twisted cycle
	\begin{equation}
	\begin{matrix}
	H_n(M,\mathcal{L}^{\vee}_{\omega})\times H^{\text{lf}}_n(M,\mathcal{L}_{\omega})&\rightarrow & \ZC\\
	([\sigma |,|\tau ])&\mapsto&[\sigma |\tau ]\,,
	\end{matrix}
	\end{equation}
	which is defined to be the intersection number \cite{KitaYoshida94} of the two cycles.
\end{itemize}
The non-degeneracy of the last two examples is a consequence of the duality of the vector spaces $H^n(M,\nabla_{\omega})$ and $H^n(M,\nabla^{\vee}_{\omega})$ as well as $H_n(M,\mathcal{L}_{\omega})$ and $H_n(M,\mathcal{L}_{\omega}^{\vee})$, which was mentioned above. Note that as a consequence of a theorem in twisted de Rham theory, the dimensions of the twisted homology and cohomology coincide $\dim\left( H^n(M,\nabla_{\omega})\right)=\dim\left( H_n(M,\mathcal{L}_{\omega})\right)$ \cite{AomotoKita11}. The same holds for the dual vector spaces, as well as the locally finite homology and the compactly supported twisted de Rham cohomology.

Since twisted cycles and twisted forms are vectors, they are, in particular, independent of the choice of a basis in the corresponding vector spaces and their representation with respect to a given basis has to change accordingly under a change of basis. Such a basis transformation can be described as follows in twisted de Rham cohomology (and similarly for the twisted homology): let $\{\langle \varphi_i|\}$ and $\{| \psi_i\rangle\}$ be bases of $H^n(M,\nabla_{\omega})$ and $H^n(M,\nabla^{\vee}_{\omega})$, respectively. The basis elements $\langle\varphi_i|$ can be expressed in terms of another basis $\{\langle \xi_i|\}$ of $H^n(M,\nabla_{\omega})$ by the master decomposition formula \cite{Mastrolia:2018uzb}
\begin{equation}\label{twistedTheory:basisTrafo}
\langle \varphi_i|=\sum_{j=1}^d b_{ij} \langle \xi_j|\,,\qquad b_{ij}= \sum_{k=1}^d \langle \varphi_i|\psi_k\rangle (\boldsymbol{C}^{-1})_{kj}\,,
\end{equation}
where $d=\dim\left(H^n(M,\nabla_{\omega})\right)$ and $\boldsymbol{C}$ is the matrix of intersection numbers of the twisted forms
\begin{align}\label{twistedTheory:basisTrafoC}
(\boldsymbol{C})_{ij}&=\langle \xi_i|\psi_j\rangle\,.
\end{align}

In order to distinguish between differential forms, integrals and twisted forms, we adopt the following conventions: differential forms are generally denoted by small letters $f$ and an integral of a differential form over a previously specified integration domain $\Delta$ by the corresponding capital letter $F=\int_{\Delta} f$, i.e.\ the differential form in $F$ is $f$. The twisted cohomology class of $f$ is denoted by the twisted form $\langle f |$, such that $\langle f|\Delta\rangle =\int_{\Delta\otimes u_{\Delta}} f$. Moreover, a vector of differential forms, integrals and twisted forms is denoted by the corresponding bold letter $\boldsymbol{f}$, $\boldsymbol{F}$ or $\langle \boldsymbol{f} |$, respectively. 

\subsection{Open-superstring amplitudes at tree level}
Having introduced the relevant mathematical setup for this article, in this subsection, we are going to introduce the corresponding physical objects. We review different representations of the final results of colour-ordered, tree-level open-superstring amplitudes involving $N$ massless states, calculated in \rcites{Mafra:2011nv,Mafra:2011nw} using methods from pure spinor cohomology \cite{Mafra:2010jq}, and their connection to twisted de Rham theory according to \rcites{Mizera:2017cqs,Mizera:2019gea,Frellesvig:2019uqt}. Since we mostly consider tree-level amplitudes in open-superstring theory, we generally refer to them as amplitudes without further specification and will be more specific in case amplitudes arising in other theories are considered.

The worldsheet of $N$ interacting closed strings at tree level can be mapped by conformal symmetry to a Riemann surface of genus zero $\Sigma \cong \ZC P^1$. External closed-string states are mapped to vertex operators leading to $N$ punctures on the Riemann sphere $\ZC P^1$. However, this configuration exhibits an $\SL(2,\ZC)$ redundancy, which can be used to fix three insertion points, usually chosen to be $(z_1,z_{N-1},z_N)=(0,1,\infty)$, leaving a constant factor of $(z_1-z_{N-1})(z_{N-1}-z_N)(z_1-z_N)$ in the amplitude integral due to the Faddeev--Popov Jacobian. While closed-string amplitudes are integrals over the full Riemann sphere $\ZC P^1$, open strings propagating in spacetime can be described by one hemisphere of $\ZC P^1$: a disk topology. The corresponding open-string amplitudes are integrals on the disk boundary with $N$ punctures, where the boundary is represented as the real line (plus infinity) $\ZR P^1\subset \ZC P^1$ of the Riemann sphere with the corresponding redundancy from the conformal Killing group $\SL(2,\ZR)$ of the disk topology. Therefore, the relevant geometry which includes closed- and open-string integrals at tree level is the moduli space of $N$-punctured Riemann spheres
\begin{align}\label{sec:amp:moduliSpace}
\mathcal{M}_{0,N}&=\text{Conf}_N(\ZC P^1)/\SL(2,\ZC)\nonumber\\
&=\{(z_2,z_3,\dots,z_{N-2})\in (\ZC P^1)^{N-3}|z_i\neq z_1,z_j,z_{N-1},z_N\text{ for all } i\neq j\in\{2,3,\dots,N-2\}\}
\end{align}
and the natural labelling of the insertion points is given by 
\begin{equation}\label{amplitudes:NaturalLabelling}
0=z_1<z_2<\dots<z_{N-1}=1\,.
\end{equation}

\begin{figure}[b]
	\centering
	\begin{tikzpicture}[scale=0.8]	
	\begin{scope}[yshift=-1cm,xshift=-6.5cm]
	\draw[->] (-3,0) to (3,0);
	\draw[->] (-2,0) to (-2,2.5);
	\draw[blue,line width=0.25mm] (-2,0) to (2,0);
	\draw[line width=0.25mm] (-2,0.1) to (-2,-0.1);
	\draw[line width=0.25mm,dotted] (-0.5,0.2) to (0.5,0.2); 
	\draw (-2.5,-0.35) node{{\small  $0=x_2$}};
	\draw[line width=0.25mm](-1.5,0.1) to (-1.5,-0.1);
	\draw (-1.4,-0.4) node{{\small  $x_n$}};
	\draw (-0.5,-0.4) node{{\small  $x_{n-1}$}};
	\draw[line width=0.25mm](-1,0.1) to (-1,-0.1);
	\draw (0.9,-0.4) node{{\small  $x_5$}};
	\draw[line width=0.25mm](1,0.1) to (1,-0.1);
	\draw (1.6,-0.4) node{{\small  $x_4$}};
	\draw[red,line width=0.3mm](1.5,0.1) to (1.5,-0.1);
	\draw[red,line width=0.25mm,<->](1.2,0.4) to (1.8,0.4);
	\draw (2.8,-0.35) node{{\small  $x_3=1$}};
	\draw[line width=0.25mm](2,0.1) to (2,-0.1);
	\draw (2.5,2) node{{\small  $x_{1}=\infty$}};
	\draw[dashed] (2.3,0) arc (0:180:2.3cm and 2.3cm); 
	\end{scope}
	\end{tikzpicture}
	\caption{The non-standard labelling convention used for the $n$ punctures on the Riemann sphere. The variable position $x_4$ is the auxiliary marked point in the amplitude recursion.}
	\label{figRiemannSphere}
\end{figure}
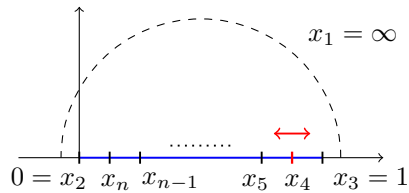

In order to formulate the amplitude recursion for open tree-level amplitudes in \secref{sec:recursion}, an auxiliary point $z_0$ at the position $z_{N-2}<z_0<z_{N-1}$ has been introduced in \rcite{Broedel:2013aza}, leading to $N+1$ punctures on the boundary of the disk. If this puncture $z_0$ is included, it turns out to be more convenient to introduce another labelling convention than the one given in \eqn{amplitudes:NaturalLabelling}. This second labelling is adapted to the recursive differential equations satisfied by Selberg integrals associated to the $n=N+1$ times punctured boundary of the disk and is denoted by $x_i$ with the gauge fixing $(x_1,x_2,x_3)=(\infty,0,1)$ and the ordering $\prec$ defined by
\begin{equation}\label{sec:rec:ordering}
0=x_2<x_n<x_{n-1}<\dots<x_3=1\,,\qquad i\prec j\Rightarrow x_i<x_j\,,
\end{equation}
as depicted in \figref{figRiemannSphere}. 

Since $x_4=z_0$ is the auxiliary point parametrising the integration region of the iterated integrals in the tree-level amplitude recursion, $x_4$ will serve as the variable in the relevant differential equation. The two labellings are identified as follows:
\begin{equation}\label{amplitudes:BothCoords}
0=z_1=x_2< z_2 = x_n< z_3=x_{n-1}<\dots<z_{N-2}=x_5< z_0=x_4< z_{N-1}=x_3=1
\end{equation}
and $z_N=x_1=\infty$. Thus the corresponding permutation is
\begin{equation*}
\sigma_{\text{label}} (4,2,n,n-1,\dots,5,3,1)=(0,1,2,3,\dots,N-2,N-1,N)\,,
\end{equation*}
such that $ x_i=\sigma_{\text{label}}z_{i}=z_{\sigma_{\text{label}}(i)}$. Similarly, we denote the dimensionless Mandelstam variables corresponding to the labelling $z_i$ defined in \eqn{amplitudes:NaturalLabelling} by
\begin{equation*}
s_{i_1 i_2\dots i_m}=s_{i_1, i_2,\dots, i_m}=\alpha'(k_{i_1}+k_{i_2}+\dots+k_{i_m})^2
\end{equation*}
where $k_{i}$ denotes the external on-shell momentum corresponding to the insertion point $z_i$ and where $\alpha'$ is the universal Regge slope, proportional to the inverse string tension. In terms of the labelling $x_i$ or the ordering $\prec$, respectively, the Mandelstam variables are denoted by
\begin{equation}\label{sec:rec:orderingM}
t_{i_1 i_2\dots i_m}=\sigma_{\text{label}}s_{i_1 i_2\dots i_m}\,.
\end{equation}

\subsubsection{Colour-ordered amplitudes}
Colour-ordered, tree-level superstring amplitudes of $N$ massless, open-string states are given by \cite{Mafra:2011nv,Mafra:2011nw}
\begin{align}\label{amplitudes:generalForm}
A_{\open}(\Pi,\ap)=\sum_{\sigma\in S_{N-3}} F_{\Pi}^{\sigma}(\ap)A_{\YM}\left(1,\sigma\left(2,\dots,N-2\right),N-1,N \right)\,,
\end{align}
where the amplitudes $A_{\YM}$ constitute a basis of Yang--Mills amplitudes and $F_{\Pi}^{\sigma}$ denotes the string corrections, which are given by a generalised Euler integral, a linear combination of Selberg integrals \cite{Selberg44},
\begin{align}\label{amplitudes:FSigmaPi}
F_{\Pi}^{\sigma}&=(-1)^{N-3}\prod_{i=2}^{N-2}\int_{D(\Pi)} dz_i \, \KN\,\sigma \left(\prod_{k=2}^{N-2}\sum_{m=1}^{k-1}\frac{s_{mk}}{z_{mk}} \right)\,,
\end{align}
where $z_{ij}=z_{i,j}=z_i-z_j$ and the Koba--Nielsen factor is denoted by 
\begin{equation}\label{amplitudes:KobaNielsen}
\KN=\prod _{1\leq i<j\leq N-1} |z_{ij}|^{s_{ij}}\,.
\end{equation}
Note that the Koba--Nielsen factor corresponds to the multi-valued factor $u(z)$ in the string amplitudes mentioned at the beginning of \subsecref{ssec:twistedDeRhamCohomology}, where the relevant branch for the string corrections is chosen to be the real-valued function defined in \eqn{amplitudes:KobaNielsen}. The permutation $\sigma \in S_{N-3}$ in \eqn{amplitudes:FSigmaPi} acts on all the indices $2\leq i\leq N-2$ within the brackets to the right of $\sigma$. The integration domain $D(\Pi)$ is determined by
$\Pi\in S_{N-3}$ according to
$z_{\Pi(i)}<z_{\Pi(i+1)}$ for $2\leq i\leq N-2$.

Using integration by parts, the integrals $F_{\Pi}^{\sigma}$ can be represented in $N-2$ equivalent ways. These representations are parametrised by $1\leq \nu\leq N-2$ and given by the integrals \cite{Broedel:2013tta}
\begin{align}\label{amplitudes:FSigmaPinu}
F_{\Pi,\nu}^{\sigma}&=(-1)^{N-3}\prod_{i=2}^{N-2}\int_{D(\Pi)} dz_i \, \KN\,\sigma \left(\prod_{k=2}^{\nu}\sum_{j=1}^{k-1}\frac{s_{jk}}{z_{jk}}\prod_{m=\nu+1}^{N-2}\sum_{l=m+1}^{N-1}\frac{s_{ml}}{z_{ml}} \right)\,,
\end{align}
such that the original integral corresponds to the representation labelled by $\nu=N-2$
\begin{align*}
F_{\Pi}^{\sigma}&=F_{\Pi,N-2}^{\sigma}\,.
\end{align*}

\subsubsection{Parke--Taylor forms and $Z$-theory amplitudes}
The integrals $F_{\Pi}^{\sigma}$ can be expressed in terms of another basis of integrals. It consists of integrals of Parke--Taylor forms \cite{PhysRevLett.56.2459}
\begin{align*}
\PT(\sigma)&=\frac{d\mu_N}{\prod_{i=1}^N z_{\sigma (i), \sigma(i+1)}}\,,
\end{align*}
where $\sigma \in S_{N}$, with the notational simplification $\sigma(N+1)=\sigma (1)$. The measure is given by
\begin{equation*}
d \mu_N = \frac{\bigwedge_{i=1}^N d z_i}{\vol\left(\SL(2,\ZR)\right)}=z_{1,N-1}z_{N-1,N}z_{1,N}\bigwedge_{i=2}^{N-2} d z_i
\end{equation*}
where the three punctures $(z_1,z_{N-1},z_N)=(0,1,\infty)$ have been fixed to get rid of the redundancy from the conformal Killing group $\SL(2,\ZR)$ of the disk topology. The integrals are given by  
\begin{align}\label{sec:amp:Z}
Z_{\Pi}(\sigma)&=\int_{D(\Pi)}\KN\, \PT(\sigma)\,,
\end{align}
which are the amplitudes appearing in a certain $Z$-theory \cite{Mafra:2016mcc}. 

The invertible transformation to the integrals $F_{\Pi}^{\sigma}$ is
\begin{align}\label{amplitudes:trafoFtoZ}
F_{\Pi}^{\sigma}&=(-1)^{N-3}\sum_{\rho\in S_{N-3}}S\left[\rho \left(2,\dots,\nu\right)|\sigma\left(2,\dots,\nu\right)\right]_1Z_{\Pi}(1,\rho(2,\dots,N-2),N,N-1)\,,
\end{align}
where the so-called momentum kernel \cite{BjerrumBohr:2010hn, Stieberger:2009hq} is given by \cite{Broedel:2013tta}
\begin{align}\label{amplitudes:Smatrix}
S\left[\rho\left(2,\dots,N-2\right)|\sigma\left(2,\dots,N-2\right)\right]_1&=\prod_{j=2}^{N-3}\left(s_{1,\rho(j)}+\sum_{k=2}^{j-1}\theta(\rho(j),\rho(k))s_{\rho(j),\rho(k)}\right)
\end{align}
and $\theta$ equals one if the ordering of $\rho(j)$ and $\rho(k)$ is the same in the ordered sets $\rho\left(2,\dots,N-2\right)$ and $\sigma\left(2,\dots,N-2\right)$, and zero if it is reversed. As in \eqn{amplitudes:trafoFtoZ} for the integrals $F_{\Pi}^{\sigma}$, any representation $F_{\Pi,\nu}^{\sigma}$ may be expressed in terms of the disk integrals $Z_{\Pi}$ using the identity
\begin{align*}
\sum_{\rho\in S_{\nu-1}}\frac{S\left[\rho\sigma\left(2,\dots,\nu\right)|\sigma\left(2,\dots,\nu\right)\right]_1}{\rho\sigma\left(z_{1,2}\dots z_{\nu-1,\nu}\right)}&=\sigma \left(\prod_{k=2}^{\nu}\sum_{m=1}^{k-1}\frac{s_{mk}}{z_{mk}}\right)\,,
\end{align*}
such that
\begin{align}\label{amplitudes:trafoFNuSigmaToZ}
F_{\Pi,\nu}^{\sigma}&=(-1)^{N-3}\sum_{\rho\in S_{\nu-1}}S\left[\rho\sigma\left(2,\dots,\nu\right)|\sigma\left(2,\dots,\nu\right)\right]_1\nonumber\\
&\phantom{=(-1)^{N-3}}\sum_{\tau\in S_{N-2-\nu}}S\left[\tau\sigma\left(N-2,\dots,\nu+1\right)|\sigma\left(N-2,\dots,\nu+1\right)\right]_{N-1}\nonumber\\
&\phantom{=(-1)^{N-3}\sum_{\tau\in S_{N-2-\nu}}}Z_{\Pi}(1,\rho\sigma \left(2,\dots,\nu\right),N,\tilde{\tau}\sigma\left(N-1,\dots,\nu+1\right),N-1)\,,
\end{align}
where $\tilde{\tau}(1,\dots,m)=\tau(m,\dots,1)$. Using relations such as \cite{Broedel:2013tta}
\begin{align*}
Z_{\Pi}(1,\alpha,N-1,\beta)&=(-1)^{|\beta|}\sum_{\sigma\in\alpha\shuffle \tilde{\beta} }Z_{\Pi}(1,\sigma,N-1)\,,
\end{align*}
the string corrections to open-superstring amplitudes can be expressed as some linear combinations
\begin{align*}
F_{\Pi}^{\sigma}&=\sum_{\gamma} n^{\sigma}_{\Pi}(\rho)\, Z_{\Pi}(\rho)
\end{align*}
over a set of $(N-3)!$ permutations $\gamma$.

\subsubsection{Amplitudes in twisted de Rham theory}
Rephrasing amplitudes in terms of Parke--Taylor forms admits a convenient formulation in terms of twisted de Rham theory. While the differential forms are defined on the moduli space $\mathcal{M}_{0,N}$, the function $u$ defining the integrable connection $\nabla_{\omega}$ in \eqn{preliminaries:nabla} with $\omega = d\log u$ is given by the multi-valued function
\begin{equation}
u(z)=\prod _{1\leq i<j\leq N-1} z_{ji}^{s_{ij}}
\end{equation}
with real branch the Koba--Nielsen factor, i.e.\ \eqn{amplitudes:KobaNielsen}. The twisted cycles corresponding to the integration domain $D(\Pi)$ are denoted
by \cite{Mizera:2017cqs}
\begin{equation}\label{sec:amp:cycles}
\mathcal{C}(\Pi)=\Delta^{o}_{N-3}(\Pi)\otimes u_{\Delta^{o}_{N-3}(\Pi)}\,,\qquad \Delta_{N-3}(\Pi)=\overline{\{0<z_{\Pi (2)}<z_{\Pi (3)}<\dots< z_{\Pi (N-2)}<1\}}\,,
\end{equation}
where
\begin{align*}
u_{\Delta^{o}_{N-3}(\Pi)}(z)&=\KN=\prod _{1\leq i<j\leq N-1} z_{\Pi (j),\Pi (i)}^{s_{\Pi (i),\Pi (j)}}
\end{align*}
is the real branch of $u(z)$ on $\Delta^{o}_{N-3}(\Pi)$. Moreover, the Mandelstam variables $s_{ij}$ are assumed to meet the conditions in \rcite{aomoto1975}, i.e.\ that they are sufficiently generic, see also \rcite{Brown:2018omk1}, such that the only non-vanishing cohomology is $H^{k}(\mathcal{M}_{0,N},\nabla_{\omega})$ with $k=\dim(\mathcal{M}_{0,N})=N-3$. In order to simplify notation, the abbreviations 
\begin{alignat*}{7}
H^{N-3}_{\omega}&=H^{N-3}(\mathcal{M}_{0,N}\,,\nabla_{\omega})\,,\qquad{}&{}H^{N-3}_{-\omega}&=H^{N-3}(\mathcal{M}_{0,N},\nabla^{\vee}_{\omega})\,, \nonumber\\
H_{N-3}^{\omega}&=H^{\text{lf}}_{N-3}(\mathcal{M}_{0,N},\mathcal{L}_{\omega})\,,\qquad{}&{}H_{N-3}^{-\omega}&=H^{\text{lf}}_{N-3}(\mathcal{M}_{0,N},\mathcal{L}^{\vee}_{\omega})
\end{alignat*}
are used. The dimensions of $H_{\omega}^{N-3}$ and $H^{\omega}_{N-3}$ coincide \cite{AomotoKita11} and are given by \cite{aomoto1987}
\begin{align}\label{eqn:dimH}
\dim\left(H_{\omega}^{N-3}\right)&=\dim\left(H^{\omega}_{N-3}\right)=(N-3)!\,.
\end{align}
We define the intersection number of two twisted forms $\langle \varphi |\in H^{N-3}_{\omega}$ and $ |\psi \rangle\in H^{N-3}_{-\omega}$ according to the intersection number \eqref{intNumb:Forms} by
\begin{equation*}
\begin{matrix}
H^{N-3}_{\omega} \times H^{N-3}_{-\omega}&\rightarrow & \ZC\\
(\langle \varphi |,|\psi\rangle)&\mapsto&\langle \varphi |\psi\rangle=\frac{1}{(-2\pi i)^{N-3}}\int_{\mathcal{M}_{0,N}} \psi\wedge \iota_{\omega}(\varphi)\,
\end{matrix}\,,
\end{equation*}
where $\iota_{\omega}:H^{N-3}_{\omega}\to H^{N-3}_{\text{c}}(\mathcal{M}_{0,N}\,,\nabla_{\omega})$ is the map constructed in \rcite{Mizera:2019gea} such that $i_{\omega}(\varphi)$ has compact support and defines the same twisted cohomology class as $\varphi$.

Using the above definitions, the $Z$-theory amplitudes \eqref{sec:amp:Z} can be expressed as the pairing
\begin{align}\label{PTtwistedForm}
Z_{\Pi}(\sigma)&=\langle \PT(\sigma)|\mathcal{C}(\Pi)]
\end{align}
and \eqn{eqn:dimH} ensures that for a fixed permutation $\Pi$, there are $(N-3)!$ linearly independent amplitudes $Z_{\Pi}(\sigma)$ which correspond to a basis of twisted forms $\langle \PT(\sigma)|\in H_{\omega}^{N-3}$ labelled by $(N-3)!$ distinct permutations $\sigma$. Practically, this means that for any additional permutation $\rho$, the amplitude $Z_{\Pi}(\rho)$ is a linear combination of the amplitudes $Z_{\Pi}(\sigma)$ obtained by partial fractioning and integration by parts.

\subsubsection{Fibration bases}
It turns out that for the discussion in the next section yet another basis of $H_{\omega}^{N-3}$ than the one spanned by the Parke--Taylor forms is useful. This is the so-called fibration basis \cite{Mizera:2019gea} which belongs to a more general class of bases, all of which we simply call fibration bases and which define for each $p$ in $3\leq p\leq n$ a basis of the twisted cohomology of the configuration space of $n-p$ points on $\ZC P^1\setminus\{x_1,x_2,\dots,x_p\}$ with the $p$ fixed coordinates $\{x_1,x_2,\dots,x_p\}$:
\begin{equation}
\mathcal{F}_{n,p}=\{(x_{p+1},x_{p+2},\dots,x_{n})\in (\ZC P^1)^{n-p}|\forall i\neq j: x_i\neq x_1,x_2,\dots,x_{p},x_j\}\,.
\end{equation}
The fibration in $p$ is defined by an inclusion map $\iota_p: \mathcal{F}_{n,p}\hookrightarrow \mathcal{F}_{n,p-1}$ which forgets the fixation of points enlarging the configuration space as $p$ decreases: beginning with all the $n$ punctures on the Riemann sphere being fixed, the configuration space $\mathcal{F}_{n,n}$ is a single point. Forgetting the fixing of $x_n$ yields the larger configuration space $\mathcal{F}_{n,n-1}$ and repeating the application of this forgetful map $n-3$ times leads to $\mathcal{F}_{n,3}$, where the definition \eqref{sec:amp:moduliSpace} of the moduli space of $n$-punctured Riemann spheres can be recovered: $\mathcal{F}_{n,3}=\mathcal{M}_{0,n}$. As shown in the next section, the fibration bases for $p=3$ and $p=4$ are well-suited for the study of the amplitude recursion established in \rcite{Broedel:2013aza}. The fibration bases can be introduced using the coordinates $x_i$ of the $n=N+1$ punctures and by arranging the representative differential forms of the twisted forms which constitute the basis of the twisted de Rham cohomology of $\mathcal{F}_{n,p}$ in a single vector, which is recursively defined as follows: the recursion starts with $\boldsymbol{f}^{n,+}=(f^{n,+})=(1)$ and iterates for $p$ and $q$ such that $3\leq q<p\leq n$ by defining the $q$-th subvector of $\boldsymbol{f}^{p-1,+}$ in terms of the vector $\boldsymbol{f}^{p,+}$ as
\begin{equation}\label{sec:amp:defFPlus}
	(\boldsymbol{f}^{p-1,+})_q=\boldsymbol{f}^{p-1,+}_q=\frac{d x_p}{x_{p,q}}\wedge \boldsymbol{f}^{p,+}\,.
\end{equation}
Therefore, the entries of the vector $\boldsymbol{f}^{p-1,+}$ can be labelled as follows
\begin{equation}\label{amplitudes:fibrationBasisElements}
(\boldsymbol{f}^{p-1,+})_{i_p, i_{p+1},\dots, i_{n}}=f^{p-1,+}_{i_p, i_{p+1},\dots, i_{n}}=\frac{d x_p}{x_{p,i_p}}\wedge\frac{d x_{p+1}}{x_{p+1,i_{p+1}}}\wedge\dots\wedge\frac{d x_{n}}{x_{n,i_{n}}}\,,
\end{equation}
where $3\leq i_k < k$.

The vector $\langle\boldsymbol{f}^{p,+}|$ contains the fibration basis of the twisted cohomology of $\mathcal{F}_{n,p}$ and satisfies for $3\leq p\leq n$ the differential equation 
\begin{align}\label{amplitudes:deqFibrationBasis}
d \langle\boldsymbol{f}^{p,+}|&=\tilde{\boldsymbol{\omega}}_p^{+}\wedge\langle\boldsymbol{f}^{p,+}|\,, 
\end{align}
where 
\begin{align*}
\tilde{\boldsymbol{\omega}}_p^{+}&=\sum_{2\leq i<j\leq p}\boldsymbol{\Omega}_p^{ij}d\log(x_{ij})
\end{align*}
and $\boldsymbol{\Omega}_p^{ij}$ are called braid matrices, which satisfy for distinct $i,j,k,l$ the infinitesimal pure braid relations~\cite{aomotoBraid,KohnoBraid}
\begin{equation}
[\boldsymbol{\Omega}_p^{ij},\boldsymbol{\Omega}_p^{kl}]=0\,,\qquad [\boldsymbol{\Omega}_p^{ij}+\boldsymbol{\Omega}_p^{jk},\boldsymbol{\Omega}_p^{ik}]=0\,.
\end{equation}
These matrices contain the information about the braiding of different fibres of the moduli space of punctured Riemann spheres and are recursively defined as follows \cite{Mizera:2019gea}: the recursion starts with $\boldsymbol{\Omega}_n^{ij}=s_{ij}$ and iterates according to 
\begin{align}\label{amplitudes:BraidMatrices}
(\boldsymbol{\Omega}_{p-1}^{ij})_{qr}&=\begin{cases}
\boldsymbol{\Omega}_p^{ij}&\text{if }q=r, q\neq i,j,\\
\boldsymbol{\Omega}_p^{pj}+\boldsymbol{\Omega}_p^{qj}&\text{if }q=r=i,  j\neq 1,2,\\
\boldsymbol{\Omega}_p^{pj}+\boldsymbol{\Omega}_p^{qj}+\boldsymbol{\Omega}_p^{pq}&\text{if }q=r=i, j= 1,2,\\
\boldsymbol{\Omega}_p^{ip}+\boldsymbol{\Omega}_p^{iq}&\text{if }q=r=j,\\
-\boldsymbol{\Omega}_p^{pj}&\text{if }q=i, r=j,\\
-\boldsymbol{\Omega}_p^{ip}&\text{if }q=j, r=i,\\
\boldsymbol{\Omega}_p^{ip}&\text{if }j=1,r=i, q\neq i,\\
\boldsymbol{\Omega}_p^{pr}&\text{if }j=2,q=i, r\neq i,\\
0&\text{otherwise}\,.
\end{cases}
\end{align}
Simple examples are for $n=5$ the following results
\begin{equation}\label{sec:amp:Omega4}
\boldsymbol{\Omega}_4^{42}=\begin{pmatrix}
t_{42}&0\\
t_{53}&t_{542}
\end{pmatrix}\,,\quad \boldsymbol{\Omega}_4^{43}=\begin{pmatrix}
t_{43}+t_{54}&-t_{54}\\
-t_{53}&t_{43}+t_{53}
\end{pmatrix}
\end{equation}
and for $n=6$ the matrices
\begin{align*}
\boldsymbol{\Omega}_4^{42}&=
\begin{pmatrix}
t_{42} & 0 & 0 & 0 & 0 & 0 \\
t_{63} & t_{642} & t_{65} & 0 & 0
& 0 \\
0 & 0 & t_{42} & 0 & 0 & 0 \\
t_{53}+t_{65} & 0 & -t_{65} &
t_{542} & 0 & 0 \\
0 & t_{53} & 0 & t_{63} &
t_{6542} &
0 \\
-t_{63} & 0 & t_{53}+t_{63} & t_{63} & 0 &t_{6542} 
\end{pmatrix}
\end{align*}
as well as
\begin{small}
\begin{align*}
&\boldsymbol{\Omega}_4^{43}=\nonumber\\
&\begin{pmatrix}
t_{43}\!+\!t_{54}\!+\!t_{64} & -t_{64} & 0 & -t_{54} &
0 & 0 \\
-t_{63} & t_{43}\!+\!t_{54}\!+\!t_{63}\!+\!t_{65} &
-t_{65} & 0 & -t_{54}-t_{65} & t_{65} \\
0 & -t_{64} & t_{43}\!+\!t_{54}\!+\!t_{64} & 0 & t_{64}
& -t_{54}-t_{64} \\
-t_{53}-t_{65} & 0 & t_{65} &
t_{43}\!+\!t_{53}\!+\!t_{64}\!+\!t_{65} & -t_{64} &
-t_{65} \\
0 & -t_{53} & 0 & -t_{63} &
t_{43}\!+\!t_{53}\!+\!t_{63} & 0 \\
t_{63} & 0 & -t_{53}-t_{63} & -t_{63} & 0 &
t_{43}\!+\!t_{53}\!+\!t_{63}
\end{pmatrix}\,.
\end{align*}
\end{small}
The differential equation with respect to $x_4$ satisfied by the vector $\langle\boldsymbol{f}^{4,+}|$ is of particular importance for our investigation below (recall that $x_4=z_0$ is the auxiliary insertion point in the context of the $N$-point amplitude recursion in \rcite{Broedel:2013aza})
\begin{align}\label{amplitudes:KZfibrationBasis}
\frac{\partial}{\partial x_4}\langle\boldsymbol{f}^{4,+}(x_4)|&=\left(\frac{\boldsymbol{\Omega}_4^{42}}{x_4}+\frac{\boldsymbol{\Omega}_4^{43}}{x_4-1}\right)\langle\boldsymbol{f}^{4,+}(x_4)|\,.
\end{align}
A differential equation of this form is called Knizhnik-Zamolodchikov (KZ) equation \cite{KNIZHNIK198483}. 

\subsection{KZ equation}\label{sec:KZ}
The KZ equation is not only the backbone of the amplitude recursion of \rcite{Broedel:2013aza} but has some remarkable mathematical properties and, in particular, a beautiful connection to polylogarithms. In this section, some of its properties are reviewed following the lines of \rcite{Brown:2013qva}. These will be the last mathematical preliminaries required to state the amplitude recursion in the following section.

Let $\boldsymbol{e}_0$ and $\boldsymbol{e}_1$ be representations of two Lie algebra generators and $\bF$ a function of $z \in \ZC\setminus \{0,1\}$ with values $\bF(z)$ in the vector space the representations $\boldsymbol{e}_0$, $\boldsymbol{e}_1$ act on, such that $\bF$ satisfies the KZ equation
\begin{align}\label{KZ:KZeq}
\frac{d \bF(z)}{d z }&=\left(\frac{\boldsymbol{e}_0}{z}+\frac{\boldsymbol{e}_1}{z-1}\right)\bF(z)\,.
\end{align}
Due to the singularities in the KZ equation at $z=0,1$, the boundary values of $\bF$ as $z\to 0$ and $z\to 1$ need to be regularised
\begin{equation}\label{KZ:C0C1}
\boldsymbol{C}_0=\lim_{z\rightarrow 0}z^{-\boldsymbol{e}_0}\bF(z),\qquad \boldsymbol{C}_1=\lim_{z\rightarrow 1}(1-z)^{-\boldsymbol{e}_1}\bF(z)\,.
\end{equation}
As reviewed in the remaining part of this subsection, these two regularised boundary values are related by the so-called Drinfeld associator $\boldsymbol{\Phi}(\boldsymbol{e}_0,\boldsymbol{e}_1)$ \cite{Drinfeld:1989st,Drinfeld2} according to the associator equation
\begin{align}\label{sec:KZ:DrinfeldEq}
\boldsymbol{C}_1&=\boldsymbol{\Phi}(\boldsymbol{e}_0,\boldsymbol{e}_1)\,\boldsymbol{C}_0\,.
\end{align}
The Drinfeld associator may be expressed in terms of a series involving commutators of $\boldsymbol{e}_0$ and $\boldsymbol{e}_1$ with the coefficients being multiple zeta values, which was originally shown in \rcite{Le} and which is reviewed in this paragraph following the lines of \rcite{Brown:2013qva}. Multiple zeta values $\zeta_{\boldsymbol{w}}$ are multiple polylogarithms evaluated at $z=1$, if they converge. Multiple polylogarithms\footnote{Note that this convention differs from the usual definition in terms of sums on $|z|<1$ by a sign: for $\boldsymbol{w}=\boldsymbol{e}_0^{n_r-1}\boldsymbol{e}_1\dots \boldsymbol{e}_0^{n_1-1}\boldsymbol{e}_1$, $n_i\geq 1$, they are related according to $G_{\boldsymbol{w}}(z)=(-1)^r\sum_{1\leq k_1<\dots<k_r}\frac{z^{k_r}}{k_1^{n_1}\dots k_r^{n_r}}=(-1)^r\Li_{n_1,\dots,n_r}(z)$. Due to this close relation, we call the subclass $G_{\boldsymbol{w}}(z)$ of the Goncharov polylogarithms simply multiple polylogarithms, while multiple zeta values are the values $\Li_{n_1,\dots,n_r}(1)$.} in one variable $G_{\boldsymbol{w}}$, in turn, are a subclass of the Goncharov polylogarithms \cite{Goncharov:2001iea} and multi-valued functions on $\ZC\setminus \{0,1\}$, indexed by words $\boldsymbol{w}\in\{\boldsymbol{e}_0,\boldsymbol{e}_1\}^{\times}$ generated by the letters $\boldsymbol{e}_0$ and $\boldsymbol{e}_1$, which satisfy for $i=0,1$ the differential equations
\begin{equation}\label{KZ:deqLi}
d G_{\boldsymbol{e}_i \boldsymbol{w}}(z)=\omega_i G_{\boldsymbol{w}}(z)\,,\qquad \omega_0=\frac{dz}{z}\,,\qquad \omega_1=\frac{dz}{z-1}\,.
\end{equation}
The boundary values at $z=0$ are determined by
\begin{equation}\label{KZ:bc0}
\lim_{z\rightarrow 0}G_{\boldsymbol{w}}(z)=0\,,\qquad G_{\boldsymbol{e}_0^n}(z)=\frac{\log^n(z)}{n!}\,,
\end{equation}
where $\boldsymbol{w}$ is a word not beginning with $\boldsymbol{e}_0$, and by the shuffle product
\begin{align*}
G_{\boldsymbol{w}'} (z) G_{\boldsymbol{w}''}(z)&=G_{\boldsymbol{w}'\shuffle \, \boldsymbol{w}''}(z)\,,
\end{align*}
where $\boldsymbol{w}',\boldsymbol{w}''\in\{\boldsymbol{e}_0,\boldsymbol{e}_1\}^{\times}$, which can be used to relate the remaining cases to the two boundary values in \eqn{KZ:bc0}.

Using the above definitions, multiple zeta values are labelled by words of the form
\begin{equation}
\boldsymbol{w}=\boldsymbol{e}_0^{n_r-1}\boldsymbol{e}_1\dots \boldsymbol{e}_0^{n_1-1}\boldsymbol{e}_1
\end{equation}
with $n_r\geq 2$, i.e.\ not beginning with $\boldsymbol{e}_1$, which lead to convergent values, defined by
\begin{equation}\label{sec:amp:mzv}
\zeta_{\boldsymbol{w}}=(-1)^r G_{\boldsymbol{w}}(1)=\sum_{0<k_1<\dots < k_r}\frac{1}{k_1^{n_1}\cdots k_r^{n_r}}=\zeta_{n_1,\dots,n_r}\,.
\end{equation}
This definition can be generalised to any word $\boldsymbol{w}\in\{\boldsymbol{e}_0,\boldsymbol{e}_1\}^{\times}$ using the following regularisation, which is the tangential base point regularisation \cite{Deligne89} pointing in the positive direction at $z=0$ and pointing in the opposite direction at $z=1$, respectively \cite{Brown:2013qva}. The regularisation as $z\to 0$ corresponds to the choices of the boundary values \eqref{KZ:bc0}, while the regularisation as $z\to 1$ is required to tame the pole of the differential form $dz/(z-1)$ in the outermost integration at $z=1$. This effectively results\footnote{Using the shuffle algebra to extract the divergent contributions for $z=1$ appearing in the form of $G_{\boldsymbol{e}_1}(z)=\log(1-z)$ in $G_{\boldsymbol{w}}(z)$, any multiple polylogarithm $G_{\boldsymbol{w}}(z)$ can be written on $z\in(0,1)$ such that it takes the form $G_{\boldsymbol{w}}(z)=\sum_{k=0}^{|\boldsymbol{w}|}c_k(z) \log(1-z)^k$, where $c_k (z)$ are holomorphic functions of $z$ in a neighbourhood of $z=1$. Thus, for any word $\boldsymbol{w}\in\{\boldsymbol{e}_0,\boldsymbol{e}_1\}^{\times}$, the multiple zeta value $\zeta_{\boldsymbol{w}}$ can be defined by the regularised value of $G_{\boldsymbol{w}}(z)$ (up to a sign) at $1$, which, in turn, is the coefficient $c_0(z)$, i.e.\ $\zeta_{\boldsymbol{w}}=(-1)^r\text{Reg}_{z=1}\left(G_{\boldsymbol{w}}(z)\right)=(-1)^rc_0(1)$. This leads to the results in \eqn{sec:genus0:regularisedMZV}.} for any words $\boldsymbol{w},\boldsymbol{w}'$ and $n_r\geq 2$ in the definitions \cite{Brown:2013qva}
\begin{align}\label{sec:genus0:regularisedMZV}
\zeta_{\boldsymbol{e}_0}&=\zeta_{\boldsymbol{e}_1}=0\,,\nonumber\\
\zeta_{\boldsymbol{e}_0^{n_r-1}\boldsymbol{e}_1\dots \boldsymbol{e}_0^{n_1-1}}&=\zeta_{n_1,\dots,n_r}\,,\nonumber\\
\zeta_{\boldsymbol{w}}\zeta_{\boldsymbol{w}'}&=\zeta_{\boldsymbol{w}\shuffle \boldsymbol{w}'}\,.
\end{align}

The above definitions can be related to the KZ equation by considering the following generating function of the multiple polylogarithms
\begin{align*}
\bL(z)&=\sum_{\boldsymbol{w}\in\{\boldsymbol{e}_0,\boldsymbol{e}_1\}^{\times}} \boldsymbol{w}\,G_{\boldsymbol{w}}(z)\,.
\end{align*}
By the differential equations \eqref{KZ:deqLi}, this function satisfies the KZ equation \eqref{KZ:KZeq}. Furthermore, the boundary conditions \eqref{KZ:bc0} close to $z=0$ imply the asymptotic behaviour 
\begin{equation}\label{KZ:asymptL}
\bL(z)\sim z^{\boldsymbol{e}_0}\qquad \text{as }  z\rightarrow 0\,,
\end{equation}
i.e.\ that there exists some function $h(z)$ with $h(0)=1$ and which is holomorphic close to $z_0$, such that in a neighbourhood of the origin $\bL(z)=h(z) z^{\boldsymbol{e}_0}$. By the symmetry $z\mapsto 1-z$ of the KZ equation, there is another solution $\bL_1$ which satisfies 
\begin{equation}\label{KZ:asymptL1}
\bL_1(z)\sim (1-z)^{\boldsymbol{e}_1}\qquad \text{as }  z\rightarrow 1\,.
\end{equation} 
Since for two solutions $\bF_0$ and $\bF_1$ of the KZ equation \eqref{KZ:KZeq}, the product $(\bF_1)^{-1}\bF_0$ is independent of $z$ and by the definitions \eqref{KZ:C0C1} as well as the asymptotics \eqref{KZ:asymptL}, \eqref{KZ:asymptL1} of $\bL(z)$ and $\bL_1(z)$, respectively, the calculation 
\begin{equation*}
(\bL_1(z))^{-1}\bL(z)\boldsymbol{C}_0=\lim_{z\rightarrow 0}(\bL_1(z))^{-1}\bF(z)=\lim_{z\rightarrow 1}(\bL_1(z))^{-1}\bF(z)=\boldsymbol{C}_1
\end{equation*}
shows that the Drinfeld associator defined in terms of the solutions $\bL(z)$ and $\bL_1(z)$
\begin{align}\label{KZ:PhiL}
\boldsymbol{\Phi}(\boldsymbol{e}_0,\boldsymbol{e}_1)&=(\bL_1(z))^{-1}\bL(z)
\end{align}
indeed relates the regularised boundary values \eqref{KZ:C0C1} according to \eqn{sec:KZ:DrinfeldEq}. Using the \mbox{$z$-independence} of $(\bL_1(z))^{-1}\bL(z)$  and evaluating \eqn{KZ:PhiL} for $z\rightarrow 1$ finally leads to an expression of the Drinfeld associator in terms of the multiple zeta values \cite{Le}
\begin{align}\label{KZ:PhiMZV}
\boldsymbol{\Phi}(\boldsymbol{e}_0,\boldsymbol{e}_1)&=\lim_{z\rightarrow 1}(1-z)^{-\boldsymbol{e}_1}\bL(z)\nonumber\\
&=\sum_{\boldsymbol{w}\in\{\boldsymbol{e}_0,\boldsymbol{e}_1\}^{\ast}} \boldsymbol{w}\, \zeta_{\boldsymbol{w}}\nonumber\\
&=1+\zeta_2 [\boldsymbol{e}_1,\boldsymbol{e}_0]+\zeta_3\left([\boldsymbol{e}_0+\boldsymbol{e}_1,[\boldsymbol{e}_1,\boldsymbol{e}_0]]\right)+\dots\,,
\end{align}
showing that the Drinfeld associator is the generating series of the multiple zeta values. The limit $z\rightarrow 1$ is chosen to correspond to applying the tangential base point regularisation, such that the prefactor $(1-z)^{-\boldsymbol{e}_1}$ leads to the regularisation \eqref{sec:genus0:regularisedMZV} of the divergent terms in $\bL(z)$ \cite{Brown:mmv}.

\subsection{Examples of simple open-string amplitudes}\label{amplitudes:ex}
In this section, the simplest examples of tree-level amplitudes of open-string states are reviewed in terms of the different descriptions introduced in the previous subsections. For the sake of simplicity, the ordering of the domain of integration $D(\Pi)$ is chosen to be the natural one, i.e.\ $\Pi=\id$.

\subsubsection{Four-point amplitude}
The lowest non-trivial amplitude at genus zero is found at $N=4$. It is given according to \eqn{amplitudes:generalForm} by \cite{Mafra:2011nw}
\begin{align*}
A_{\open}(\id,\ap)&= F_{\id}^{\id}(\ap)A_{\YM}\left(1,2,3,4 \right)\,,
\end{align*}
where
\begin{align}\label{amplitudes:4pt}
F_{\id}^{\id}&=-\int_{0}^1 dz_2\, |z_{12}|^{s_{12}}|z_{23}|^{s_{23}}\frac{s_{12}}{z_{12}}=\frac{\Gamma(1+s_{12})\Gamma(1+s_{23})}{\Gamma(1+s_{12}+s_{23})}\nonumber\\
&=1-\zeta_2 s_{12}s_{23}+\zeta_3 s_{12}s_{23}(s_{12}+s_{23})+\mathcal{O}((\alpha')^4)
\end{align}
has the form of the Veneziano amplitude \cite{Veneziano1968}. Its representation in terms of $Z$-amplitudes reads
\begin{equation*}
F_{\id}^{\id}=-s_{12}Z_{\id}(34)=-s_{12}\langle \PT(34)|\mathcal{C}(\id)]\,,
\end{equation*}
which is in agreement with the definitions \eqref{amplitudes:trafoFtoZ,PTtwistedForm}, because $S[2|2]_1=s_{12}$. Therefore, the colour-ordered four-point amplitude is determined by the Parke--Taylor form $\PT(34)$, which in turn is the following linear combination of elements of the fibration basis $\boldsymbol{f}^{4,+}$ for $n=5$:
\begin{equation}
\PT(34)=\frac{d z_2}{z_{12}}=\frac{d x_5}{x_{25}}=-f^{4,+}_2\,.
\end{equation}

\subsubsection{Five-point amplitude}
The five-point amplitude is given by \cite{Mafra:2011nw}
\begin{align*}
A_{\open}(\id,\ap)&= F_{\id}^{\id}(\ap)A_{\YM}\left(1,2,3,4 \right)+F^{(23)}_{\id}(\ap)A_{\YM}\left(1,3,2,4 \right)\,,
\end{align*}
where
\begin{align*}
F_{\id}^{\id}&=F_{\id,3}^{\id}\nonumber\\
&=\int_{0<z_2<z_3<1}dz_2 dz_3\, \prod_{i<j}|z_{ij}|^{s_{ij}}\frac{s_{12}}{z_{12}}\left(\frac{s_{13}}{z_{13}}+\frac{s_{23}}{z_{23}}\right)\nonumber\\
&=\int_{0<z_2<z_3<1}dz_2 dz_3\, \prod_{i<j}|z_{ij}|^{s_{ij}}\frac{s_{12}}{z_{12}}\frac{s_{34}}{z_{34}}\nonumber\\
&=F_{\id,2}^{\id}\nonumber\\
&=1+\zeta_2 (s_{12}s_{34}-s_{34}s_{45}-s_{12}s_{51})\nonumber\\
&\phantom{=}-\zeta_3 (s_{12}^2 s_{34}+2 s_{12}s_{23}s_{34}+s_{12}s_{34}^2-s_{34}^2 s_{45}-s_{34}s_{45}^2-s_{12}^2s_{51}-s_{12}s_{51}^2)+\mathcal{O}((\alpha')^4)
\end{align*}
and
\begin{align*}
F_{\id}^{(23)}&=F_{\id,3}^{(23)}\nonumber\\
&=\int_{0<z_2<z_3<1}dz_2 dz_3\, \prod_{i<j}|z_{ij}|^{s_{ij}}\frac{s_{13}}{z_{13}}\left(\frac{s_{12}}{z_{12}}+\frac{s_{32}}{z_{32}}\right)\nonumber\\
&=\int_{0<z_2<z_3<1}dz_2 dz_3\, \prod_{i<j}|z_{ij}|^{s_{ij}}\frac{s_{13}}{z_{13}}\frac{s_{24}}{z_{24}}\nonumber\\
&=F_{\id,2}^{(23)}\nonumber\\
&=\zeta_2 s_{12}s_{24}-\zeta_3 s_{13}s_{24}(s_{12}+s_{23}+s_{34}+s_{45}+s_{51})+\mathcal{O}((\alpha')^4)\,.
\end{align*}
These amplitudes can be expressed in terms of the $Z$-amplitudes 
\begin{align*}
Z_{\id}(\rho)&=\int_{0<z_2<z_3<1}dz_2dz_3 \prod_{1\leq i<j\leq 4}|z_{ij}|^{s_{ij}}\frac{1}{z_{1,\rho (2)}z_{\rho (2),\rho (3)}} 
\end{align*}
for $\sigma\in\{\id,(2\,3)\}$ as the linear combination
\begin{align*}
\begin{pmatrix}
F^{\id}_{\id}\\
F^{(23)}_{\id}
\end{pmatrix}&=\begin{pmatrix}
s_{12}(s_{13}+s_{23})&s_{12}s_{13}\\
s_{12}s_{13}&s_{13}(s_{12}+s_{23})
\end{pmatrix}\begin{pmatrix}
Z_{\id}(\id)\\
Z_{\id}(23)
\end{pmatrix}\,.
\end{align*}
The matrix above is in agreement with the definition \eqref{amplitudes:Smatrix} of $S[\rho(2,3)|\sigma(2,3)]_1$.

\section{Amplitude recursion}
\label{sec:recursion}
Having introduced the necessary preliminaries in the previous section, we can finally state and investigate the amplitude recursion described in \rcite{Broedel:2013aza}, which is based upon the results of \rcites{aomoto1987,Terasoma}. The origin of the recursion is the differential equation \eqref{amplitudes:deqFibrationBasis} satisfied by the fibration basis which in turn is determined by the braid matrices \eqref{amplitudes:BraidMatrices}. This relation of the differential structure of Selberg integrals to the geometric structure of the moduli space (encoded in the braid matrices) has been described before and the corresponding differential equation, called Gauss--Manin connection, has explicitly been given in \rcite{aomoto1987} in terms of so-called admissible forms. A more recent investigation of Selberg integrals, their differential structure and, in particular, their connection to the Drinfeld associator can be found in \rcite{Terasoma}. 

Even though we use a similar notion of admissible forms as introduced in the latter two references and they are the main reference for the recursion in \rcite{Broedel:2013aza}, the primary reference for our reformulation of the amplitude recursion is \rcite{Mizera:2019gea}. The reason for this choice is that this reference formulates the central objects describing the integrals occurring in the recursion in terms of twisted de Rham theory and the fibration basis introduced therein is compatible with a convenient gauge choice for the $\SL(2,\ZC)$ redundancy of the moduli space $\mathcal{M}_{0,N}$ in \eqn{sec:amp:moduliSpace}. 

We start in \subsecref{rec:old} by reviewing the recursive construction of \rcite{Broedel:2013aza} and rephrase it in terms of twisted de Rham theory in \subsecref{rec:new} and \subsecref{sec:derGraph}. Furthermore, from here on unless specified otherwise, we use the ordering defined in \eqn{sec:rec:ordering} and in particular the notation $(n,x_i,t_{ij})$ rather than $(N,z_i,s_{ij})$ for the number of insertion points, their positions and the Mandelstam variables, respectively. 

\subsection{Review of the amplitude recursion}\label{rec:old}
The amplitude recursion proposed in \rcite{Broedel:2013aza} is based on the construction of a solution $\bhF$ of the KZ equation, such that the regularised boundary values $\boldsymbol{C}_0$ and $\boldsymbol{C}_1$ encode the $(n-2)$-point and the $(n-1)$-point string corrections\footnote{From here on unless specified otherwise, we use the ordering $\Pi=\id$ for the integration domain in \eqn{sec:amp:cycles} and usually omit the corresponding subscript $\id$.}
\begin{align*}
F^{\sigma}&=F^{\sigma}_{\id}\,.
\end{align*}
Using the sum expansion \eqref{KZ:PhiMZV} of the Drinfeld associator, the $\alpha'$-expansion of the $(n-1)$-point amplitude in $\boldsymbol{C}_1$ can be calculated by $\boldsymbol{\Phi}(\boldsymbol{e}_0,\boldsymbol{e}_1)\,\boldsymbol{C}_0$ at all orders in $\alpha'$.

Concretely, the solution $\bhF$ is similar to the equivalent representations $F_{\id,\nu}^{\sigma}$ of $F^{\sigma}$ defined in \eqn{amplitudes:FSigmaPinu}, however, an additional puncture $x_4$ (recall that $x_4=z_0$) at $x_{5}<x_4<x_{3}=1$ is introduced. The solution is explicitly given by the vector
\begin{align}\label{sec:rec:defBHF}
\bhF(x_4)&=(\bhF_{n-3},\bhF_{n-4},\dots,\bhF_{1})
\end{align} 
of length $(n-3)!$, where the subvectors $\bhF_{\nu}$ are of length $(n-4)!$ and defined\footnote{The exact conversion from the original definition in \rcite{Broedel:2013aza} using the labelling $(N,z_i,s_{ij})$ is presented in \appref{app:transl:label}} by the elements
\begin{align}\label{amplitudes:FhatSigmanu}
\hat{F}_{\nu}^{\sigma}&=(-1)^{n}\prod_{i=5}^{n}\int_0^{x_{i-1}} dx_i\, \hat{u}(x)\,\sigma \left(\prod_{k=n-\nu+2}^{n}\sum_{j=3}^{k-1}\frac{t_{kj}}{x_{kj}}\prod_{m=5}^{n-\nu+1}\left(\sum_{l=5}^{m-1}\frac{t_{ml}}{x_{ml}}+\frac{t_{m3}}{x_{m3}}\right)\right)\,,
\end{align}
labelled by the permutations $\sigma\in S_{n-4}$ acting on the indices $\{5,6,\dots, n\}$, where 
\begin{equation*}
\hat{u}(x)=\prod_{2\leq i\prec j\leq n}x_{ji}^{t_{ij}}\,,\qquad \prod_{i=5}^{n}\int_0^{x_{i-1}} dx_i=\int_0^{x_4}dx_5\int_0^{x_5}dx_6\dots \int_0^{n-1}dx_n\,.
\end{equation*}
A comparison with the definition of the integrals $F_{\id,\nu}^{\sigma}$ defined in \eqn{amplitudes:FSigmaPinu} shows that in the limit $x_4\rightarrow 1$ and for $t_{4i}\rightarrow 0$, the vector $\bhF(x_4)$ encodes these representations
\begin{align}\label{sec:rec:lim1}
\lim_{t_{4i\to 0}}\lim_{x_4\rightarrow 1}\hat{F}^{\sigma}_{\nu}(x_4)|_{t_{43=0}}&=F_{\id,\nu}^{\sigma}\,,
\end{align} 
where $t_{43}=0$ is required since, otherwise, the factor $\lim_{x_4\to 1=x_3}\hat{u}(x)=0$ would render the integral zero. In \rcite{Broedel:2013aza} it is stated that $\bhF(x_4)$ satisfies the KZ equation 
\begin{align}\label{amplitudes:KZhatF}
\frac{d \bhF(x_4)}{d x _4}&=\left(\frac{\boldsymbol{e}_0}{x_4}+\frac{\boldsymbol{e}_1}{x_4-1}\right)\bhF(x_4)
\end{align}
for some matrices $\boldsymbol{e}_0$ and $\boldsymbol{e}_1$ with the non-vanishing entries being homogeneous polynomials of degree one in $t_{ij}$ and integer coefficients. In particular, the first $(n-3)$ rows of $\boldsymbol{e}_1$ are given by\footnote{See \appref{app:transl:e1} for the derivation.} 
\begin{align}\label{sec:rec:e1}
\boldsymbol{e}_1&=\begin{pmatrix}
t_{43}\,\mathbb{I}_{(n-4)!\times (n-4)!}&0_{(n-4)!\times (n-4)(n-4)!}\\
\vdots&\vdots
\end{pmatrix}\,.
\end{align}
Therefore, the theory of the KZ equation reviewed in \secref{sec:KZ} can be applied to $\bhF(x_4)$. The connection of the regularised boundary values of $\bhF(x_4)$ for $x_4\rightarrow 0,1$ by means of the Drinfeld associator as given in \eqn{sec:KZ:DrinfeldEq} yields the tree-level amplitude recursion, since the first $(n-4)!$ entries of the regularised boundary value 
\begin{align*}
\boldsymbol{C}_1&=\lim_{x_4\rightarrow 1}(1-x_4)^{-\boldsymbol{e}_1}\bhF(x_4)
\end{align*}
are related to the $(n-1)$-point string corrections $\bF|_{n-1}=(F^{\sigma})_{\sigma\in S_{n-4}}$ according to
\begin{align*}
\boldsymbol{C}_1&=(\bF|_{n-1}+\mathcal{O}(t_{4i}),\dots)
\end{align*}
due to equation\footnote{Note that the requirement $t_{43}=0$ is implemented in the regularisation by \eqn{sec:rec:e1}: the prefactor $(1-x_4)^{-\boldsymbol{e}_1}$ removes the factor $x_{43}^{t_{43}}$ in $\hat{u}(x)$ and hence, prevents the factor $x_{43}^{-t_{43}}\hat{u}(x)$ from vanishing as $x_4\to x_3$.} \eqref{sec:rec:lim1}. The lower regularised boundary value
\begin{align*}
\boldsymbol{C}_0&=\lim_{x_4\rightarrow 0}x_4^{-\boldsymbol{e}_0}\bhF(x_4)
\end{align*}
is slightly more delicate. As calculated in \appref{app:c0}, it turns out that 
\begin{align*}
\boldsymbol{C}_0&= (\bF\!|_{n-2}+\mathcal{O}(t_{4i}),0_{(n-4)(n-4)!})\,,
\end{align*}
where $\bF\!|_{n-2}$ is the vector of the $(n-2)$-point string corrections. Using the above properties of $\bhF(x_4)$, the recursion proposed in \subsecref{amplitudes:ex} is the following algorithm: 
\begin{enumerate}
	\item The vector $d\bhF(x_4)/d x_4$ is expressed in the form of the KZ equation \eqref{KZ:KZeq} using integration by parts and partial fractioning.
	\item The matrices $\boldsymbol{e}_0$ and $\boldsymbol{e}_1$ are read off from the resulting equation, such that the $\alpha'$-expansion of the Drinfeld associator $\boldsymbol{\Phi}(\boldsymbol{e}_0,\boldsymbol{e}_1)$ can be calculated using \eqn{KZ:PhiMZV}.
	\item The $(n-1)$-point string corrections $\bF|_{n-1}$ are determined by the $(n-2)$-point string corrections $ \bF|_{n-2}$ using the limit $t_{4i}\rightarrow 0$ of $\boldsymbol{C}_1=\boldsymbol{\Phi}(\boldsymbol{e}_0,\boldsymbol{e}_1)\,\boldsymbol{C}_0$, i.e.
	\begin{align}\label{KZ:calculation}
	\begin{pmatrix}
	\bF|_{n-1}\\
	\vdots
	\end{pmatrix}&=\left(\mathbb{I}+\zeta_2 [\boldsymbol{e}_1,\boldsymbol{e}_0]+\zeta_3\left([\boldsymbol{e}_0+\boldsymbol{e}_1,[\boldsymbol{e}_1,\boldsymbol{e}_0]]+\dots\right)\right)|_{t_{4i}=0}\begin{pmatrix}
	\bF|_{n-2}\\
	0_{(n-4)(n-4)!}
	\end{pmatrix}\,.
	\end{align}
\end{enumerate}

In \rcite{Broedel:2013aza}, the recursion is explicitly shown to hold for the examples from \subsecref{amplitudes:ex} and the examples up to the nine-point amplitudes are given on the webpage \cite{MZVWebsite}. The first example is the four-point amplitude for $n=5$, where the derivative of the vector of integrals
\begin{equation}\label{sec:rec:4ptZ}
\bhF =\begin{pmatrix}
\hat{F}_2^{\id}\\
\hat{F}_1^{\id}
\end{pmatrix}=-\int_0^{x_4} dx_5\, |x_{25}|^{t_{25}}|x_{54}|^{t_{54}} |x_{53}|^{t_{53}} \begin{pmatrix}
\frac{t_{25}}{z_{25}}\\\frac{t_{53}}{z_{53}}
\end{pmatrix}
\end{equation} 
satisfies the KZ equation
\begin{equation}\label{sec:rec:4PtEx}
\frac{d \bhF(x_4)}{d x_4}=\left(\frac{\boldsymbol{e}_0}{x_4}+\frac{\boldsymbol{e}_1}{x_4-1}\right)\bhF(x_4)\,,\qquad \boldsymbol{e}_0=\begin{pmatrix}
t_{25}&-t_{25}\\
0&0
\end{pmatrix}\,,\qquad 
\boldsymbol{e}_1=\begin{pmatrix}
0&0\\
-t_{53}&t_{53}
\end{pmatrix}
\end{equation}
for $t_{4i}=0$. The regularised boundary values are 
\begin{equation*}
\boldsymbol{C}_0=\begin{pmatrix}
1\\0
\end{pmatrix}\,,\qquad
\boldsymbol{C}_1=\begin{pmatrix}
F^{\id}\\\dots
\end{pmatrix}\,,
\end{equation*}
where $F^{\id}$ is the Veneziano amplitude given in \eqn{amplitudes:4pt}. Note that the three-point string correction is just one. Calculating the right-hand side of \eqn{KZ:calculation} yields
\begin{align*}
&\left( \mathbb{I}+\zeta_2 [\boldsymbol{e}_1,\boldsymbol{e}_0]+\zeta_3\left([\boldsymbol{e}_0+\boldsymbol{e}_1,[\boldsymbol{e}_1,\boldsymbol{e}_0]]\right)+\dots \right)\begin{pmatrix}
1\\
0
\end{pmatrix}\nonumber\\
&=\begin{pmatrix}
1\\
0
\end{pmatrix}-\zeta_2 \begin{pmatrix}
t_{25}t_{53}\\
t_{25}t_{53}
\end{pmatrix}
+\zeta_3 \begin{pmatrix}
	t_{25}^2t_{53}+t_{25}t_{53}^2\\
	t_{25}^2t_{53}+t_{25}t_{53}^2
\end{pmatrix}+\dots\,,
\end{align*}
such that the first entry indeed reproduces the $\alpha'$-expansion of the four-point string correction in \eqn{amplitudes:4pt}. 

\subsection{Reformulation in twisted de Rham theory}\label{rec:new}
The amplitude recursion of \rcite{Broedel:2013aza} presented in the previous subsection can be understood and optimised in terms of twisted de Rham theory. In particular, we will provide recursive expressions for the matrices $\boldsymbol{e}_0$ and $\boldsymbol{e}_1$ at any level $n$ using techniques from intersection theory.

The integrals in $\bhF(x_4)$ defined in \eqn{amplitudes:FhatSigmanu} are determined by the $n$-th twisted de Rham cohomology of the configuration space $\mathcal{F}_{n,4}$ with the local coefficient of the twisted cycles given by
\begin{align}\label{sec:algo:localCoef}
\hat{u}(x)&=\prod_{2\leq i\prec j\leq n} x_{ji}^{s_{ij}}\,.
\end{align}
Denoting the differential forms in the integrals $\bhF=(\hat{F}_{\nu}^{\sigma})_{\sigma\in S_{n-4},\nu=1,\dots,n-3}$ defined in \eqn{amplitudes:FhatSigmanu} by 
\begin{align*}
\hat{f}_{\nu}^{\sigma}=\sigma \left(\prod_{k=n-\nu+2}^{n}\sum_{j=3}^{k-1}\frac{t_{kj}}{x_{kj}}\prod_{m=5}^{n-\nu+1}\left(\sum_{l=5}^{m-1}\frac{t_{ml}}{x_{ml}}+\frac{t_{m3}}{x_{m3}}\right)\right)dx_5\wedge dx_6\wedge\dots\wedge dx_n\,,
\end{align*}
the corresponding $(n-3)!$ twisted forms $\langle \hat{f}_{\nu}^{\sigma}| $ form a basis\footnote{According to \eqn{sec:rec:lim1}, in the limit $x_4\to x_3$ and $t_{4i}\to 0$, for each $\nu$ the $(n-4)!$ twisted forms $\langle \hat{f}_{\nu}^{\sigma}| $ form a basis of $H^{n-1}_{\omega}$ parametrised by $\sigma$, since the relation \eqref{amplitudes:trafoFNuSigmaToZ} is invertible and the twisted Parke--Taylor forms are such a basis. The non-vanishing Mandelstam variables $t_{4i}$ and the distinction $x_4\neq x_3$ ensure that the twisted forms $\langle \hat{f}_{\nu}^{\sigma}| $ are also linearly independent for different $\nu$, forming a basis of $H^{n-4}(\mathcal{F}_{n,4},\nabla_{\hat{\omega}})$.} of the twisted de Rham cohomology of the configuration space of four fixed coordinates $H^{n-4}(\mathcal{F}_{n,4},\nabla_{\hat{\omega}})$, where $\hat{\omega}=d\log(\hat{u})$. The original integrals $\bhF(x_4)$ can be recovered by
\begin{equation}\label{sec:rec:coefsTrafo}
\bhF=(\hat{F}_{\nu}^{\sigma})_{\sigma\in S_{n-4},\nu=1,\dots,n-3}=(\langle \hat{f}_{\nu}^{\sigma}|\mathcal{C}])_{\sigma\in S_{n-4},\nu=1,\dots,n-3}\,,
\end{equation}
where 
\begin{align}\label{sec:algo:intDomain}
\mathcal{C}=\{x_2<x_n<x_{n-1}<\dots <x_4\}\otimes \hat{u}
\end{align}
is the cycle corresponding to the natural ordering on the disk, where $\hat{u}$ is real-valued. Using the basis transformation~\eqref{twistedTheory:basisTrafo}, the entries $\hat{F}_{\nu}^{\sigma}=\langle \hat{f}_{\nu}^{\sigma}|\mathcal{C}]$ of $\bhF$ can be expressed in terms of the fibration basis $\boldsymbol{f}^{4,+}$
\begin{align}\label{add:basisTrafoEntries}
\langle \hat{f}_{\nu}^{\sigma}|&=\sum_{3\leq i_k<k} b_{\nu;i_5, i_{6},\dots, i_{n}}^{\sigma} \langle f^{4,+}_{i_5, i_6,\dots, i_n}|\,,
\end{align}
where $\langle f^{4,+}_{i_5, i_6,\dots, i_n}|\in H^{n-4}(\mathcal{F}_{n,4},\nabla_{\hat{\omega}})$ is the twisted cohomology class of the entry
\begin{align*}
(\boldsymbol{f}^{4,+})_{i_5, i_{6},\dots, i_{n}}=f^{4,+}_{i_5, i_6,\dots, i_n}=\frac{d x_5}{x_{5,i_5}}\wedge\frac{d x_{6}}{x_{6,i_{6}}}\wedge\dots\wedge\frac{d x_{n}}{x_{n,i_{n}}}
\end{align*} 
of the vector $\boldsymbol{f}^{4,+}$ which constitutes the fibration basis for $p=4$ defined in \eqn{sec:amp:defFPlus}. In the following subsection, we show how to combinatorially calculate the intersection numbers $b_{\nu;i_5, i_{6},\dots, i_{n}}^{\sigma}$ of the twisted forms. Then, the basis transformation \eqref{add:basisTrafoEntries} can be written in matrix form 
\begin{align}\label{sec:rec:BasisTrafo}
\langle\boldsymbol{\hat{f}}(x_4)|&=\bB \langle\boldsymbol{f}^{4,+}(x_4)|\,,
\end{align}
where $\bB\in \mathrm{GL}_{(n-3)!}(\ZZ[t_{ij}])$, with the non-vanishing entries $b_{\nu;i_5, i_{6},\dots, i_{n}}^{\sigma}$ being polynomials of degree one in the Mandelstam variables $t_{ij}$ with integer coefficients, and $\langle\boldsymbol{\hat{f}}|=(\langle \hat{f}_{\nu}^{\sigma}|)_{\sigma,\nu}$ is the vector of the twisted forms in $\bhF$. Therefore, the KZ equation \eqref{amplitudes:KZhatF} satisfied by $\bhF(x_4)$ can be related to the KZ equation \eqref{amplitudes:KZfibrationBasis} satisfied by the fibration basis $\boldsymbol{f}^{4,+}(x_4)$ on the level of twisted forms according to
\begin{equation}\label{sec:algo:KZEqfhat}
\frac{d}{d x_4}\langle \boldsymbol{\hat{f}}(x_4)|=\bB \frac{d}{d x_4}\langle\boldsymbol{f}^{4,+}(x_4)|=\bB \left(\frac{\boldsymbol{\Omega}_4^{42}}{x_4}+\frac{\boldsymbol{\Omega}_4^{43}}{x_4-1} \right)\bB^{-1}\langle\boldsymbol{\hat{f}}(x_4)|\,,
\end{equation}
such that 
\begin{equation}\label{sec:rec:ETrafoOmega}
\boldsymbol{e}_0=\bB \boldsymbol{\Omega}_4^{42}\bB^{-1}\,,\qquad \boldsymbol{e}_1=\bB \boldsymbol{\Omega}_4^{43}\bB^{-1}\,.
\end{equation}
This actually proves that $\frac{d}{d x_4}\bhF(x_4)$ in \eqref{amplitudes:KZhatF} can indeed be cast in the form of the KZ equation. To summarise the content of the next subsections, calculating the basis transformation $\bB$ leads in combination with the recursive construction of $\boldsymbol{\Omega}_4^{42}$ and $\boldsymbol{\Omega}_4^{43}$ to explicit expressions for the matrices $\boldsymbol{e}_0$ and $\boldsymbol{e}_1$. Note that as shown in \subsecref{sec:derGraph}, alternatively, the braid matrices $\boldsymbol{\Omega}_4^{42}$ and $\boldsymbol{\Omega}_4^{43}$ can conveniently be calculated using a graphical procedure in terms of directed trees.

The coefficients $b_{\nu;i_5, i_{6},\dots, i_{n}}^{\sigma}$ can be calculated using \eqn{twistedTheory:basisTrafo}, which simplifies by a certain choice of the basis of the dual space $H^{n-4}(\mathcal{F}_{n,4},\nabla_{-\hat{\omega}})$. As shown in \rcite{Mizera:2019gea}, a dual basis $\{|f^{4,-}_{i_5,i_6,\dots,i_n}\rangle\} \subset H^{n-4}(\mathcal{F}_{n,4},\nabla_{-\hat{\omega}})$ orthonormal to $\{\langle f^{4,+}_{i_5,i_6,\dots,i_n}|\} \subset H^{n-4}(\mathcal{F}_{n,4},\nabla_{\hat{\omega}})$, where  $3\leq i_k < k$, is given by the twisted forms represented by the elements of the recursively constructed vector
\begin{align*}
(\boldsymbol{f}^{p-1,-})_q&=\left(\frac{d x_p}{x_{p,q}}-\frac{d x_p}{x_{p,2}}\right)(\boldsymbol{\Omega}^{pq}_p)^{T}\wedge \boldsymbol{f}^{p,-}\,,
\end{align*}
where $3\leq q<p\leq n$ and $\boldsymbol{f}^{n,-}=(f^{n,-})=(1)$. The orthonormality condition \cite{Mizera:2019gea}
\begin{align}\label{sec:ampl:orthonormFibration}
\langle \boldsymbol{f}^{4,+}|(\boldsymbol{f}^{4,-})^{T}\rangle &=\mathbb{I}
\end{align} 
implies that according to \eqn{twistedTheory:basisTrafoC} the coefficients in the basis transformation \eqref{add:basisTrafoEntries} are the intersection numbers
\begin{align}\label{add:c}
b_{\nu;i_5, i_{6},\dots, i_{n}}^{\sigma}&=\langle \hat{f}^{\sigma}_{\nu}|f^{4,-}_{i_5,i_6,\dots,i_n}\rangle\,.
\end{align}
Furthermore, using the transformation \eqref{amplitudes:trafoFNuSigmaToZ} the coefficients can be expressed in terms of the intersection numbers $
\langle \PT(\rho_{\nu})|f^{4,-}_{i_5,i_6,\dots,i_n}\rangle$,
where $\rho_{\nu}\in S_{N}$ is a permutation of the form
\begin{align*}
\rho_{\nu}:(1,2,\dots,\nu,\nu+1,\dots,N)\mapsto (1,\rho_{\nu}(2),\dots,\rho_{\nu}(\nu),N,\rho_{\nu}(\nu+1),\dots,\rho_{\nu}(N-1),N-1)\,.
\end{align*}
There are several ways to recursively compute the coefficients $b_{\nu;i_5, i_{6},\dots, i_{n}}^{\sigma}$, i.e.\ the intersection numbers $\langle \hat{f}^{\sigma}_{\nu}|f^{4,-}_{i_5,i_6,\dots,i_n}\rangle$ or $ \langle \PT(\rho_{\nu})|f^{4,-}_{i_5,i_6,\dots,i_n}\rangle$, respectively. Two methods are described in the following subsections. The first is purely combinatorial and the second originates in the recently proposed recursion for intersection numbers in \rcite{Mizera:2019gea} and will be shown to be equivalent to the former.

\subsubsection{Partial-fractioning algorithm using directed tree graphs}\label{sec:rec:algo}
In contrast to usual calculations of intersection numbers of twisted forms, it is possible to avoid consideration of any pole structures of the twisted forms involved to calculate the coefficients $b_{\nu;i_5, i_{6},\dots, i_{n}}^{\sigma}$ of the basis transformation \eqref{add:basisTrafoEntries} and instead employ an algorithm defined by partial fractioning.

Recall that $ \hat{f}_{\nu}^{\sigma}$ is the form in the integrand of $ \hat{F}_{\nu}^{\sigma}$ given in \eqn{amplitudes:FhatSigmanu}, and thus it is a linear combination
\begin{align}\label{sec:rec:twistedFormsHFSigmaPerturbed}
\hat{f}_{\nu}^{\sigma}&= \sigma \left(\prod_{k=n-\nu+2}^{n}\sum_{j=3}^{k-1}\frac{t_{kj}}{x_{kj}}\prod_{m=5}^{n-\nu+1}\left(\sum_{l=5}^{m-1}\frac{t_{ml}}{x_{ml}}+\frac{t_{m3}}{x_{m3}}\right)\right)dx_5\wedge dx_6\wedge\dots\wedge dx_n\nonumber\\
&=\sum_{(i_5,i_6,\dots,i_n)\in I_{\nu}}\left(\prod_{k=5}^n t_{ k, i^{\sigma}_k}\right)  \varphi^{\sigma}_{i_5,i_6,\dots,i_n}
\end{align}
of the differential forms
\begin{align}\label{add:sigmaPertFormFrac}
\varphi^{\sigma}_{i_5,i_6,\dots,i_n} 
&=\frac{dx_5\wedge dx_6\wedge\dots\wedge dx_n}{x_{ 5, i^{\sigma}_5}x_{ 6, i^{\sigma}_6}\cdots x_{ n, i^{\sigma}_n} }\,,
\end{align}
where 
\begin{align}\label{sec:algo:Inu}
I_{\nu}&=\lbrace (i_5,i_6,\dots,i_n)\in\ZN^{n-4}\,|\,3\leq i_k<k\text{ for all }k \text{ and } i_k\neq 4\text{ for } 5\leq k\leq n-\nu+1\rbrace
\end{align}
and
\begin{equation}
i^{\sigma}_k=\sigma (i_{\sigma^{-1} (k)})\,,
\end{equation}
which in general does not satisfy $3\leq i^{\sigma}_k<k$. 

Let us call an index $i_k$ labelled by $k$ satisfying
\begin{equation}
3 \leq i_k<k
\end{equation}
\textit{admissible}, and \textit{non-admissible} otherwise. A variable $x_{i_k}$ with admissible index $i_k$ is called admissible as well, which upon comparing with \figref{figRiemannSphere} simply means that $x_k<x_{i_k}\leq 1$, and $x_{i_k}$ is called non-admissible if $i_k$ is non-admissible. Similarly, we call a sequence $(i_5,i_6,\dots,i_n)$ admissible if all the indices $i_k$ are admissible, and non-admissible otherwise. Furthermore, if $(i_5,i_6,\dots,i_n)$ is admissible, the sequence $(i^{\sigma}_5,i^{\sigma}_6,\dots,i^{\sigma}_n)$ is called \textit{$\sigma$-permuted admissible}. 

In order to conveniently formulate the algorithm below, let us introduce the following graphical notation\footnote{The graphical notation is introduced for three purposes: first, the calculations involving iterative applications of partial fractioning can be displayed intuitively. Second, this notation is adapted to the analogous genus-one string integrals in a forthcoming project \cite{BroedelGeneralizedGraphFunctions}, where only an additional weight for each edge has to be introduced to fully describe the corresponding integrands. There, the graphs and manipulations thereon will be essential to capture the complexity of the derivations of various identities, which include the genus-one extensions to this article. Third, this representation facilitates efficient computer implementations (in various programming languages, not necessarily computer algebra systems needed for symbolic manipulations) of the algorithm below and calculations using adjacency matrices only.} for products of fractions in terms of directed graphs. For a single factor $\frac{1}{x_{ji}}$ we write
\begin{align*}
\frac{1}{x_{ji}}&=\grEdgeReverse{j}{i}=\grEdge{i}{j}\,,
\end{align*}
where the arrow points in the direction of the first index of $x_{ji}$. By definition, reversing an arrow introduces a minus sign:
\begin{equation*}
\frac{1}{x_{ij}}=\grEdgeReverse{i}{j}=\grEdge{j}{i}=-\grEdgeReverse{j}{i}=-\frac{1}{x_{ji}}\,.
\end{equation*}
A graph $\grEdge{i}{j}$ is called admissible if the arrow points from a smaller number $i$ to a larger number $j$ and non-admissible otherwise. More generally, a fraction of a product of $\prod_{k=5}^n x_{k,i_k}$ can be represented by a directed graph
\begin{equation}\label{add:formToTuple}
\frac{1}{x_{ 5, i_5}x_{ 6, i_6}\cdots x_{ n, i_n} }= \prod_{k=5}^n \grEdge{i_k}{k}\,,
\end{equation}
where the product of two edges with a coinciding vertex is defined by concatenation 
\begin{equation*}
\frac{1}{x_{ji}}\frac{1}{x_{kj}}=\grEdge{i}{j}\, \grEdge{j}{k}=\grDoubleEdge{i}{j}{k}\,.
\end{equation*}
For example, for $n=8$ and the admissible sequence $(i_5,i_6,i_7,i_8)=(3,5,4,5)$, we can write the following product $g$ as
\begin{equation*}
g=\frac{1}{x_{53}x_{65}x_{74}x_{85}}={\tiny\begin{tikzpicture}
	\tikzset{vertex/.style = {shape=circle,draw,minimum size=0.5em}}
	\tikzset{edge/.style = {->,> = latex'}}
	\tikzset{baseline={(0, -0.4em)}}
	\node[vertex] (3) at (0,0.75) {3};
	\node[vertex] (5) at (0,0) {5};
	\node[vertex] (6) at (-0.5,-0.75) {6};
	\node[vertex] (8) at (0.5,-0.75) {8};
	\node[vertex] (4) at (1,0.75) {4};
	\node[vertex] (7) at (1,0) {7};
	\draw[edge] (3) to (5);
	\draw[edge] (5) to (6);
	\draw[edge] (5) to (8);
	\draw[edge] (4) to (7);
	\end{tikzpicture}}\,.
\end{equation*}
Using this example, more notation may be introduced following the established convention for directed (tree) graphs. The graph $g$ consists of the two \textit{subgraphs} \\
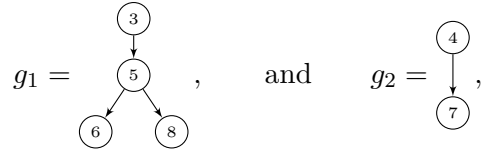
\begin{equation*}
g_1={\tiny\begin{tikzpicture}
	\tikzset{vertex/.style = {shape=circle,draw,minimum size=0.5em}}
	\tikzset{edge/.style = {->,> = latex'}}
	\tikzset{baseline={(0, -0.4em)}}
	\node[vertex] (3) at (0,0.75) {3};
	\node[vertex] (5) at (0,0) {5};
	\node[vertex] (6) at (-0.5,-0.75) {6};
	\node[vertex] (8) at (0.5,-0.75) {8};
	\draw[edge] (3) to (5);
	\draw[edge] (5) to (6);
	\draw[edge] (5) to (8);
	\end{tikzpicture}}\,,\qquad \text{and}\qquad 	g_2={\tiny\begin{tikzpicture}
\tikzset{vertex/.style = {shape=circle,draw,minimum size=0.5em}}
	\tikzset{edge/.style = {->,> = latex'}}
	\tikzset{baseline={(0, -0.4em)}}
	\node[vertex] (4) at (0,0.5) {4};
	\node[vertex] (7) at (0,-0.5) {7};
	\draw[edge] (4) to (7);
	\end{tikzpicture}}\,,
\end{equation*}
which are the two independent factors in the fraction $g=g_1 g_2$. The first subgraph $g_1$ has a \textit{branch point} at the vertex $5$ and some subgraphs with no branch point, i.e.\ \textit{branches}, for example the two branches
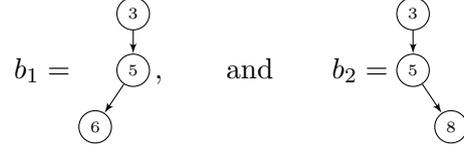
\begin{equation*}
b_1={\tiny\begin{tikzpicture}
	\tikzset{vertex/.style = {shape=circle,draw,minimum size=0.5em}}
	\tikzset{edge/.style = {->,> = latex'}}
	\tikzset{baseline={(0, -0.4em)}}
	\node[vertex] (3) at (0,0.75) {3};
	\node[vertex] (5) at (0,0) {5};
	\node[vertex] (6) at (-0.5,-0.75) {6};
	\draw[edge] (3) to (5);
	\draw[edge] (5) to (6);
	\end{tikzpicture}}\,,\qquad \text{and}\qquad 	b_2={\tiny\begin{tikzpicture}
	\tikzset{vertex/.style = {shape=circle,draw,minimum size=0.5em}}
	\tikzset{edge/.style = {->,> = latex'}}
	\tikzset{baseline={(0, -0.4em)}}
	\node[vertex] (3) at (0,0.75) {3};
	\node[vertex] (5) at (0,0) {5};
	\node[vertex] (8) at (0.5,-0.75) {8};
	\draw[edge] (3) to (5);
	\draw[edge] (5) to (8);
	\end{tikzpicture}}
\end{equation*}
with \textit{root} vertex $3$. The graph $g$ has two roots: $3$ and $4$. Moreover, two vertices $i$ and $j$ are called \textit{branch-connected} if there exists a branch which contains $i$ and $j$.

Thus, upon identifying the coefficient of the differential forms $f^{4,+}_{i_5,i_6,\dots,i_n}$ and $\varphi^{\sigma}_{i_5,i_6,\dots,i_n}$ with such a directed tree graph, the fibration basis corresponds to the admissible sequences
\begin{equation}\label{add:fibBasis}
f^{4,+}_{i_5,i_6,\dots,i_n}=\left(\prod_{k=5}^n \grEdge{i_k}{k}\right) dx_5\wedge dx_6\wedge\dots\wedge dx_n\,,\qquad (i_5,i_6,\dots,i_n)\text{ admissible}
\end{equation}
and the differential forms in $\hat{F}^{\sigma}_{\nu}$ to the $\sigma$-permuted admissible sequences 
\begin{equation}\label{add:sigmaPertForm}
\varphi^{\sigma}_{i_5,i_6,\dots,i_n}=\left(\prod_{k=5}^n \grEdge{i^{\sigma}_k}{k}\right) dx_5\wedge dx_6\wedge\dots\wedge dx_n\,,\qquad (i_5,i_6,\dots,i_n)\text{ admissible}.
\end{equation}
Such graphs $\prod_{k=5}^n \grEdge{i_k}{k}$ and $\prod_{k=5}^n \grEdge{i^{\sigma}_k}{k}$, where $(i_5,i_6,\dots,i_n)$ is admissible, are called admissible and $\sigma$-permuted admissible, respectively. Going the other way around, a graph is admissible if and only if for all vertices $5\leq v\leq n$, there is exactly one vertex pointing from a lower vertex to $v$ and the vertices $v=2,3,4$ have no incoming arrows. If the vertices $v=2,3,4$ have no outgoing arrow either, they are often omitted and not shown, which can be justified by defining a vertex without arrows to equal unity, i.e.\  ${\tiny\begin{tikzpicture}
	\tikzset{vertex/.style = {shape=circle,draw,minimum size=0.5em}}
	\tikzset{edge/.style = {->,> = latex'}}
	\tikzset{baseline={(0, -0.4em)}}
	\node[vertex] (4) at (0,0) {$v$};
	\end{tikzpicture}}=1$. Correspondingly, the differential forms in \eqn{add:sigmaPertForm} are called \textit{$\sigma$-permuted admissible forms}. Note that since $i^{\sigma}_k=\sigma (i_{\sigma^{-1}( k)})$, upon comparing \eqns{add:fibBasis}{add:sigmaPertForm}, we see that for $\sigma=\id$ the $\sigma$-permuted admissible forms are exactly the elements of the fibration basis
\begin{align}\label{sec:rec:idPermutedForms}
\varphi^{\id}_{i_5,i_6,\dots,i_n}&=f^{4,+}_{i_5,i_6,\dots,i_n}\,.
\end{align}

Below, we will show that the $\sigma$-permuted forms can combinatorially be expressed as a linear combination of the fibration basis only using the partial-fractioning identity
\begin{equation}\label{app:pf:id}
\frac{1}{x_{k,l}}\frac{1}{x_{k,m}}=\frac{1}{x_{k,l}}\frac{1}{x_{l,m}}-\frac{1}{x_{k,m}}\frac{1}{x_{l,m}}=\left(\frac{1}{x_{k,l}}-\frac{1}{x_{k,m}}\right)\frac{1}{x_{l,m}}\,,
\end{equation}
where $m<l<k$. This identity can be expressed in terms of an operation on the directed trees
\begin{equation}\label{app:pfIdGraphs}
\grTriangleSink{m}{l}{k}=\grTriangleClock{m}{l}{k}-\grTriangleSource{m}{l}{k}=\left(\grEdge{l}{k}-\grEdge{m}{k}\right)\grEdge{m}{l}\,,
\end{equation}
where, as for the multiplication, the distributivity and additivity of the graphs follows directly from their definition \eqref{add:formToTuple} as fractions. Note that since $m<l<k$ the graph on the left-hand side and the graphs on the right-hand side are admissible. Therefore, the application of the partial-fractioning identity in the form \eqref{app:pf:id} for the ordering $m<l<k$ preserves admissibility and defines a structure-preserving operation on the space of admissible graphs and forms, respectively. Or, turning the reasoning around, the representation of the partial fractioning identity in \eqns{app:pf:id}{app:pfIdGraphs} is the unique choice preserving the admissibility for $m<l<k$. First, this allows to reconnect the vertices in a given branch keeping the admissibility. Second, this reconnecting of a branch will allow us to rewrite non-admissible branches as linear combinations of admissible ones. Consecutive applications of the partial-fractioning identity can conveniently be described using double arrows for $m<l<k$
\begin{equation}\label{sec:algo:pfGraphs}
\grTriangleDSink{m}{l}{k}=\grTriangleClock{m}{l}{k}-\grTriangleSource{m}{l}{k}\,,
\end{equation}
where the sign on the right-hand side is determined by the single arrow on the left-hand side: the diagram where the two single arrows begin on the same vertex picks up a negative sign. Using this notation, the partial-fractioning identity \eqref{app:pf:id} is expressed in terms of the following identity of graphs:
\begin{equation}
\grTriangleSink{m}{l}{k}=\grTriangleDSink{m}{l}{k}\qquad \Leftrightarrow\qquad \frac{1}{x_{k,l}}\frac{1}{x_{k,m}}=\frac{1}{x_{k,l}}\frac{1}{x_{l,m}}-\frac{1}{x_{k,m}}\frac{1}{x_{l,m}}\,.
\end{equation}
Recursively, we denote for $n<m<l<k$ the successive application of the Fay identity, which always starts at the highest vertex, as follows
\begin{equation}\label{sec:algo:ex2PF}
\grSquareSink{m}{k}{l}{n}=\grSquareClock{m}{k}{l}{n}-\grSquareSource{m}{k}{l}{n}=\grSquareDSink{m}{k}{l}{n}=\grSquareDClock{m}{k}{l}{n}-\grSquareDSource{m}{k}{l}{n}=\grSquareDDSink{m}{k}{l}{n}\,.
\end{equation}
Thus, writing out the double arrows in terms of sums and beginning at the smallest vertex, the graph on the right-hand side is defined to be the linear combination
\begin{equation*}
\grSquareDDSink{m}{k}{l}{n}=\grSquareDClock{m}{k}{l}{n}-\grSquareDSource{m}{k}{l}{n}=\grSquareClockClock{m}{k}{l}{n}-\grSquareClockSouce{m}{k}{l}{n}-\grSquaren{m}{k}{l}{n}+\grSquarec{m}{k}{l}{n}\,,
\end{equation*}
where the sign of the single graphs on the right-hand side is determined by the direction of the double arrows on the left-hand side, as for the original definition \eqref{sec:algo:pfGraphs}. Thus, the identity
\begin{align*}
\grSquareSink{m}{k}{l}{n}&=\grSquareDDSink{m}{k}{l}{n}
\end{align*}
is the partial-fractioning identity
\begin{align*}
\frac{1}{x_{ln}x_{km}x_{kl}}&=\frac{1}{x_{mn}}\left(\left(\frac{1}{x_{kl}}- \frac{1}{x_{km}}\right)\left(\frac{1}{x_{lm}}- \frac{1}{x_{ln}}\right)\right)\nonumber\\
&=\frac{1}{x_{mn}x_{lm}x_{kl}}-\frac{1}{x_{mn}x_{lm}x_{km}}-\frac{1}{x_{mn}x_{ln}x_{kl}}+\frac{1}{x_{mn}x_{ln}x_{km}}\,.
\end{align*}

Using these definitions, the algorithm\footnote{A related algorithm to convert non-admissible to sums of admissible sequences is used in a similar context in \rcite{aomoto1987}.} explained on an example in \appref{app:ex} to express a $\sigma$-permuted admissible form $	\varphi^{\sigma}_{i_5,i_6,\dots,i_n}$ in terms of the fibration basis and hence, to determine the entries $b_{\nu;i_5,\dots,i_n}^{\sigma}$ of the basis transformation $\bB$ can be summarised as follows:
\begin{enumerate}
	\item Express the form $	\varphi^{\sigma}_{i_5,i_6,\dots,i_n}$ in terms of its graph $\prod_{k=5}^n \grEdge{i^{\sigma}_k}{k}$ using \eqn{add:sigmaPertForm}.
	\item If existing, consider the highest vertex $h\in \{5,6,\dots,n-1\}$ of $\prod_{k=5}^n \grEdge{i^{\sigma}_k}{k}$ with a non-admissible subgraph $\grEdgeLong{i^{\sigma}_{h}}{h}$, i.e.\
	\begin{equation}
	h=\max\lbrace k\in \{5,6,\dots,n-1\}\,|\,k<i_k^{\sigma} \rbrace\,.
	\end{equation}
	If no such $h$ exists, the graph $\prod_{k=5}^n \grEdge{i^{\sigma}_k}{k}$ is admissible, hence, the $\sigma$-permuted admissible form $\varphi^{\sigma}_{i_5,i_6,\dots,i_n}=f^{4,+}_{i^{\sigma}_5,i^{\sigma}_6,\dots,i^{\sigma}_n}$ is admissible and an element of the fibration basis, and we are done. Otherwise, there exists\footnote{See \appref{app:proofAlgo} for an explicit proof.} a positive integer $l$ and vertices
	\begin{equation*}
	h^l<h<h^{l-1}<h^{l-2}<\dots < h^{1}<i_{h}^{\sigma}
	\end{equation*}
	such that the graph
	\begin{align}\label{sec:algo:branchNonAdm}
	b_h&={\tiny\begin{tikzpicture}
		\tikzset{vertex/.style = {shape=circle,draw,minimum size=0.5em}}
		\tikzset{edge/.style = {->,> = latex'}}
		\tikzset{baseline={(0, -0.4em)}}
		\node[vertex] (1) at (0,0) {$h^l$};
		\node[vertex] (2) at (1.5,0) {$h^{l-1}$};
		\node[vertex] (3) at (3,0) {$h^{l-2}$};
		\node[] (4) at (4.5,0) {\dots};
		\node[vertex] (5) at (5.5,0) {$h^1$};
		\node[vertex] (6) at (6.5,0) {$i^{\sigma}_h$};
		\node[vertex] (7) at (7.5,0) {h};
		\draw[edge] (1) to (2);
		\draw[edge] (2) to (3);
		\draw[edge] (3) to (4);
		\draw[edge] (4) to (5);
		\draw[edge] (5) to (6);
		\draw[edge] (6) to (7);
		\end{tikzpicture}}\nonumber\\
	&=\frac{1}{x_{h, i_h^{\sigma}}}\frac{1}{x_{i_h^{\sigma}, h^1}}\prod_{k=1}^{l-1}\frac{1}{x_{h^k, h^{k+1}}}
	\end{align}
	is a subgraph/factor of the branch containing the vertex $h$, i.e.\ a subbranch. Using the partial-fractioning identity  \eqref{sec:algo:pfGraphs} iteratively, as in the example \eqref{sec:algo:ex2PF}, this subbranch can be written as a linear combination of admissible graphs only
	\begin{align}\label{sec:algo:admissibleSubbranch}
		b_h&=-{\tiny\begin{tikzpicture}
			\tikzset{vertex/.style = {shape=circle,draw,minimum size=0.5em}}
			\tikzset{edge/.style = {->,> = latex'}}
			\tikzset{baseline={(0, -0.4em)}}
			\node[vertex] (1) at (0,0.25) {$h^l$};
			\node[vertex] (2) at (1.5,0.25) {$h^{l-1}$};
			\node[vertex] (3) at (3,0.25) {$h^{l-2}$};
			\node[] (4) at (4.5,0.25) {\dots};
			\node[vertex] (5) at (5.5,0.25) {$h^1$};
			\node[vertex] (6) at (6.5,0.25) {$i^{\sigma}_h$};
			\node[vertex] (7) at (6,-0.5) {h};
			\draw[edge] (1) to (2);
			\draw[edge] (2) to (3);
			\draw[edge] (3) to (4);
			\draw[edge] (4) to (5);
			\draw[edge] (5) to (6);
			\draw[edge] (7) to (6);
			\end{tikzpicture}}\nonumber\\
		&=-{\tiny\begin{tikzpicture}
			\tikzset{vertex/.style = {shape=circle,draw,minimum size=0.5em}}
			\tikzset{edge/.style = {->,> = latex'}}
			\tikzset{baseline={(0, -0.4em)}}
			\node[vertex] (1) at (0,0.25) {$h^l$};
			\node[vertex] (2) at (1.5,0.25) {$h^{l-1}$};
			\node[vertex] (3) at (3,0.25) {$h^{l-2}$};
			\node[] (4) at (4.5,0.25) {\dots};
			\node[vertex] (5) at (5.5,0.25) {$h^1$};
			\node[vertex] (6) at (6.5,0.25) {$i^{\sigma}_h$};
			\node[vertex] (7) at (6,-0.5) {h};
			\draw[edge] (1) to (2);
			\draw[edge] (2) to (3);
			\draw[edge] (3) to (4);
			\draw[edge] (4) to (5);
			\draw[double,edge] (5) to (6);
			\draw[double,edge] (7) to (6);
			\draw[edge] (7) to (5);
			\end{tikzpicture}}\nonumber\\
	&=-{\tiny\begin{tikzpicture}
		\tikzset{vertex/.style = {shape=circle,draw,minimum size=0.5em}}
		\tikzset{edge/.style = {->,> = latex'}}
		\tikzset{baseline={(0, -0.4em)}}
		\node[vertex] (1) at (0,0.25) {$h^l$};
		\node[vertex] (2) at (1.5,0.25) {$h^{l-1}$};
		\node[vertex] (3) at (3,0.25) {$h^{l-2}$};
		\node[] (4) at (4.5,0.25) {\dots};
		\node[vertex] (5) at (5.5,0.25) {$h^1$};
		\node[vertex] (6) at (6.5,0.25) {$i^{\sigma}_h$};
		\node[vertex] (7) at (3.5,-1) {h};
		\draw[double,edge] (1) to (2);
		\draw[double,edge] (2) to (3);
		\draw[double,edge] (3) to (4);
		\draw[double,edge] (4) to (5);
		\draw[double,edge] (5) to (6);
		\draw[edge] (1) to (7);
		\draw[double,edge] (7) to (6);
		\draw[double,edge] (7) to (5);
		\draw[double,edge] (7) to (4);
		\draw[double,edge] (7) to (3);
		\draw[double,edge] (7) to (2);
		\end{tikzpicture}}\nonumber\\
	&=-\frac{1}{x_{h, h^l}}\left(\frac{1}{x_{i_h^{\sigma}, h^1}}-\frac{1}{x_{i_h^{\sigma}, h}}\right)\prod_{k=1}^{l-2}\left(\frac{1}{x_{h^k, h^{k+1}}}-\frac{1}{x_{h^k, h}}\right)\left(\frac{1}{x_{h^{l-1}, h}}-\frac{1}{x_{h^{l-1}, h^l}}\right)\,,
	\end{align}
	where the right-hand side is indeed a linear combination of admissible graphs only, since all the arrows point from a lower number to a higher number, no vertex has two incoming single arrows and the vertex $h^l$ from which a single arrow points to $h$ is also smaller than $h$, i.e.\ each term has an admissible subgraph $\grEdge{h^l}{h}$, unlike the original graph in \eqn{sec:algo:admissibleSubbranch} with $\grEdge{i_h^{\sigma}}{h}$ and $i_h^{\sigma}>h$. Thus, replacing the subbranch $b_h$ in the graph $\prod_{k=5}^n \grEdge{i^{\sigma}_k}{k}$ by the right-hand side of \eqn{sec:algo:admissibleSubbranch} yields a linear combination of graphs, where for each graph, the highest vertex $j$ with non-admissible subgraph $\grEdge{i^{\sigma}_j}{j}$ is smaller than $h$.
	\item Repeat step 2 for each graph in the linear combination obtained above.
\end{enumerate}
This algorithm ends after finitely many repetitions of the second step and yields the linear combination of $\varphi^{\sigma}_{i_5,i_6,\dots,i_n}$ in terms of the fibration basis given by the entries $f^{4,+}_{i_5,i_6,\dots,i_n}$ of $\boldsymbol{f}^{4,+}$. Therefore, this algorithm defines a vector-valued map $\boldsymbol{\adm}=(\adm_{i'_5,i'_6,\dots,i'_n})_{3\leq i'_k<k }$ acting on a $\sigma$-permuted admissible (twisted) form and mapping it to $\ZZ^{(n-3)!}$, with entries $\adm_{i'_5,i'_6,\dots,i'_n}$ given by the coefficients of this unique linear combination:
\begin{equation*}
\varphi^{\sigma}_{i_5,i_6,\dots,i_n}=\sum_{(i'_5,i'_6,\dots,i'_n)}\adm_{i'_5,i'_6,\dots,i'_n}\left(\varphi^{\sigma}_{i_5,i_6,\dots,i_n}\right) f^{4,+}_{i'_5,i'_6,\dots,i'_n}=\boldsymbol{\adm}^T\left(\varphi^{\sigma}_{i_5,i_6,\dots,i_n}\right) \boldsymbol{f}^{4,+}\,,
\end{equation*}
where the sum runs over all the admissible sequences $(i'_5,i'_6,\dots,i'_n)$, i.e.\ $3\leq i'_k<k$ for all $k=5,6,\dots,n$. The intersection numbers $b_{\nu;i_5, i_{6},\dots, i_{n}}^{\sigma}=\langle f^{\sigma}_{\nu}|f^{4,-}_{i_5,i_6,\dots,i_n}\rangle$ can then be obtained using \eqn{sec:rec:twistedFormsHFSigmaPerturbed}: the twisted form $\langle \hat{f}_{\nu}^{\sigma}|$ can be calculated as follows
\begin{align*}
\langle \hat{f}_{\nu}^{\sigma}|&=\sum_{(i_5,i_6,\dots,i_n)\in I_{\nu}}\left(\prod_{k=5}^n t_{ k, i^{\sigma}_k}\right) \langle \varphi^{\sigma}_{i_5,i_6,\dots,i_n}|\nonumber\\
&=\sum_{(i_5,i_6,\dots,i_n)\in I_{\nu}}\left(\prod_{k=5}^n t_{ k, i^{\sigma}_k}\right) \boldsymbol{\adm}^T\left(\varphi^{\sigma}_{i_5,i_6,\dots,i_n}\right) \langle\boldsymbol{f}^{4,+}|\,,
\end{align*}
where the set $I_{\nu}$ is defined in \eqn{sec:algo:Inu}, such that the intersection numbers are the coefficients in the above linear combination 
\begin{align}\label{sec:algo:IntCoef}
b_{\nu;i'_5, i'_{6},\dots, i'_{n}}^{\sigma}&=\sum_{(i_5,i_6,\dots,i_n)\in I_{\nu}}\left(\prod_{k=5}^n t_{ k, i^{\sigma}_k}\right)\adm_{i'_5,i'_6,\dots,i'_n}\left(\varphi^{\sigma}_{i_5,i_6,\dots,i_n}\right)\,.
\end{align}
They are in particular homogeneous polynomials of degree $n-4$ in the Mandelstam variables. Note that if $\sigma=\id$, the algorithm is trivial since $\varphi^{\id}_{i_5,i_6,\dots,i_n}$ is already admissible, such that
\begin{align*}
\adm_{i'_5,i'_6,\dots,i'_n}\left(\varphi^{\id}_{i_5,i_6,\dots,i_n}\right)&=\prod_{k=5}^n \delta_{i_k,i'_k}
\end{align*}
and \eqn{sec:algo:IntCoef} simplifies to 
\begin{align}\label{sec:algo:IntCoefId}
b_{\nu;i_5, i_{6},\dots, i_{n}}^{\id}&=\begin{cases}
\prod_{k=5}^n t_{ k, i_k}&\text{if } (i_5,i_6,\dots,i_n)\in I_{\nu} \,,\\
0&\text{otherwise}\,.
\end{cases}
\end{align}
Since the intersection numbers are the coefficients of the basis transformation \eqref{add:basisTrafoEntries}, the rows of the transformation matrix in \eqn{sec:rec:BasisTrafo}, i.e.\
\begin{align}\label{sec:algo:B}
\langle\boldsymbol{\hat{f}}(x_4)|&=\bB \langle\boldsymbol{f}^{4,+}(x_4)|\,,
\end{align}
are given by 
\begin{align}\label{sec:algo:BRow}
\bB^{\sigma}_{\nu}&=\sum_{(i_5,i_6,\dots,i_n)\in I_{\nu}}\left(\prod_{k=5}^n t_{ k, i^{\sigma}_k}\right)\boldsymbol{\adm}^T\left(\varphi^{\sigma}_{i_5,i_6,\dots,i_n}\right)\,.
\end{align}

The above algorithm only uses partial fractioning, which is an identity on the level of the differential forms and not only an identity of their twisted cohomology class (unlike integration by parts). This implies that \eqns{sec:algo:B}{sec:algo:BRow} also hold on the level of the differential forms, i.e.\ 
\begin{align*}
\boldsymbol{\hat{f}}(x_4)=\bB \boldsymbol{f}^{4,+}(x_4)\,.
\end{align*}

\subsubsection{Examples: the four- and five-point string integrals}\label{sec:rec:ExamplesAlgo}
The above algorithm is applied to some examples in \appref{app:ex}, in particular to derive the transformation matrix $\bB$ for $n=5$ and $n=6$, i.e.\ for the four- and five-point string integrals. 

In the four-point case, the transformation matrix in
\begin{align*}
\langle \boldsymbol{\hat{f}}|&=\bB \langle\boldsymbol{f}^{4,+}|
\end{align*}
is given by 
\begin{align*}
\bB&=\begin{pmatrix}-t_{53}& -t_{54}\\ -t_{53}&0	
\end{pmatrix}\,.
\end{align*}
Thus, the matrices $\boldsymbol{e}_0$ and $\boldsymbol{e}_1$ appearing in the KZ equation of $\langle \boldsymbol{\hat{f}}|$ can immediately be obtained using the braid matrices for $n=5$  given in \eqn{sec:amp:Omega4} and the transformation in \eqn{sec:rec:ETrafoOmega}. They read
\begin{equation*}
\boldsymbol{e}_0=\bB \boldsymbol{\Omega}_4^{42}\bB^{-1}=\begin{pmatrix}
t_{542}&-t_{52}\\0&t_{42}
\end{pmatrix}\,,\quad \boldsymbol{e}_1=\bB \boldsymbol{\Omega}_4^{43}\bB^{-1}=\begin{pmatrix}
t_{43}&0\\-t_{53}&t_{543}
\end{pmatrix}
\end{equation*}
and degenerate in the limit $t_{4i}\rightarrow 0$ to the matrices found in \rcite{Broedel:2013aza} and given in \eqn{sec:rec:4PtEx}.

The calculation of the five-point string integrals, which corresponds to $n=6$, requires non-trivial applications of the algorithm. The resulting transformation matrix for
\begin{align*}
\langle \boldsymbol{\hat{f}}|&=\bB \langle\boldsymbol{f}^{4,+}|
\end{align*}
turns out to be
\begin{align*}
\bB&=\begin{pmatrix}
t_{63}t_{53}&t_{64}t_{53}&t_{65}t_{53}&t_{63}t_{54}&t_{64}t_{54}&t_{65}t_{54}\\
t_{63}t_{53}+t_{65}t_{63}&t_{64}t_{53}&-t_{65}t_{63}&t_{63}t_{54}&t_{64}t_{54}+t_{64}t_{65}&-t_{65}t_{64}\\
t_{63}t_{53}&t_{64}t_{53}&t_{65}t_{53}&0&0&0\\
t_{63}t_{53}+t_{65}t_{63}&0&-t_{65}t_{63}&t_{63}t_{54}&0&0\\
t_{63}t_{53}&0&t_{65}t_{53}&0&0&0\\
t_{63}t_{53}+t_{65}t_{63}&0&-t_{65}t_{63}&0&0&0
\end{pmatrix}\,.
\end{align*}
Therefore, the matrices $\boldsymbol{e}_0$ and $\boldsymbol{e}_1$ are given by
\begin{equation*}
\boldsymbol{e}_0=\bB \boldsymbol{\Omega}^4_{42}\bB^{-1}=
\begin{pmatrix}
t_{6542} & 0 & -t_{52}-t_{65} &
-t_{62} & -t_{62} & t_{62} \\
0 & t_{6542} & -t_{52} &
-t_{62}-t_{65} & t_{52} & -t_{52} \\
0 & 0 & t_{642} & 0 & -t_{62} & 0 \\
0 & 0 & 0 & t_{542} & 0 & -t_{52} \\
0 & 0 & 0 & 0 & t_{42} & 0 \\
0 & 0 & 0 & 0 & 0 & t_{4 2} 
\end{pmatrix}
\end{equation*}
and
\begin{equation*}
\boldsymbol{e}_1=\bB \boldsymbol{\Omega}^4_{43}\bB^{-1}=
\begin{pmatrix}
t_{43} & 0 & 0 & 0 & 0 & 0 \\
0 & t_{43} & 0 & 0 & 0 & 0 \\
-t_{53} & 0 & t_{543} & 0 & 0 & 0 \\
0 & -t_{63} & 0 & t_{643} & 0 & 0 \\
-t_{53} & t_{53} & -t_{63}-t_{65} & -t_{53} &
t_{6543} & 0 \\
t_{63} & -t_{63} & -t_{63} & -t_{53}-t_{65} & 0 &
t_{6543}
\end{pmatrix}\,.
\end{equation*}
Indeed, in the limit $t_{4i}\rightarrow 0$ the matrices of \rcite{Broedel:2013aza} are recovered. The same behaviour has been checked explicitly for the examples up to $n=9$. 

\subsubsection{Recursive algorithm for intersection numbers of twisted forms}\label{sec:algo:recMizera}
Another approach to recursively calculate the intersection numbers \eqref{add:c} is the application of the recently proposed recursive formula to calculate intersection numbers of twisted forms in \rcite{Mizera:2019gea}. It is based on expressing the differential forms in terms of the fibration basis $\boldsymbol{f}^{p,+}$ and its dual $\boldsymbol{f}^{p,-}$, using their orthonormality \eqref{sec:ampl:orthonormFibration} valid for any $p\in \{3,4,\dots,n\}$ and the behaviour of the fibration bases close to the punctures. In our case, we need to calculate the row vector of intersection numbers $\boldsymbol{B}_{\nu}^{\sigma}=\langle \hat{f}_{\nu}^{\sigma}|(\boldsymbol{f}^{4,-})^T\rangle$. In order to do so, let us define $\boldsymbol{B}_{\nu}^{\sigma,n}=(\hat{f}_{\nu}^{\sigma})$ and for $4 \leq q< n$
\begin{equation*}
\boldsymbol{B}_{\nu}^{\sigma,p}=\langle \hat{f}_{\nu}^{\sigma}|\boldsymbol{f}^{p,-}\rangle\,,
\end{equation*}
such that $(\boldsymbol{B}_{\nu}^{\sigma,4})^T=\boldsymbol{B}_{\nu}^{\sigma}$ is the row \eqref{sec:algo:BRow} of the transformation matrix $\boldsymbol{B}$. The recursion in \rcite{Mizera:2019gea} applied to $\boldsymbol{B}_{\nu}^{\sigma,p}$ is given by
\begin{align}\label{add:recursionIntersectionNumber}
(\boldsymbol{B}_{\nu}^{\sigma,p-1})_{r}&=\sum_{q=2}^{p-1}\Res_{x_p=x_q}\left(\boldsymbol{\mathrm{M}}_{pqr}\boldsymbol{B}_{\nu}^{\sigma,p}\right)\,,
\end{align}
where the matrix
\begin{align*}
\boldsymbol{\mathrm{M}}_{pqr}&=\sum_{k=0}^{\infty}\boldsymbol{\mathrm{M}}^k_{pqr}(x_p-x_q)^k
\end{align*}
is defined at the zeroth order by 
\begin{align*}
\boldsymbol{\mathrm{M}}^0_{pqr}&=\begin{cases}
-(\boldsymbol{\Omega}^{pr}_q)^{T}\left((\boldsymbol{\Omega}^{pr}_q)^{T}\right)^{-1}&\text{if } q=2\,,\\
\mathbb{I}&\text{if } q=r\,,\\
0&\text{otherwise}
\end{cases}
\end{align*}
and at higher orders according to the recursion
\begin{align*}
\boldsymbol{\mathrm{M}}^k_{pqr}&=\left((-1)^{k-1}\left(\frac{\delta_{q\neq r}}{(z_q-z_r)^k}-\frac{\delta_{q\neq 2}}{z_q^k}\right)(\boldsymbol{\mathrm{\Omega}}^{pr}_p)^{T}+\sum_{j\neq q,j=1}^{p-1}\sum_{i=0}^{k-1}\frac{\boldsymbol{\mathrm{M}}^i_{pqr}}{z_j-z_q}(\boldsymbol{\mathrm{\Omega}}^{pr}_p)^{T}\right)\left((\boldsymbol{\mathrm{\Omega}}^{pr}_p)^{T}-k\mathbb{I}\right)^{-1}\,.
\end{align*}
Therefore, using \eqn{sec:rec:twistedFormsHFSigmaPerturbed}, we find
\begin{align*}
(\boldsymbol{B}_{\nu}^{\sigma,n-1})_{i'_n}&=\sum_{q=2}^{n-1}\Res_{x_n=x_q}\left(\boldsymbol{\mathrm{M}}_{nqi'_n}\boldsymbol{B}_{\nu}^{\sigma,n}\right)\nonumber\\
&=\sum_{q=2}^{n-1}\Res_{x_n=x_q}\left(\boldsymbol{\mathrm{M}}_{nqi'_n}\hat{f}^{\sigma}_{\nu}\right)\nonumber\\
&=\sum_{q=2}^{n-1}\Res_{x_n=x_q}\left(\boldsymbol{\mathrm{M}}_{nqi'_n}\sum_{(i_5,i_6,\dots,i_n)\in I_{\nu}}\frac{t_{5,i^{\sigma}_5}t_{6,i^{\sigma}_6}\cdots t_{n,i^{\sigma}_n}}{x_{5,i^{\sigma}_5}x_{6,i^{\sigma}_6}\cdots x_{n,i^{\sigma}_n} }dx_5\wedge dx_6\wedge\dots\wedge dx_n\right)\,.
\end{align*}
But since $i_k^{\sigma}\neq 2$ and the differential forms have only simple poles, i.e.\ are logarithmic, we have by the definition of $\boldsymbol{\mathrm{M}}^0_{pqr}$
\begin{align}\label{sec:algo:recMizeraEquiv}
(\boldsymbol{B}_{\nu}^{\sigma,n-1})_{i'_n}
&=\sum_{(i_5,i_6,\dots,i_n)\in I_{\nu}}\Res_{x_n=x_{i'_n}}\left(\frac{t_{5,i^{\sigma}_5}t_{6,i^{\sigma}_6}\cdots t_{n,i^{\sigma}_n}}{x_{5,i^{\sigma}_5}x_{6,i^{\sigma}_6}\cdots x_{n,i^{\sigma}_n} }dx_5\wedge dx_6\wedge\dots\wedge dx_n\right)\nonumber\\
&=\sum_{(i_5,i_6,\dots,i_{n})\in I_{\nu}}\delta_{i_n^{\sigma},i'_n}t_{n,i^{\sigma}_n}\left(\frac{t_{5,i^{\sigma}_5}t_{6,i^{\sigma}_6}\cdots t_{n-1,i^{\sigma}_{n-1}}}{x_{5,i^{\sigma}_5}x_{6,i^{\sigma}_6}\cdots x_{n-1,i^{\sigma}_{n-1}} }dx_5\wedge dx_6\wedge\dots\wedge dx_{n-1}\right)\,.
\end{align}
The residuum extracts the appropriate Mandelstam variable $t_{n,i^{\sigma}_n}$. In order to proceed with the recursion \eqref{add:recursionIntersectionNumber} and to take the residuum at $x_{n-1}=x_{i'_{n-1}}$, the form has to be expressed in the coordinate $x_{n-1,i'_{n-1}}$ by potentially applying partial fractioning to uncover the entire dependences on $x_{n-1,i'_{n-1}}$ and eliminate redundant variables. However, this leads to exactly the same procedure as described in the previous subsection and, hence, the two recursions are equivalent.

\subsection{Braid matrices: a graphical derivation}\label{sec:derGraph}
The graphical notation introduced in \subsecref{sec:rec:algo} can also be used to calculate the braid matrices $\boldsymbol{\Omega}_4^{42}$ and $\boldsymbol{\Omega}_4^{43}$. Even though their recursive construction \eqref{amplitudes:BraidMatrices} is known, such a graphical derivation may in particular be beneficial once similar amplitude recursions for higher genera are considered. Therefore, we show in this subsection how the derivative of the basis elements of the fibration basis with respect to $x_4$ can be calculated and put into the form of a KZ equation using directed tree graphs. 

From \eqn{add:fibBasis}, we know how to describe the differential forms in the fibration basis in terms of directed graphs. In order to simplify the notation, we denote the corresponding twisted form by the graph defining a representative of its twisted cohomology class. Hence, the fibration basis is given by the elements 
\begin{align*}
\left\langle\prod_{k=5}^n \grEdge{i_k}{k} \,\right|&=\langle f^{4,+}_{i_5,i_6,\dots,i_n}|\,,
\end{align*}
for all the admissible sequences $(i_5,i_6,\dots,i_n)$. Before the graphical calculation of $\frac{\partial}{\partial x_4}\langle f^{4,+}_{i_5,i_6,\dots,i_n}|$ is given, this derivative acting on the integrand of the fibration basis element is rewritten using integration by parts, such that it only acts on the local coefficient $\hat{u}(x)$ of the twisted cycle $\mathcal{C}$ from \eqn{sec:algo:intDomain} in the integral
\begin{align}\label{sec:deriv:Integral}
\langle f^{4,+}_{i_5,i_6,\dots,i_n}|\mathcal{C}\rangle &=\int_{0< x_n<\dots<x_4}\hat{u}(x)f^{4,+}_{i_5,i_6,\dots,i_n}\nonumber\\
&=\int_{0< x_n<\dots<x_4}\hat{u}(x)\prod_{k=5}^n \frac{1}{x_{k, i_k}} dx_5\wedge dx_6\wedge\dots\wedge dx_n\,.
\end{align}
This can conveniently be described using the following definitions: for a graph $g$, we define 
\begin{align*}
V_r(g)&=\{v\in\ZN| v=r \text{ or }v\text{ is branch-connected to }r \}\,.
\end{align*}
Using integration by parts and
\begin{equation}
\frac{\partial}{\partial x_i}\frac{1}{x_{ij}}=-\frac{\partial}{\partial x_j}\frac{1}{x_{ij}}
\end{equation}
to move any derivative acting on the product $\prod_{k=5}^n \frac{1}{x_{k, i_k}}$ to the factor $\hat{u}(x)$ in the integrand of \eqn{sec:deriv:Integral}, this results for $g=\prod_{k=5}^n \frac{1}{x_{k, i_k}}=\prod_{k=5}^n \grEdge{i_k}{k}$ and $\mathcal{C}$ the iterated integration domain over the punctures $x_5,x_6,\dots,x_n$ loaded with $\hat{u}(x)$ in the expression
\begin{align}\label{sec:deriv:derivInt}
\frac{\partial}{\partial x_4}\langle f^{4,+}_{i_5,i_6,\dots,i_n}|\mathcal{C}\rangle&=\int_{0< x_n<\dots<x_4}\left(\sum_{j\in V_4(g)}\frac{\partial}{\partial x_j}\hat{u}(x)\right)\prod_{k=5}^n \frac{1}{x_{k, i_k}} dx_5\wedge dx_6\wedge\dots\wedge dx_n\nonumber\\
&=\int_{0< x_n<\dots<x_4}\hat{u}(x)\left(\sum_{j\in V_4(g)}\sum_{i\neq j}\frac{t_{ji}}{x_{ji}}\right)\prod_{k=5}^n \frac{1}{x_{k, i_k}} dx_5\wedge dx_6\wedge\dots\wedge dx_n\nonumber\\
&=\int_{0< x_n<\dots<x_4}\hat{u}(x)\left(\sum_{j\in V_4(g)}\left(\sum_{l\in V_2(g)}+\sum_{m\in V_3(g)}+\sum_{i\in V_4(g),i\neq j}\right)\frac{t_{ji}}{x_{ji}}\right)\nonumber\\
&\phantom{=\int_{0< x_n<\dots<x_4}\hat{u}(x)(}\cdot \prod_{k=5}^n \frac{1}{x_{k, i_k}} dx_5\wedge dx_6\wedge\dots\wedge dx_n\nonumber\\
&=\int_{0< x_n<\dots<x_4}\hat{u}(x)\left(\sum_{j\in V_4(g)}\left(\sum_{l\in V_2(g)}+\sum_{m\in V_3(g)}\right)\frac{t_{ji}}{x_{ji}}\right)\nonumber\\
&\phantom{=\int_{0< x_n<\dots<x_4}\hat{u}(x)(}\cdot\prod_{k=5}^n \frac{1}{x_{k, i_k}} dx_5\wedge dx_6\wedge\dots\wedge dx_n\,,
\end{align}
where we have used the antisymmetry of $\frac{t_{ji}}{x_{ji}}$ in the last step. \Eqn{sec:deriv:derivInt} can be expressed in terms of twisted forms as
\begin{align}\label{sec:deriv:derivGraphs}
\frac{\partial}{\partial x_4}\left\langle\prod_{k=5}^n \grEdge{i_k}{k} \,\right|&=\sum_{j\in V_4(g)}\sum_{l\in V_2(g)\cup V_3(g)} t_{jl}\left\langle\left(\prod_{k=5}^n \grEdge{i_k}{k}\right)\, \grEdge{l}{j} \,\right|\,,
\end{align}
where the graph 
\begin{equation*}
g\, \grEdge{l}{j}=
\left(\prod_{k=5}^n \grEdge{i_k}{k}\right)\, \grEdge{l}{j}={\tiny\begin{tikzpicture}
		\tikzset{vertex/.style = {shape=circle,draw,minimum size=0.5em}}
		\tikzset{edge/.style = {->,> = latex'}}
		\tikzset{baseline={(0, -0.4em)}}
		\node[vertex] (1) at (0,1) {$4$};
		\node[] (2) at (1,1) {$\dots$};
		\node[vertex] (3) at (2,1) {$i_j$};
		\node[vertex] (4) at (3,1) {$j$};
		\node[] (5) at (4,1) {$\dots$};
		\node[vertex] (6) at (0,0) {$x$};
		\node[] (7) at (1,0) {$\dots$};
		\node[vertex] (8) at (2,0) {$i_l$};
		\node[vertex] (9) at (3,0) {$l$};
		\node[] (10) at (4,0) {$\dots$};
		\node[vertex] (11) at (0,-1) {$5-x$};
		\node[] (12) at (1,-1) {$\dots$};
		\draw[edge] (1) to (2);
		\draw[edge] (2) to (3);
		\draw[edge] (3) to (4);
		\draw[edge] (4) to (5);
		\draw[edge] (6) to (7);
		\draw[edge] (7) to (8);
		\draw[edge] (8) to (9);
		\draw[edge] (9) to (10);
		\draw[edge] (9) to (4);
		\draw[edge] (11) to (12);
		\end{tikzpicture}}=g_{jl}
\end{equation*}
is the graph obtained by connecting the vertex $l$ to the vertex $j$ in the graph $g=\prod_{k=5}^n \grEdge{i_k}{k}$, which we denote by $g_{jl}$, and $x$ is $2$ or $3$ if $l\in V_2(g)$ or $l\in V_3(g)$, respectively. Since in $g_{jl}$, the branch with the root $4$ is connected to the branch with the root $x=2,3$, while the branch with the root $5-x=3,2$ remains disconnected, iterative applications of the Fay identity can be used to lower this connection, such that a linear combination of admissible graphs with a factor $\grEdge{x}{4}$ are left. These factors can be pulled out of the integral in \eqn{sec:deriv:derivInt} and yield the fractions $\frac{1}{x_4}$ and $\frac{1}{x_4-1}$ in the KZ equation \eqn{amplitudes:KZfibrationBasis} for $x=2$ and $x=3$, respectively. The corresponding coefficients obtained from this factorisation on the right-hand side of \eqn{sec:deriv:derivGraphs} are the linear combinations of the Mandelstam variables which constitute the coefficients in the braid matrices $\boldsymbol{\Omega}^4_{4x}$. At each step, the Fay identity has to be applied in the form of \eqn{app:pf:id} such that the admissibility is preserved.

As an example, let us graphically derive the braid matrices $\boldsymbol{\Omega}^4_{42}$ and $\boldsymbol{\Omega}^4_{43}$ in \eqn{sec:amp:Omega4}. While the full calculation can be found in \appref{app:ex:deriv}, we only show the crucial steps here. The two twisted forms which constitute the fibration basis are 
\begin{equation*}
\langle f^{4,+}_{3}|=\left\langle \frac{d x_5}{x_{53}}\right|=\left\langle \grEdge{3}{5}\right|\,,\quad \text{and}\quad \langle f^{4,+}_{4}|=\left\langle \frac{d x_5}{x_{54}}\right|=\left\langle \grEdge{4}{5}\right|\,.
\end{equation*}
Beginning with the former, we find that for $g_3=\grEdge{3}{5}$
\begin{equation*}
V_2(g_3)=\{2\}\,,\quad V_3(g_3)=\{3,5\}\,,\quad V_4(g_3)=\{4\}\,.
\end{equation*}
According to \eqn{sec:deriv:derivGraphs}, the derivative of $\langle f^{4,+}_{3}|$ with respect to $x_4$ is therefore given by
\begin{align}\label{sec:deriv:ex1}
\frac{\partial}{\partial x_4}\langle f^{4,+}_{3}|&=\frac{\partial}{\partial x_4}\left\langle\grEdge{3}{5}\right|\nonumber\\
&=t_{42}\left\langle\grEdge{2}{4}\,\grEdge{3}{5}\right|+t_{43}\left\langle\grDoubleEdgeSource{4}{3}{5}\right|+t_{45}\left\langle\grDoubleEdge{3}{5}{4}\right|\nonumber\\
&=t_{42}\left\langle\grEdge{2}{4}\,\grEdge{3}{5}\right|+t_{43}\left\langle\grDoubleEdgeSource{4}{3}{5}\right|+t_{45}\left\langle\grTriangleSource{3}{4}{5}-\grTriangleClock{3}{4}{5}\right|\nonumber\\
&=\left(\frac{\begin{pmatrix}
	t_{42}&0
	\end{pmatrix}}{x_4}+\frac{\begin{pmatrix}
	t_{43}+t_{45}&-t_{45}
	\end{pmatrix}}{x_4-1}\right)\langle \boldsymbol{f}^{4,+}|\,,
\end{align}
where the row vectors $\begin{pmatrix}
	t_{42}&0
	\end{pmatrix}$ and $\begin{pmatrix}
	t_{43}+t_{45}&-t_{45}
	\end{pmatrix}$ are indeed the first rows of $\boldsymbol{\Omega}^4_{42}$ and $\boldsymbol{\Omega}^4_{43}$, respectively, as given in \eqn{sec:amp:Omega4}. The second equality in \eqn{sec:deriv:ex1} follows from \eqn{sec:deriv:derivGraphs} and the third equality is the application of the Fay identity to recover admissible graphs as described below \eqn{sec:deriv:derivGraphs}. Similarly, we find for $g_4=\grEdge{4}{5}$
	\begin{equation*}
	V_2(g_4)=\{2\}\,,\quad V_3(g_4)=\{3\}\,,\quad V_4(g_4)=\{4,5\}\,,
	\end{equation*}
	such that \eqn{sec:deriv:derivGraphs} implies
\begin{align*}
\frac{\partial}{\partial x_4}\langle f^{4,+}_4|&=\frac{\partial}{\partial x_4}\left\langle\grEdge{4}{5}\right|\nonumber\\
&=t_{42}\left\langle\grDoubleEdge{2}{4}{5}\right|+,t_{43}\left\langle\grDoubleEdge{3}{4}{5}\right|+t_{52}\left\langle\grTriangleSink{2}{4}{5}\right|+t_{53}\left\langle\grTriangleSink{3}{4}{5}\right|\nonumber\\
&=t_{42}\left\langle\grDoubleEdge{2}{4}{5}\right|+,t_{43}\left\langle\grDoubleEdge{3}{4}{5}\right|+t_{53}\left\langle\grTriangleClock{3}{4}{5}-\grTriangleSource{3}{4}{5}\right|\nonumber\\
&\phantom{=}+t_{52}\left\langle\grTriangleClock{2}{4}{5}+\grEdge{2}{4}\,\left(\frac{t_{53}}{t_{52}}\grEdge{3}{5}+\frac{t_{54}}{t_{52}}\grEdge{4}{5}\right)\right|\nonumber\\
&=\left(\frac{\begin{pmatrix}
	t_{53}&t_{542}
	\end{pmatrix}}{x_4}+\frac{\begin{pmatrix}
	-t_{53}&t_{43}+t_{53}
	\end{pmatrix}}{x_4-1}\right)\langle\boldsymbol{f}^{4,+}|\,,
\end{align*}
where we used that
\begin{align*}
t_{52}\left\langle\grEdge{2}{5}\right|+t_{53}\left\langle\grEdge{3}{5}\right|+t_{54}\left\langle\grEdge{4}{5}\right|&=0\,,
\end{align*}
since the left-hand side is a total derivative, to express $\left\langle\grEdge{2}{5}\right|$ in terms of the two (fibration) basis vectors $\left\langle\grEdge{3}{5}\right|$ and $\left\langle\grEdge{4}{5}\right|$ of the twisted cohomology. And indeed, the row vectors $\begin{pmatrix}
t_{53}&t_{542}
\end{pmatrix}$ and $\begin{pmatrix}
-t_{53}&t_{43}+t_{53}
\end{pmatrix}$ obtained are the second rows of the braid matrices $\boldsymbol{\Omega}^4_{42}$ and $\boldsymbol{\Omega}^4_{43}$, respectively, as \eqn{sec:amp:Omega4} approves.

While for the above examples this graphical approach seems rather superficial, it gives a convenient tool to calculate the derivatives of the fibration basis for higher $n$. It can be implemented in any computer algebra system as a manipulation of the adjacency matrices of the directed graphs defining the fibration basis using matrix operations only. This procedure to evaluate the derivatives and obtain the matrices appearing in the differential equation of the given basis turns out to be a convenient tool for similar amplitude recursions involving vector-valued differential equations with matrix-valued connections at higher genera such as for example the elliptic KZB equation in the one-loop recursion of \rcite{BroedelKaderli}, where recursive definitions such as the construction of the braid matrices for genus zero in \eqn{amplitudes:BraidMatrices} are not available.

\section{Conclusions}
\label{sec:conclusion}
In this article, we have reviewed the tree-level amplitude recursion of open-superstring states introduced in \rcite{Broedel:2013aza} and pointed out its relation to twisted de Rham theory. This investigation led to the following results:
\begin{itemize}
	\item The vector of string integrals with an auxiliary point introduced in \rcite{Broedel:2013aza}, which interpolates between the $N$- and the $(N-1)$-point string corrections and which satisfies a KZ equation, has been related to the fibration basis constructed in \rcite{Mizera:2019gea}. The transformation matrix can recursively be determined using \eqn{sec:algo:BRow}. In \eqn{sec:algo:recMizeraEquiv}, this recursion was shown to be equivalent to the recursion of intersection numbers of twisted forms stated in \rcite{Mizera:2019gea}. 
	
	\item The transformation matrix is given by the intersection numbers of the twisted forms appearing in the string integrals with an auxiliary point and the dual fibration basis. Thus, the recursion \eqref{sec:algo:BRow} gives a purely combinatorial derivation of these intersection numbers in terms of directed tree graphs, which is based on the partial-fractioning algorithm described in \subsecref{sec:rec:algo}. This allows for a convenient implementation in computer algebra systems using (weighted) adjacency matrices and matrix operations thereon.
	
	\item While the vector of string integrals with an auxiliary point is the relevant solution of the KZ equation in the amplitude recursion of \rcite{Broedel:2013aza}, the representations of the Lie algebra generators in the KZ equation \eqref{amplitudes:KZfibrationBasis} satisfied by the fibration basis are braid matrices with a well-known recursive definition. Therefore, the above basis transformation leads to a recursive construction of the matrix representations appearing in the KZ equation \eqref{sec:algo:KZEqfhat} of the vector of string integrals with an auxiliary point, which constitute the letters for the Drinfeld associator used in the amplitude recursion. This shows in particular, that the matrices occurring in the amplitude recursion are braid matrices as well. 
	
	\item In \eqn{sec:deriv:derivGraphs}, the derivatives of the twisted forms which constitute the fibration basis has been expressed graphically in terms of directed tree graphs. Starting from this expression using the graphical algorithm described in \subsecref{sec:derGraph}, the braid matrices can be derived in an alternative way to the recursion of \rcite{Mizera:2019gea}. On the one hand, this completes the graphical derivation of the matrix representations in the KZ equation of the string integrals with an auxiliary point. On the other hand, this procedure may be used in similar constructions at higher genera, where no alternative derivation of the relevant matrices are available. 
	
	\item As discussed below, this analysis reveals the essential features of the amplitude recursion. This may lead to similar recursions for loop amplitudes or higher-genera Riemann surfaces, respectively. A first result in this direction is described by the one-loop recursion in \rcite{BroedelKaderli}.
	
	\item Moreover, formulating the recursion of \rcite{Broedel:2013aza} in terms of twisted de Rham theory proves various statements about the feasibility of the recursion. For example the fact that the differential equation of the vector of string integrals with an auxiliary point can indeed be written in the form of a KZ equation.
\end{itemize}

These results do not only allow for an efficient implementation of the tree-level amplitude recursion and a description in terms of twisted de Rham theory, but offer insights in the essential features allowing for such a recursion. The differential one-forms $ dx_k/x_{k,i_k} $ in the string corrections span the logarithmic derivatives of the genus-zero Koba--Nielsen factor. Defining iterated integrals over the punctures with integration kernels the admissible one-forms $ dx_k/x_{k,i_k}$, i.e.\ $3\leq i_k<k$,  and the empty integral being the Koba--Nielsen factor leads to a recursive construction of the representations of the Lie algebra generators in the corresponding KZ equation. This is exactly how the fibration basis is defined and how the braid matrices come up. 

It may be expected that a similar construction for higher genera is possible. The relevant differential one-forms are determined by the logarithmic derivatives of the higher-genus Koba--Nielsen factor. These one-forms define the higher-genus class of iterated integrals with the empty integral being the corresponding Koba--Nielsen factor and the integration kernels of these iterated integrals naturally satisfy an admissibility condition. Therefore, the differential equation with respect to the insertion point defining the outermost integration boundary satisfied by this iterated integral can be cast into a sum over all admissible differential one-forms with coefficients some linear combination of the admissible iterated integrals. In order to recover admissible iterated integrals at this point, a similar mechanism to manipulate the labels of a product of differential one-forms as partial fractioning is required, for example a Fay identity. These linear combinations constitute the matrices, which serve as letters in a Drinfeld-like associator construction, which itself is determined by the singularities occurring in the differential one-forms and relates some limits of the iterated integrals. These limits, in turn, should contain the amplitudes at the current genus and (possibly) amplitudes at lower genera. In \rcite{BroedelKaderli}, this construction has been carried out for the one-loop open-string corrections defined on genus-one Riemann surfaces. The generalisation to higher genera and possibly other theories remains an open task.

\subsection*{Acknowledgments}
We are grateful to Johannes Broedel for the numerous helpful advices, insightful discussions and useful comments on the draft of this article. Furthermore, we would like to thank Oliver Schlotterer for various interesting discussions and Sebastian Mizera for a helpful conversation about the draft of this article. AK would
like to thank the IMPRS for Mathematical and Physical Aspects of Gravitation,
Cosmology and Quantum Field Theory, of which he is a member and which renders
his studies possible.  Furthermore, he is supported by the Swiss Studies
Foundation, to which he would like to express his gratitude.


\section*{Appendix}
\appendix

\section{Notes on the solution of the KZ equation for string corrections}
In this section, we investigate some properties of the solution $\hat{F}_{\nu}^{\sigma}(x)$ of the KZ equation given in \eqn{amplitudes:FhatSigmanu}, which is the backbone of the amplitude recursion for the string corrections proposed in \rcite{Broedel:2013aza}.

\subsection{Translation between different labellings}\label{app:transl:label}
In this subsection, the integral \eqref{amplitudes:FhatSigmanu} as originally\footnote{The original definition is actually defined with $s_{01}=s_{0,N-1}=0$, however, this does not change the subvector of $\boldsymbol{C}_1$ containing the string corrections: this simply leads to the exponential contribution in \eqn{eqn:appE1}, which cancels the additional factor $z_{0,N-1}^{s_{0,N-1}}$ in $\boldsymbol{C}_1$ from $\hat{F}_{\nu}^{\sigma}(z)$. Similarly, it does not change $\boldsymbol{C}_0$.} defined in \rcite{Broedel:2013aza} in terms of the labelling $(N,z_i,s_{ij})$ and the auxiliary puncture $z_{N-2}<z_0<z_{N-1}$, i.e.\
\begin{align}\label{app:transl:originalDefHatF}
\hat{F}_{\nu}^{\sigma}(z)&=(-1)^{N-3}\int_0^{z_0}dz_{N-2}\prod_{i=2}^{N-3}\int_{0}^{z_{i+1}} dz_i \, u(z)\prod_{i=1}^{N-1}|z_{0i}|^{s_{0i}}\sigma \left(\prod_{k=2}^{\nu}\sum_{j=1}^{k-1}\frac{s_{jk}}{z_{jk}}\prod_{m=\nu+1}^{N-2}\sum_{l=m+1}^{N-1}\frac{s_{ml}}{z_{ml}} \right)\,,
\end{align}
is expressed in terms of the labelling $(n,x_i,t_{ij})$ without the appearance of $z_1=x_2=0$ in the expression on which the permutation $\sigma$ acts. The result is the integral $\hat{F}_{\nu}^{\sigma}(x)$ as defined in \eqn{amplitudes:FhatSigmanu}. This can be achieved by an iterative application of integration by parts with respect to the variable with the highest label in the first product until the product is empty
\begin{align*}
&\hat{F}_{\nu}^{\sigma}(z)\nonumber\\
&=(-1)^{N-3}\int_0^{z_0}dz_{N-2}\prod_{i=2}^{N-2}\int_{0}^{z_{i+1}} dz_i \, u(z)\prod_{i=1}^{N-1}|z_{0i}|^{s_{0i}}\sigma \left(\prod_{k=2}^{\nu}\sum_{j=1}^{k-1}\frac{s_{jk}}{z_{jk}}\prod_{m=\nu+1}^{N-2}\sum_{l=m+1}^{N-1}\frac{s_{ml}}{z_{ml}} \right)\nonumber\\
&=(-1)^{N-3}\int_0^{z_0}dz_{N-2}\prod_{i=2}^{N-2}\int_{0}^{z_{i+1}} dz_i \, u(z)\prod_{i=1}^{N-1}|z_{0i}|^{s_{0i}}\nonumber\\
&\phantom{=(-1)^{N-3}\int_0^{z_0}}\sigma \left(\left(\prod_{k=2}^{\nu-1}\sum_{j=1}^{k-1}\frac{s_{jk}}{z_{jk}}\right)\left(\sum_{s=\nu+1}^{N-1}\frac{s_{\nu s}}{z_{\nu s}}+\frac{s_{\nu 0}}{z_{\nu 0}}\right)\left(\prod_{m=\nu+1}^{N-2}\sum_{l=m+1}^{N-1}\frac{s_{ml}}{z_{ml}}\right) \right)\nonumber\\
&=(-1)^{N-3}\int_0^{z_0}dz_{N-2}\prod_{i=2}^{N-2}\int_{0}^{z_{i+1}} dz_i \, u(z)\prod_{i=1}^{N-1}|z_{0i}|^{s_{0i}}\nonumber\\
&\phantom{=(-1)^{N-3}\int_0^{z_0}}\sigma \left(\prod_{k=2}^{\nu-2}\sum_{j=1}^{k-1}\frac{s_{jk}}{z_{jk}}\prod_{r=\nu-1}^{\nu}
\left(\sum_{s=r+1}^{N-1}\frac{s_{r s}}{z_{r s}}+\frac{s_{\nu 0}}{z_{\nu 0}}\right)\prod_{m=\nu+1}^{N-2}\sum_{l=m+1}^{N-1}\frac{s_{ml}}{z_{ml}} \right)\nonumber\\
&=(-1)^{N-3}\int_0^{z_0}\!dz_{N-2}\prod_{i=2}^{N-2}\int_{0}^{z_{i+1}}\! dz_i u(z)\prod_{i=1}^{N-1}|z_{0i}|^{s_{0i}}\sigma\! \left(\prod_{r=2}^{\nu}\left(\sum_{s=r+1}^{N-1}\frac{s_{r s}}{z_{r s}}\!+\!\frac{s_{\nu 0}}{z_{\nu 0}}\right)\!\prod_{m=\nu+1}^{N-2}\sum_{l=m+1}^{N-1}\frac{s_{ml}}{z_{ml}}\right)\nonumber\\
&=(-1)^{n}\prod_{i=5}^{n}\int_0^{x_{i-1}} dx_i\, \prod_{2\leq i\prec j\leq n}|x_{ij}|^{t_{ij}}\sigma \left(\prod_{k=n-\nu+2}^{n}\sum_{j=3}^{k-1}\frac{t_{kj}}{x_{kj}}\prod_{m=5}^{n-\nu+1}\left(\sum_{l=5}^{m-1}\frac{t_{ml}}{x_{ml}}+\frac{t_{m3}}{x_{m3}}\right)\right)\nonumber\\
&=\hat{F}^{\sigma}_{\nu}(x)\,,
\end{align*}
where in the second last line, the labelling has been changed from $(N,z_i,s_{ij})$ to $(n,x_i,t_{ij})$ according to eqs.\ \eqref{sec:rec:ordering}-\eqref{sec:rec:orderingM}. A comparison with the original definition \eqref{app:transl:originalDefHatF} and \eqn{amplitudes:FSigmaPinu} shows that the latter integrals are recovered for $s_{0,N-1}=0$ in the limit $z_{0}=x_4\rightarrow 1$ and $s_{0i}=t_{4, \sigma_{\text{label}}^{-1} (i)}\rightarrow 0$ of the former, thus
\begin{align}\label{sec:transl:Lim1}
\lim_{t_{4i}\to 0}\lim_{x_4\rightarrow 1}\hat{F}^{\sigma}_{\nu}(x_4)|_{t_{43=0}}&=F_{\id,\nu}^{\sigma}\,.
\end{align} 

\subsection{The first rows of $\boldsymbol{e}_1$}\label{app:transl:e1}
The condition $t_{43}=0$ in \eqn{sec:transl:Lim1} is incorporated for $\nu=n-3$ in the first $(n-4)!$ entries of $\boldsymbol{e}_1$: for $\nu=n-3$ there is neither a $z_{N-1}$ nor a $z_0$ appearing in the factor the permutation $\sigma$ acts on in \eqn{app:transl:originalDefHatF}. Thus, in the derivative
\begin{align*}
\frac{\partial}{\partial z_0}\hat{F}_{N-2}^{\sigma}(z_0)&=(-1)^{N-3}\int_0^{z_0}dz_{N-2}\prod_{i=2}^{N-2}\int_{0}^{z_{i+1}} dz_i \, u(z)\prod_{i=2}^{N-2}z_{0i}^{s_{0i}}\left(\sum_{i=1}^{N-1} \frac{s_{0i}}{z_{0i}}\right)\,\sigma \left(\prod_{k=2}^{N-2}\sum_{j=1}^{k-1}\frac{s_{jk}}{z_{jk}}\right)\,,
\end{align*}
the quotient $\frac{1}{z_{0i}}$ with $1\leq i<N-1$ can be traded using partial fractioning with the other quotients in $\sigma \left(\prod_{k=2}^{N-2}\sum_{j=1}^{k-1}\frac{s_{jk}}{z_{jk}}\right)$, which does not contain any variable $z_{N-1}$, for $\frac{1}{z_{01}}=\frac{1}{z_{0}}$ which contribute to the matrix $\boldsymbol{e}_0$. Thus, the only quotient of the form $\frac{1}{z_{0,N-1}}$ comes from differentiating the factor $u(z)$ and can simply be pulled out of the integral together with the corresponding coefficient $s_{0,N-1}$. This is the only contribution to $\boldsymbol{e}_1$ in the KZ equation of $\bhF(z_0)$ (see e.g.\ \eqn{app:rec:e1N6}), such that
\begin{align}\label{eqn:appE1}
\boldsymbol{e}_1&=\begin{pmatrix}
s_{0,N-1}\mathbb{I}_{(n-4)!\times (n-4)!}&0_{(n-4)!\times (n-4)(n-4)!}\\
\vdots&\vdots
\end{pmatrix}\,,
\end{align}
which proves \eqn{sec:rec:e1}. 

\subsection{Regularised boundary value $\boldsymbol{C}_0$}\label{app:c0}
In this section, the regularised boundary value 
\begin{align*}
\boldsymbol{C}_0&=\lim_{x_4\rightarrow 0}x_4^{-\boldsymbol{e}_0}\bhF(x_4)
\end{align*} 
is calculated and shown to contain the $(n-2)$-point tree-level string corrections. This derivation is closely related to the proofs in \rcite{Terasoma}. The calculation is shown in terms of the labelling $(N,z_i,s_{ij})$, since in terms of this labelling, the components of $\bhF(x_4)$ given by the integrals $\hat{F}_{\nu}^{\sigma}(z_0)$ defined in \eqn{app:transl:originalDefHatF} only depend in the factor $u(z)$ on $z_0=x_4$. Using the substitution $z_i=z_0 w_i$ for $0\leq i\leq N-2$ with $w_{0}=1$ and the definition $s_{\text{max}}=s_{12\dots,N-2}+\sum_{j=2}^{N-2}s_{0j}$, we find for $\nu=N-2$ and $\sigma\in S_{N-3}$ 
\begin{align*}
&\lim_{z_0\rightarrow 0}z_0^{-s_{\max}}\hat{F}_{\id,N-2}^{\sigma}(z_0)\nonumber\\
&=(-1)^{N-3}\lim_{z_0\rightarrow 0}z_0^{-s_{\max}}\int_0^{z_0}d z_{N-2}\prod_{i=2}^{N-3}\int_{0}^{z_{i+1}} dz_i \, u(z)\prod_{k=2}^{N-2}z_{0k}^{s_{0k}}\sigma \left(\prod_{k=2}^{N-2}\sum_{j=1}^{k-1}\frac{s_{jk}}{z_{jk}}\right)\nonumber\\
&=(-1)^{N-3}\lim_{z_0\rightarrow 0}\int_0^{1}d w_{N-2}\prod_{i=2}^{N-3}\int_{0}^{w_{i+1}} dw_i \, \prod_{1\leq i<j\leq N-2}|w_{ij}|^{s_{ij}}\nonumber\\
&\phantom{=(-1)^{N-3}\lim_{z_0\rightarrow 0}\int_0^{1}d w_{N-2}\prod_{i=2}^{N-3}\int_{0}^{w_{i+1}} dw_i}\, \prod_{k=1}^{N-2}(1-z_0 w_k )^{s_{k,N-1}}\prod_{k=2}^{N-2}(1-w_k)^{s_{0k}}\sigma \left(\prod_{k=2}^{N-2}\sum_{j=1}^{k-1}\frac{s_{jk}}{w_{jk}}\right)\nonumber\\
&=(-1)^{N-3}\int_0^{1}d w_{N-2}\prod_{i=2}^{N-3}\int_{0}^{w_{i+1}} dw_i \, \prod_{1\leq i<j\leq N-2}|w_{ij}|^{s_{ij}}\prod_{k=2}^{N-2}(1-w_k)^{s_{0k}}\sigma \left(\prod_{k=2}^{N-2}\sum_{j=1}^{k-1}\frac{s_{jk}}{w_{jk}}\right)\nonumber\\
&=F^{\sigma}|_{s_{i,N-1}=s_{0i}}\,,
\end{align*}
which indeed corresponds to the $N$-point string corrections with $s_{i,N-1}=s_{0i}$. If $\nu<N-2$, there would appear $N-2-\nu$  less factors of $z_0$ in the denominator than in the integration measure after the change of variables $z_i=z_0 w_i$, leading to vanishing integrals. Thus, only the integrals $\hat{F}^{\sigma}_{\id,\nu}$ with $\nu=N-2$ do not vanish in the regularised limit $\lim_{z_0\rightarrow 0}z_0^{-s_{\max}}\hat{F}_{\id,\nu}^{\sigma}(z_0)$ giving the above result. However, this limit does not yet yield $(N-1)$-point string corrections. As observed for the limit $z_0\to 1$, the Mandelstam variables $s_{0,i}$ had to be set to zero before the $N$-point amplitudes could be recovered. Applying this limit $s_{0i}\to 0$ for the present boundary value, where $z_0\to 0$, effectively removes one external state leaving $(N-1)$-point integrals. Concretely, assuming that $\sigma(N-2)=N-2$, using integration by parts with respect to $w_{N-2}$ and the Dirac delta function in the form
\begin{align*}
\lim_{\alpha\rightarrow 0} \int_{p}^{1} dx\,\alpha (1-x)^{\alpha-1}F(x)
&=F(1)
\end{align*}
for a function $F$ which is integrable on $]p,1]$,
the additional limit $s_{0i}\rightarrow 0$ yields
\begin{align*}
&\lim_{s_{0i}\rightarrow 0}\lim_{z_0\rightarrow 0}z_0^{-s_{\max}}F_{\id,N-2}^{\sigma}(z_0,s_{0i})\nonumber\\
&=(-1)^{N-3}\lim_{s_{0i}\rightarrow 0}\int_0^{1}d w_{N-2}\prod_{i=2}^{N-3}\int_{0}^{w_{i+1}} dw_i \, \prod_{1\leq i<j\leq N-2}|w_{ij}|^{s_{ij}}\prod_{k=2}^{N-2}(1-w_k)^{s_{0k}}\sigma \left(\prod_{k=2}^{N-2}\sum_{j=1}^{k-1}\frac{s_{jk}}{w_{jk}}\right)\nonumber\\
&=(-1)^{N-3}\lim_{s_{0i}\rightarrow 0}\int_0^{1}d w_{N-2}\prod_{i=2}^{N-3}\int_{0}^{w_{i+1}} dw_i \, \prod_{1\leq i<j\leq N-2}|w_{ij}|^{s_{ij}}\left(\sum_{j=1}^{N-3}\frac{s_{j,N-2}}{w_{j,N-2}}\right)\prod_{k=2}^{N-2}(1-w_k)^{s_{0k}}\nonumber\\
&\phantom{=(-1)^{N-3}\lim_{s_{0i}\rightarrow 0}\int_0^{1}d w_{N-2}\prod_{i=2}^{N-3}\int_{0}^{w_{i+1}} dw_i \,}\sigma \left(\prod_{k=2}^{N-3}\sum_{j=1}^{k-1}\frac{s_{jk}}{w_{jk}}\right)\nonumber\\
&=(-1)^{N-3}\lim_{s_{0i}\rightarrow 0}\int_0^{1}d w_{N-2}\prod_{i=2}^{N-3}\int_{0}^{w_{i+1}} dw_i \, \left(-\frac{\partial}{\partial_{w_{N-2}}}\prod_{1\leq i<j\leq N-2}|w_{ij}|^{s_{ij}}\right)\prod_{k=2}^{N-2}(1-w_k)^{s_{0k}}\nonumber\\
&\phantom{=(-1)^{N-3}\lim_{s_{0i}\rightarrow 0}\int_0^{1}d w_{N-2}\prod_{i=2}^{N-3}\int_{0}^{w_{i+1}} dw_i \,}\sigma \left(\prod_{k=2}^{N-3}\sum_{j=1}^{k-1}\frac{s_{jk}}{w_{jk}}\right)\nonumber\\
&=(-1)^{N-4}\lim_{s_{0i}\rightarrow 0}\int_0^{1}d w_{N-2}\,s_{0,N-2}(1-w_{N-2})^{s_{0,N-2}-1}\nonumber\\
&\phantom{=(-1)^{N-4}\lim_{s_{0i}\rightarrow 0}}\prod_{i=2}^{N-3}\int_{0}^{w_{i+1}} dw_i \, \prod_{1\leq i<j\leq N-2}|w_{ij}|^{s_{ij}}\prod_{k=2}^{N-3}(1-w_k)^{s_{0k}}\sigma \left(\prod_{k=2}^{N-3}\sum_{j=1}^{k-1}\frac{s_{jk}}{w_{jk}}\right)\nonumber\\
&=(-1)^{N-4}\prod_{i=2}^{N-3}\int_{0}^{w_{i+1}} d w_i \, \prod_{1\leq i<j\leq N-2}|w_{ij}|^{s_{ij}}\sigma \left(\prod_{k=2}^{N-3}\sum_{j=1}^{k-1}\frac{s_{jk}}{w_{jk}}\right)|_{w_{N-2}=1}\nonumber\\
&=F^{\sigma}|_{N-1}\,,
\end{align*}
where $F^{\sigma}|_{N-1}$ is the string correction for $N-1$ external states with $(w_1,w_{N-2},w_{N-1})=(0,1,\infty)$.

\section{Partial-fractioning algorithm: applications}\label{app:ex}
In this section, the algorithm from \subsecref{sec:rec:algo} is applied to some examples. First, we consider a $\sigma$-permuted admissible form for $n=8$ and use the above algorithm to rewrite it in terms of the fibration basis. The second and third examples are the four-point and five-point amplitudes $\bhF(x_4)$ for $n=5$ and $n=6$, respectively, for which we calculate the basis transformation $\bB$, cf.\ \eqn{sec:rec:BasisTrafo}, to the fibration basis $\langle \boldsymbol{f}^{4,+}(x_4)|$ following \subsecref{sec:rec:algo}.

\subsection{From $\sigma$-permuted admissible to admissible}
In order to exemplify the partial-fractioning algorithm, let us consider the admissible sequence $(i_5,i_6,i_7,i_8)=(4,5,5,7)$ and the transposition $\tau=(5\, 7)$. The corresponding $\tau$-permuted admissible sequence is $(i^{\tau}_5,i^{\tau}_6,i^{\tau}_7,i^{\tau}_8)=(7,7,4,5)$, where $i_k^{\tau}=\tau (i_{\tau (k)})$, and, according to \eqns{add:sigmaPertFormFrac}{add:sigmaPertForm}, the $\tau$-permuted admissible form is given by
\begin{align}\label{sec:algo:exForm}
\varphi^{\tau}_{4,5,5,7}&=\frac{dx_5\wedge dx_6\wedge dx_7\wedge dx_8}{x_{57}x_{67}x_{74}x_{85}}\nonumber\\
&=\left({\tiny\begin{tikzpicture}
	\tikzset{vertex/.style = {shape=circle,draw,minimum size=0.5em}}
	\tikzset{edge/.style = {->,> = latex'}}
	\tikzset{baseline={(0, -0.4em)}}
	\node[vertex] (4) at (0,0.5) {$4$};
	\node[vertex] (7) at (1,0.5) {$7$};
	\node[vertex] (6) at (2,0.5) {$6$};
	\node[vertex] (5) at (1,-0.5) {$5$};
	\node[vertex] (8) at (2,-0.5) {$8$};
	\draw[edge] (4) to (7);
	\draw[edge] (7) to (6);
	\draw[edge] (7) to (5);
	\draw[edge] (5) to (8);
	\end{tikzpicture}}\right)dx_5\wedge dx_6\wedge dx_7\wedge dx_8\,,
\end{align}
which is not admissible since even though no vertex larger than four has two incoming arrows, some arrows point from a higher number to a lower number. Following the algorithm from \subsecref{sec:rec:algo}, the graph appearing in the form $\varphi^{\tau}_{4,5,5,7}$ can be written in terms of admissible graphs as follows: first, we consider the highest vertex $h$ with a non-admissible factor $\grEdge{i_h^{\tau}}{h}$, which is $h=6$ with $i_h^{\tau}=7$, and apply the partial-fractioning identity \eqref{sec:algo:pfGraphs} to the corresponding branch as given in \eqn{sec:algo:branchNonAdm}. Here, this branch is $b_6=\grDoubleEdge{4}{7}{6}$, such that according to \eqn{sec:algo:admissibleSubbranch}
\begin{equation*}
\grDoubleEdge{4}{7}{6}=-{\tiny\begin{tikzpicture}
	\tikzset{vertex/.style = {shape=circle,draw,minimum size=0.5em}}
	\tikzset{edge/.style = {->,> = latex'}}
	\tikzset{baseline={(0, -0.4em)}}
	\node[vertex] (4) at (0,0.25) {$4$};
	\node[vertex] (7) at (1,-0.25) {$7$};
	\node[vertex] (6) at (2,0.25) {$6$};
	\draw[edge] (4) to (6);
	\draw[double,edge] (6) to (7);
	\draw[double,edge] (4) to (7);
	\end{tikzpicture}}={\tiny\begin{tikzpicture}
	\tikzset{vertex/.style = {shape=circle,draw,minimum size=0.5em}}
	\tikzset{edge/.style = {->,> = latex'}}
	\tikzset{baseline={(0, -0.4em)}}
	\node[vertex] (4) at (0,0.25) {$4$};
	\node[vertex] (7) at (1,-0.25) {$7$};
	\node[vertex] (6) at (2,0.25) {$6$};
	\draw[edge] (4) to (7);
	\draw[edge] (4) to (6);
	\end{tikzpicture}}-{\tiny\begin{tikzpicture}
	\tikzset{vertex/.style = {shape=circle,draw,minimum size=0.5em}}
	\tikzset{edge/.style = {->,> = latex'}}
	\tikzset{baseline={(0, -0.4em)}}
	\node[vertex] (4) at (0,0.25) {$4$};
	\node[vertex] (7) at (1,-0.25) {$7$};
	\node[vertex] (6) at (2,0.25) {$6$};
	\draw[edge] (4) to (6);
	\draw[edge] (6) to (7);
	\end{tikzpicture}}\,.
\end{equation*}
The highest non-admissible vertex in the resulting linear combination of graphs
\begin{equation*}
{\tiny\begin{tikzpicture}
	\tikzset{vertex/.style = {shape=circle,draw,minimum size=0.5em}}
	\tikzset{edge/.style = {->,> = latex'}}
	\tikzset{baseline={(0, -0.4em)}}
	\node[vertex] (4) at (0,0.5) {$4$};
	\node[vertex] (7) at (1,0.5) {$7$};
	\node[vertex] (6) at (2,0.5) {$6$};
	\node[vertex] (5) at (1,-0.5) {$5$};
	\node[vertex] (8) at (2,-0.5) {$8$};
	\draw[edge] (4) to (7);
	\draw[edge] (7) to (6);
	\draw[edge] (7) to (5);
	\draw[edge] (5) to (8);
	\end{tikzpicture}}=-{\tiny\begin{tikzpicture}
	\tikzset{vertex/.style = {shape=circle,draw,minimum size=0.5em}}
	\tikzset{edge/.style = {->,> = latex'}}
	\tikzset{baseline={(0, -0.4em)}}
	\node[vertex] (4) at (0,1) {$4$};
	\node[vertex] (7) at (1,0.5) {$7$};
	\node[vertex] (6) at (2,1) {$6$};
	\node[vertex] (5) at (1,-0.5) {$5$};
	\node[vertex] (8) at (2,-0.5) {$8$};
	\draw[double,edge] (4) to (7);
	\draw[double,edge] (6) to (7);
	\draw[edge] (7) to (5);
	\draw[edge] (4) to (6);
	\draw[edge] (5) to (8);
	\end{tikzpicture}}={\tiny\begin{tikzpicture}
	\tikzset{vertex/.style = {shape=circle,draw,minimum size=0.5em}}
	\tikzset{edge/.style = {->,> = latex'}}
	\tikzset{baseline={(0, -0.4em)}}
	\node[vertex] (4) at (0,1) {$4$};
	\node[vertex] (7) at (1,0.5) {$7$};
	\node[vertex] (6) at (2,1) {$6$};
	\node[vertex] (5) at (1,-0.5) {$5$};
	\node[vertex] (8) at (2,-0.5) {$8$};
	\draw[edge] (4) to (7);
	\draw[edge] (7) to (5);
	\draw[edge] (4) to (6);
	\draw[edge] (5) to (8);
	\end{tikzpicture}}-{\tiny\begin{tikzpicture}
	\tikzset{vertex/.style = {shape=circle,draw,minimum size=0.5em}}
	\tikzset{edge/.style = {->,> = latex'}}
	\tikzset{baseline={(0, -0.4em)}}
	\node[vertex] (4) at (0,1) {$4$};
	\node[vertex] (7) at (1,0.5) {$7$};
	\node[vertex] (6) at (2,1) {$6$};
	\node[vertex] (5) at (1,-0.5) {$5$};
	\node[vertex] (8) at (2,-0.5) {$8$};
	\draw[edge] (6) to (7);
	\draw[edge] (7) to (5);
	\draw[edge] (4) to (6);
	\draw[edge] (5) to (8);
	\end{tikzpicture}}
\end{equation*}
is $h=5$ with $i_h=7$ and non-admissible factor $\grEdge{7}{5}$. This linear combination can be rewritten as before using \eqn{sec:algo:admissibleSubbranch} such that the final linear combination is given by
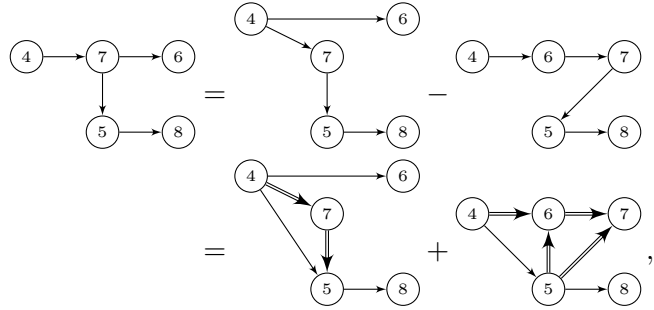
\begin{align*}
{\tiny\begin{tikzpicture}
	\tikzset{vertex/.style = {shape=circle,draw,minimum size=0.5em}}
	\tikzset{edge/.style = {->,> = latex'}}
	\tikzset{baseline={(0, -0.4em)}}
	\node[vertex] (4) at (0,0.5) {$4$};
	\node[vertex] (7) at (1,0.5) {$7$};
	\node[vertex] (6) at (2,0.5) {$6$};
	\node[vertex] (5) at (1,-0.5) {$5$};
	\node[vertex] (8) at (2,-0.5) {$8$};
	\draw[edge] (4) to (7);
	\draw[edge] (7) to (6);
	\draw[edge] (7) to (5);
	\draw[edge] (5) to (8);
	\end{tikzpicture}}&={\tiny\begin{tikzpicture}
	\tikzset{vertex/.style = {shape=circle,draw,minimum size=0.5em}}
	\tikzset{edge/.style = {->,> = latex'}}
	\tikzset{baseline={(0, -0.4em)}}
	\node[vertex] (4) at (0,1) {$4$};
	\node[vertex] (7) at (1,0.5) {$7$};
	\node[vertex] (6) at (2,1) {$6$};
	\node[vertex] (5) at (1,-0.5) {$5$};
	\node[vertex] (8) at (2,-0.5) {$8$};
	\draw[edge] (4) to (7);
	\draw[edge] (7) to (5);
	\draw[edge] (4) to (6);
	\draw[edge] (5) to (8);
	\end{tikzpicture}}-{\tiny\begin{tikzpicture}
	\tikzset{vertex/.style = {shape=circle,draw,minimum size=0.5em}}
	\tikzset{edge/.style = {->,> = latex'}}
	\tikzset{baseline={(0, -0.4em)}}
	\node[vertex] (4) at (0,0.5) {$4$};
	\node[vertex] (7) at (2,0.5) {$7$};
	\node[vertex] (6) at (1,0.5) {$6$};
	\node[vertex] (5) at (1,-0.5) {$5$};
	\node[vertex] (8) at (2,-0.5) {$8$};
	\draw[edge] (6) to (7);
	\draw[edge] (7) to (5);
	\draw[edge] (4) to (6);
	\draw[edge] (5) to (8);
	\end{tikzpicture}}\nonumber\\
&={\tiny\begin{tikzpicture}
	\tikzset{vertex/.style = {shape=circle,draw,minimum size=0.5em}}
	\tikzset{edge/.style = {->,> = latex'}}
	\tikzset{baseline={(0, -0.4em)}}
	\node[vertex] (4) at (0,1) {$4$};
	\node[vertex] (7) at (1,0.5) {$7$};
	\node[vertex] (6) at (2,1) {$6$};
	\node[vertex] (5) at (1,-0.5) {$5$};
	\node[vertex] (8) at (2,-0.5) {$8$};
	\draw[double,edge] (4) to (7);
	\draw[double,edge] (7) to (5);
	\draw[edge] (4) to (6);
	\draw[edge] (5) to (8);
	\draw[edge] (4) to (5);
	\end{tikzpicture}}+{\tiny\begin{tikzpicture}
	\tikzset{vertex/.style = {shape=circle,draw,minimum size=0.5em}}
	\tikzset{edge/.style = {->,> = latex'}}
	\tikzset{baseline={(0, -0.4em)}}
	\node[vertex] (4) at (0,0.5) {$4$};
	\node[vertex] (7) at (2,0.5) {$7$};
	\node[vertex] (6) at (1,0.5) {$6$};
	\node[vertex] (5) at (1,-0.5) {$5$};
	\node[vertex] (8) at (2,-0.5) {$8$};
	\draw[double,edge] (6) to (7);
	\draw[double,edge] (5) to (7);
	\draw[double,edge] (5) to (6);
	\draw[edge] (4) to (5);
	\draw[double,edge] (4) to (6);
	\draw[edge] (5) to (8);
	\end{tikzpicture}}\,,
\end{align*}
where each of the six terms are admissible. Writing them fully out, they are given by 
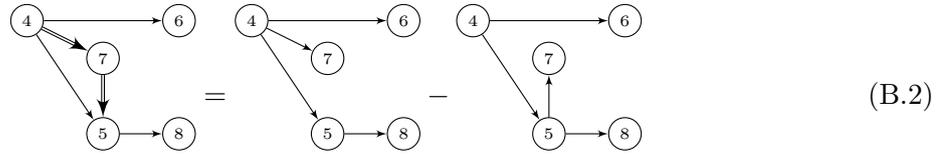
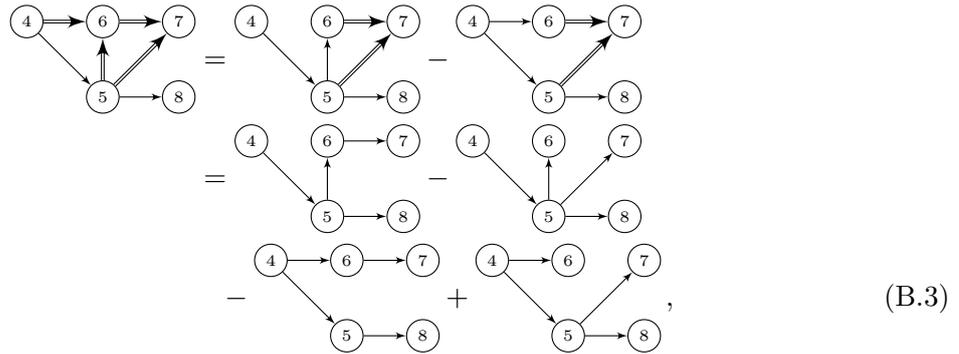
\begin{align}\label{sec:algo:ex1}
{\tiny\begin{tikzpicture}
	\tikzset{vertex/.style = {shape=circle,draw,minimum size=0.5em}}
	\tikzset{edge/.style = {->,> = latex'}}
	\tikzset{baseline={(0, -0.4em)}}
	\node[vertex] (4) at (0,1) {$4$};
	\node[vertex] (7) at (1,0.5) {$7$};
	\node[vertex] (6) at (2,1) {$6$};
	\node[vertex] (5) at (1,-0.5) {$5$};
	\node[vertex] (8) at (2,-0.5) {$8$};
	\draw[double,edge] (4) to (7);
	\draw[double,edge] (7) to (5);
	\draw[edge] (4) to (6);
	\draw[edge] (5) to (8);
	\draw[edge] (4) to (5);
	\end{tikzpicture}}&={\tiny\begin{tikzpicture}
	\tikzset{vertex/.style = {shape=circle,draw,minimum size=0.5em}}
	\tikzset{edge/.style = {->,> = latex'}}
	\tikzset{baseline={(0, -0.4em)}}
	\node[vertex] (4) at (0,1) {$4$};
	\node[vertex] (7) at (1,0.5) {$7$};
	\node[vertex] (6) at (2,1) {$6$};
	\node[vertex] (5) at (1,-0.5) {$5$};
	\node[vertex] (8) at (2,-0.5) {$8$};
	\draw[edge] (4) to (7);
	\draw[edge] (4) to (6);
	\draw[edge] (5) to (8);
	\draw[edge] (4) to (5);
	\end{tikzpicture}}-{\tiny\begin{tikzpicture}
	\tikzset{vertex/.style = {shape=circle,draw,minimum size=0.5em}}
	\tikzset{edge/.style = {->,> = latex'}}
	\tikzset{baseline={(0, -0.4em)}}
	\node[vertex] (4) at (0,1) {$4$};
	\node[vertex] (7) at (1,0.5) {$7$};
	\node[vertex] (6) at (2,1) {$6$};
	\node[vertex] (5) at (1,-0.5) {$5$};
	\node[vertex] (8) at (2,-0.5) {$8$};
	\draw[edge] (5) to (7);
	\draw[edge] (4) to (6);
	\draw[edge] (5) to (8);
	\draw[edge] (4) to (5);
	\end{tikzpicture}}
\end{align}
as well as
\begin{align}\label{sec:algo:ex2}
{\tiny\begin{tikzpicture}
	\tikzset{vertex/.style = {shape=circle,draw,minimum size=0.5em}}
	\tikzset{edge/.style = {->,> = latex'}}
	\tikzset{baseline={(0, -0.4em)}}
	\node[vertex] (4) at (0,0.5) {$4$};
	\node[vertex] (7) at (2,0.5) {$7$};
	\node[vertex] (6) at (1,0.5) {$6$};
	\node[vertex] (5) at (1,-0.5) {$5$};
	\node[vertex] (8) at (2,-0.5) {$8$};
	\draw[double,edge] (6) to (7);
	\draw[double,edge] (5) to (7);
	\draw[double,edge] (5) to (6);
	\draw[edge] (4) to (5);
	\draw[double,edge] (4) to (6);
	\draw[edge] (5) to (8);
	\end{tikzpicture}}&={\tiny\begin{tikzpicture}
	\tikzset{vertex/.style = {shape=circle,draw,minimum size=0.5em}}
	\tikzset{edge/.style = {->,> = latex'}}
	\tikzset{baseline={(0, -0.4em)}}
	\node[vertex] (4) at (0,0.5) {$4$};
	\node[vertex] (7) at (2,0.5) {$7$};
	\node[vertex] (6) at (1,0.5) {$6$};
	\node[vertex] (5) at (1,-0.5) {$5$};
	\node[vertex] (8) at (2,-0.5) {$8$};
	\draw[double,edge] (6) to (7);
	\draw[double,edge] (5) to (7);
	\draw[edge] (5) to (6);
	\draw[edge] (4) to (5);
	\draw[edge] (5) to (8);
	\end{tikzpicture}}-{\tiny\begin{tikzpicture}
	\tikzset{vertex/.style = {shape=circle,draw,minimum size=0.5em}}
	\tikzset{edge/.style = {->,> = latex'}}
	\tikzset{baseline={(0, -0.4em)}}
	\node[vertex] (4) at (0,0.5) {$4$};
	\node[vertex] (7) at (2,0.5) {$7$};
	\node[vertex] (6) at (1,0.5) {$6$};
	\node[vertex] (5) at (1,-0.5) {$5$};
	\node[vertex] (8) at (2,-0.5) {$8$};
	\draw[double,edge] (6) to (7);
	\draw[double,edge] (5) to (7);
	\draw[edge] (4) to (5);
	\draw[edge] (4) to (6);
	\draw[edge] (5) to (8);
	\end{tikzpicture}}\nonumber\\
&={\tiny\begin{tikzpicture}
	\tikzset{vertex/.style = {shape=circle,draw,minimum size=0.5em}}
	\tikzset{edge/.style = {->,> = latex'}}
	\tikzset{baseline={(0, -0.4em)}}
	\node[vertex] (4) at (0,0.5) {$4$};
	\node[vertex] (7) at (2,0.5) {$7$};
	\node[vertex] (6) at (1,0.5) {$6$};
	\node[vertex] (5) at (1,-0.5) {$5$};
	\node[vertex] (8) at (2,-0.5) {$8$};
	\draw[edge] (6) to (7);
	\draw[edge] (5) to (6);
	\draw[edge] (4) to (5);
	\draw[edge] (5) to (8);
	\end{tikzpicture}}-{\tiny\begin{tikzpicture}
	\tikzset{vertex/.style = {shape=circle,draw,minimum size=0.5em}}
	\tikzset{edge/.style = {->,> = latex'}}
	\tikzset{baseline={(0, -0.4em)}}
	\node[vertex] (4) at (0,0.5) {$4$};
	\node[vertex] (7) at (2,0.5) {$7$};
	\node[vertex] (6) at (1,0.5) {$6$};
	\node[vertex] (5) at (1,-0.5) {$5$};
	\node[vertex] (8) at (2,-0.5) {$8$};
	\draw[edge] (5) to (7);
	\draw[edge] (5) to (6);
	\draw[edge] (4) to (5);
	\draw[edge] (5) to (8);
	\end{tikzpicture}}\nonumber\\
	&\phantom{=}-{\tiny\begin{tikzpicture}
		\tikzset{vertex/.style = {shape=circle,draw,minimum size=0.5em}}
		\tikzset{edge/.style = {->,> = latex'}}
		\tikzset{baseline={(0, -0.4em)}}
		\node[vertex] (4) at (0,0.5) {$4$};
		\node[vertex] (7) at (2,0.5) {$7$};
		\node[vertex] (6) at (1,0.5) {$6$};
		\node[vertex] (5) at (1,-0.5) {$5$};
		\node[vertex] (8) at (2,-0.5) {$8$};
		\draw[edge] (6) to (7);
		\draw[edge] (4) to (5);
		\draw[edge] (4) to (6);
		\draw[edge] (5) to (8);
		\end{tikzpicture}}+{\tiny\begin{tikzpicture}
		\tikzset{vertex/.style = {shape=circle,draw,minimum size=0.5em}}
		\tikzset{edge/.style = {->,> = latex'}}
		\tikzset{baseline={(0, -0.4em)}}
		\node[vertex] (4) at (0,0.5) {$4$};
		\node[vertex] (7) at (2,0.5) {$7$};
		\node[vertex] (6) at (1,0.5) {$6$};
		\node[vertex] (5) at (1,-0.5) {$5$};
		\node[vertex] (8) at (2,-0.5) {$8$};
		\draw[edge] (5) to (7);
		\draw[edge] (4) to (5);
		\draw[edge] (4) to (6);
		\draw[edge] (5) to (8);
		\end{tikzpicture}}\,,
\end{align}
such that the equation
\begin{align}\label{sec:algo:ex3}
{\tiny\begin{tikzpicture}
	\tikzset{vertex/.style = {shape=circle,draw,minimum size=0.5em}}
	\tikzset{edge/.style = {->,> = latex'}}
	\tikzset{baseline={(0, -0.4em)}}
	\node[vertex] (4) at (0,0.5) {$4$};
	\node[vertex] (7) at (1,0.5) {$7$};
	\node[vertex] (6) at (2,0.5) {$6$};
	\node[vertex] (5) at (1,-0.5) {$5$};
	\node[vertex] (8) at (2,-0.5) {$8$};
	\draw[edge] (4) to (7);
	\draw[edge] (7) to (6);
	\draw[edge] (7) to (5);
	\draw[edge] (5) to (8);
	\end{tikzpicture}}
&={\tiny\begin{tikzpicture}
	\tikzset{vertex/.style = {shape=circle,draw,minimum size=0.5em}}
	\tikzset{edge/.style = {->,> = latex'}}
	\tikzset{baseline={(0, -0.4em)}}
	\node[vertex] (4) at (0,1) {$4$};
	\node[vertex] (7) at (1,0.5) {$7$};
	\node[vertex] (6) at (2,1) {$6$};
	\node[vertex] (5) at (1,-0.5) {$5$};
	\node[vertex] (8) at (2,-0.5) {$8$};
	\draw[double,edge] (4) to (7);
	\draw[double,edge] (7) to (5);
	\draw[edge] (4) to (6);
	\draw[edge] (5) to (8);
	\draw[edge] (4) to (5);
	\end{tikzpicture}}+{\tiny\begin{tikzpicture}
	\tikzset{vertex/.style = {shape=circle,draw,minimum size=0.5em}}
	\tikzset{edge/.style = {->,> = latex'}}
	\tikzset{baseline={(0, -0.4em)}}
	\node[vertex] (4) at (0,0.5) {$4$};
	\node[vertex] (7) at (2,0.5) {$7$};
	\node[vertex] (6) at (1,0.5) {$6$};
	\node[vertex] (5) at (1,-0.5) {$5$};
	\node[vertex] (8) at (2,-0.5) {$8$};
	\draw[double,edge] (6) to (7);
	\draw[double,edge] (5) to (7);
	\draw[double,edge] (5) to (6);
	\draw[edge] (4) to (5);
	\draw[double,edge] (4) to (6);
	\draw[edge] (5) to (8);
	\end{tikzpicture}}\nonumber\\
	&={\tiny\begin{tikzpicture}
		\tikzset{vertex/.style = {shape=circle,draw,minimum size=0.5em}}
		\tikzset{edge/.style = {->,> = latex'}}
		\tikzset{baseline={(0, -0.4em)}}
		\node[vertex] (4) at (0,1) {$4$};
		\node[vertex] (7) at (1,0.5) {$7$};
		\node[vertex] (6) at (2,1) {$6$};
		\node[vertex] (5) at (1,-0.5) {$5$};
		\node[vertex] (8) at (2,-0.5) {$8$};
		\draw[edge] (4) to (7);
		\draw[edge] (4) to (6);
		\draw[edge] (5) to (8);
		\draw[edge] (4) to (5);
		\end{tikzpicture}}-{\tiny\begin{tikzpicture}
		\tikzset{vertex/.style = {shape=circle,draw,minimum size=0.5em}}
		\tikzset{edge/.style = {->,> = latex'}}
		\tikzset{baseline={(0, -0.4em)}}
		\node[vertex] (4) at (0,1) {$4$};
		\node[vertex] (7) at (1,0.5) {$7$};
		\node[vertex] (6) at (2,1) {$6$};
		\node[vertex] (5) at (1,-0.5) {$5$};
		\node[vertex] (8) at (2,-0.5) {$8$};
		\draw[edge] (5) to (7);
		\draw[edge] (4) to (6);
		\draw[edge] (5) to (8);
		\draw[edge] (4) to (5);
		\end{tikzpicture}}\nonumber\\
	&\phantom{=}+{\tiny\begin{tikzpicture}
		\tikzset{vertex/.style = {shape=circle,draw,minimum size=0.5em}}
		\tikzset{edge/.style = {->,> = latex'}}
		\tikzset{baseline={(0, -0.4em)}}
		\node[vertex] (4) at (0,0.5) {$4$};
		\node[vertex] (7) at (2,0.5) {$7$};
		\node[vertex] (6) at (1,0.5) {$6$};
		\node[vertex] (5) at (1,-0.5) {$5$};
		\node[vertex] (8) at (2,-0.5) {$8$};
		\draw[edge] (6) to (7);
		\draw[edge] (5) to (6);
		\draw[edge] (4) to (5);
		\draw[edge] (5) to (8);
		\end{tikzpicture}}-{\tiny\begin{tikzpicture}
		\tikzset{vertex/.style = {shape=circle,draw,minimum size=0.5em}}
		\tikzset{edge/.style = {->,> = latex'}}
		\tikzset{baseline={(0, -0.4em)}}
		\node[vertex] (4) at (0,0.5) {$4$};
		\node[vertex] (7) at (2,0.5) {$7$};
		\node[vertex] (6) at (1,0.5) {$6$};
		\node[vertex] (5) at (1,-0.5) {$5$};
		\node[vertex] (8) at (2,-0.5) {$8$};
		\draw[edge] (5) to (7);
		\draw[edge] (5) to (6);
		\draw[edge] (4) to (5);
		\draw[edge] (5) to (8);
		\end{tikzpicture}}\nonumber\\
	&\phantom{=}-{\tiny\begin{tikzpicture}
		\tikzset{vertex/.style = {shape=circle,draw,minimum size=0.5em}}
		\tikzset{edge/.style = {->,> = latex'}}
		\tikzset{baseline={(0, -0.4em)}}
		\node[vertex] (4) at (0,0.5) {$4$};
		\node[vertex] (7) at (2,0.5) {$7$};
		\node[vertex] (6) at (1,0.5) {$6$};
		\node[vertex] (5) at (1,-0.5) {$5$};
		\node[vertex] (8) at (2,-0.5) {$8$};
		\draw[edge] (6) to (7);
		\draw[edge] (4) to (5);
		\draw[edge] (4) to (6);
		\draw[edge] (5) to (8);
		\end{tikzpicture}}+{\tiny\begin{tikzpicture}
		\tikzset{vertex/.style = {shape=circle,draw,minimum size=0.5em}}
		\tikzset{edge/.style = {->,> = latex'}}
		\tikzset{baseline={(0, -0.4em)}}
		\node[vertex] (4) at (0,0.5) {$4$};
		\node[vertex] (7) at (2,0.5) {$7$};
		\node[vertex] (6) at (1,0.5) {$6$};
		\node[vertex] (5) at (1,-0.5) {$5$};
		\node[vertex] (8) at (2,-0.5) {$8$};
		\draw[edge] (5) to (7);
		\draw[edge] (4) to (5);
		\draw[edge] (4) to (6);
		\draw[edge] (5) to (8);
		\end{tikzpicture}}\nonumber\\
	&={\tiny\begin{tikzpicture}
		\tikzset{vertex/.style = {shape=circle,draw,minimum size=0.5em}}
		\tikzset{edge/.style = {->,> = latex'}}
		\tikzset{baseline={(0, -0.4em)}}
		\node[vertex] (4) at (0,1) {$4$};
		\node[vertex] (7) at (1,0.5) {$7$};
		\node[vertex] (6) at (2,1) {$6$};
		\node[vertex] (5) at (1,-0.5) {$5$};
		\node[vertex] (8) at (2,-0.5) {$8$};
		\draw[edge] (4) to (7);
		\draw[edge] (4) to (6);
		\draw[edge] (5) to (8);
		\draw[edge] (4) to (5);
		\end{tikzpicture}}+{\tiny\begin{tikzpicture}
		\tikzset{vertex/.style = {shape=circle,draw,minimum size=0.5em}}
		\tikzset{edge/.style = {->,> = latex'}}
		\tikzset{baseline={(0, -0.4em)}}
		\node[vertex] (4) at (0,0.5) {$4$};
		\node[vertex] (7) at (2,0.5) {$7$};
		\node[vertex] (6) at (1,0.5) {$6$};
		\node[vertex] (5) at (1,-0.5) {$5$};
		\node[vertex] (8) at (2,-0.5) {$8$};
		\draw[edge] (6) to (7);
		\draw[edge] (5) to (6);
		\draw[edge] (4) to (5);
		\draw[edge] (5) to (8);
		\end{tikzpicture}}\nonumber\\
	&\phantom{=}-{\tiny\begin{tikzpicture}
		\tikzset{vertex/.style = {shape=circle,draw,minimum size=0.5em}}
		\tikzset{edge/.style = {->,> = latex'}}
		\tikzset{baseline={(0, -0.4em)}}
		\node[vertex] (4) at (0,0.5) {$4$};
		\node[vertex] (7) at (2,0.5) {$7$};
		\node[vertex] (6) at (1,0.5) {$6$};
		\node[vertex] (5) at (1,-0.5) {$5$};
		\node[vertex] (8) at (2,-0.5) {$8$};
		\draw[edge] (5) to (7);
		\draw[edge] (5) to (6);
		\draw[edge] (4) to (5);
		\draw[edge] (5) to (8);
		\end{tikzpicture}}-{\tiny\begin{tikzpicture}
		\tikzset{vertex/.style = {shape=circle,draw,minimum size=0.5em}}
		\tikzset{edge/.style = {->,> = latex'}}
		\tikzset{baseline={(0, -0.4em)}}
		\node[vertex] (4) at (0,0.5) {$4$};
		\node[vertex] (7) at (2,0.5) {$7$};
		\node[vertex] (6) at (1,0.5) {$6$};
		\node[vertex] (5) at (1,-0.5) {$5$};
		\node[vertex] (8) at (2,-0.5) {$8$};
		\draw[edge] (6) to (7);
		\draw[edge] (4) to (5);
		\draw[edge] (4) to (6);
		\draw[edge] (5) to (8);
		\end{tikzpicture}}\,,
\end{align}
with the right-hand side a linear combination of admissible graphs, is according to \eqns{sec:algo:ex1}{sec:algo:ex2} the partial-fractioning identity
\begin{align*}
\frac{1}{x_{57}x_{67}x_{74}x_{85}}&=\frac{1}{x_{54}x_{64}x_{74}x_{85}}-\frac{1}{x_{54}x_{64}x_{75}x_{85}}\nonumber\\
&\phantom{=}+\frac{1}{x_{54}x_{65}x_{76}x_{85}}-\frac{1}{x_{54}x_{65}x_{75}x_{85}}-\frac{1}{x_{54}x_{64}x_{76}x_{85}}+\frac{1}{x_{54}x_{64}x_{75}x_{85}}\nonumber\\
&=\frac{1}{x_{54}x_{64}x_{74}x_{85}}+\frac{1}{x_{54}x_{65}x_{76}x_{85}}-\frac{1}{x_{54}x_{65}x_{75}x_{85}}-\frac{1}{x_{54}x_{64}x_{76}x_{85}}
\end{align*}
and admissibility means that each term on the right-hand side is of the form $\prod_{k=5}^8\frac{1}{x_{k, i_k}}$, where $3\leq i_k<k$. Therefore, using eqs.\ \eqref{sec:algo:ex1,sec:algo:ex2,sec:algo:ex3} and the definitions \eqref{add:fibBasis,add:sigmaPertForm}, the differential form $\varphi^{\tau}_{4,5,5,7}$ in \eqn{sec:algo:exForm} can be expressed in terms of the fibration basis as follows
\begin{align*}
\varphi^{\tau}_{4,5,5,7}&=f^{4,+}_{4,4,4,5}+f^{4,+}_{4,5,8,5}-f^{4,+}_{4,5,5,5}-f^{4,+}_{4,4,6,5}\,.
\end{align*}

\subsection{Basis transformation for four-point string integrals}
The four-point example corresponds to $n=5$ and is based on the vector of integrals given in \eqn{sec:rec:4ptZ} expressed in terms of the labelling $(n,x_i,t_{ij})$
\begin{equation*}
\bhF(x)=\begin{pmatrix}
\hat{F}_{2}^{\id}(x)\\
\hat{F}_{1}^{\id}(x)
\end{pmatrix}
=-\int_0^{z_4} dx_5\, |x_{25}|^{t_{25}}|x_{53}|^{t_{53}} |x_{45}|^{t_{45}} \begin{pmatrix}
\frac{t_{53}}{x_{53}}+\frac{t_{54}}{x_{54}}\\\frac{t_{53}}{x_{53}}
\end{pmatrix}\,.
\end{equation*} 
The differential forms in both entries are already linear combinations of the fibration basis, since
\begin{equation*}
\boldsymbol{f}^{4,+}=\begin{pmatrix}
f^{4,+}_3\\
f^{4,+}_4
\end{pmatrix}=\begin{pmatrix}
\frac{dx_5}{x_{53}}\\
\frac{dx_5}{x_{54}}
\end{pmatrix}\,,
\end{equation*}
which is expected according to \eqn{sec:rec:idPermutedForms}. Thus, we can immediately read off the coefficients $b_{\nu;i_5}^{\id}$, which are in agreement with \eqref{sec:algo:IntCoefId}, and express the twisted forms $\langle \hat{f}^{\id}_{\nu}|$ in terms of the fibration basis
\begin{equation*}
\langle \hat{f}^{\id}_{2}|=-t_{53}\langle \varphi^{\id}_3|-t_{54}\langle \varphi^{\id}_4|=-t_{53}\langle f^{4,+}_3|-t_{54}\langle  f^{4,+}_4|\,,\qquad \langle \hat{f}^{\id}_{1}|=-t_{53}\langle f^{4,+}_3|\,,
\end{equation*}
hence, the basis transformation $\bB$ is given by 
\begin{align*}
\langle \boldsymbol{\hat{f}} | &=\begin{pmatrix}-t_{53}& -t_{54}\\ -t_{53}&0	
\end{pmatrix}\langle \boldsymbol{f}^{4,+}|\,.
\end{align*}
The matrices $\boldsymbol{e}_0$ and $\boldsymbol{e}_1$ can immediately be obtained using the braid matrices for $n=5$  given in \eqn{sec:amp:Omega4} and the transformation in \eqn{sec:rec:ETrafoOmega}
\begin{equation*}
\boldsymbol{e}_0=\bB \boldsymbol{\Omega}_4^{42}\bB^{-1}=\begin{pmatrix}
t_{542}&-t_{52}\\0&t_{42}
\end{pmatrix}\,,\qquad \boldsymbol{e}_1=\bB \boldsymbol{\Omega}_4^{43}\bB^{-1}=\begin{pmatrix}
t_{43}&0\\-t_{53}&t_{543}
\end{pmatrix}\,.
\end{equation*}
In the limit $t_{4i}\rightarrow 0$, these matrices indeed degenerate to the matrices found in \rcite{Broedel:2013aza} and given in \eqn{sec:rec:4PtEx}.

\subsection{Basis transformation for five-point string integrals}
Having calculated the basis transformation for $n=5$ in the previous subsection, we consider the example $n=6$ which corresponds to five-point amplitudes, where the vector $\bhF(x_4)$ is given by
\begin{equation*}
\bhF(x_4)=\begin{pmatrix}
\hat{F}^{\id}_3\\
\hat{F}^{(5\, 6)}_3\\
\hat{F}^{\id}_2\\
\hat{F}^{(5\, 6)}_2\\
\hat{F}^{\id}_1\\
\hat{F}^{(5\, 6)}_1
\end{pmatrix}=\int_0^{x_4}dx_5\int_0^{x_5}dx_6\, \hat{u}(x)\begin{pmatrix}
\left(\frac{t_{65}}{x_{65}}+\frac{t_{64}}{x_{64}}+\frac{t_{63}}{x_{63}}\right)
\left(\frac{t_{54}}{x_{54}}+\frac{t_{53}}{x_{53}}
\right)\\
\left(\frac{t_{56}}{x_{56}}+\frac{t_{54}}{x_{54}}+\frac{t_{53}}{x_{53}}\right)
\left(\frac{t_{64}}{x_{64}}+\frac{t_{63}}{x_{63}}
\right)\\
\left(\frac{t_{65}}{x_{65}}+
\frac{t_{64}}{x_{64}}+
\frac{t_{63}}{x_{63}}\right)\frac{t_{53}}{x_{53}}
\\
\left(\frac{t_{56}}{x_{56}}+
\frac{t_{54}}{x_{54}}+
\frac{t_{53}}{x_{53}}\right)\frac{t_{63}}{x_{63}}\\
\left(\frac{t_{65}}{x_{65}}
+\frac{t_{63}}{x_{63}}
\right)\frac{t_{53}}{x_{53}}\\
\left(\frac{t_{56}}{x_{56}}
+\frac{t_{53}}{x_{53}}
\right)\frac{t_{63}}{x_{63}}\end{pmatrix}\,.
\end{equation*}
First, note that according to the definition \eqref{sec:algo:Inu} of $I_{\nu}$
\begin{align*}
I_3=\{(3,3),(3,4),(3,5),(4,3),(4,4),(4,5)\}
\end{align*}
and 
\begin{equation*}
I_2=\{(3,3),(3,4),(3,5)\}\,,\qquad I_1=\{(3,3),(3,5)\}\,.
\end{equation*}
Furthermore, the differential forms appearing in the sums $\hat{F}_{\nu}^{\id}$ are admissible, such that we immediately obtain the corresponding rows $\bB^{\id}_i$ for $i=1,2,3$ of $\bB$ from \eqn{sec:algo:IntCoefId}
\begin{align*}
\bB&=\begin{pmatrix}
t_{63}t_{53}&t_{64}t_{53}&t_{65}t_{53}&t_{63}t_{54}&t_{64}t_{54}&t_{65}t_{54}\\
\dots& & & & &\\
t_{63}t_{53}&t_{64}t_{53}&t_{65}t_{53}&0&0&0\\
\dots& & & & &\\
t_{63}t_{53}&0&t_{65}t_{53}&0&0&0\\
\dots& & & & &
\end{pmatrix}\,.
\end{align*}
Using partial fractioning, the non-admissible forms which in general correspond to $\sigma \neq \id$, can be rewritten in terms of admissible ones. Applying the algorithm from \subsecref{sec:rec:algo}, the basis transformation can readily be determined using \eqn{sec:algo:BRow}. Starting with the last row, i.e.\ $\hat{F}_1^{(5\, 6)}$, we have to determine $\boldsymbol{\adm}^{T}(\varphi_{3,3}^{(5\, 6)})$ and $\boldsymbol{\adm}^{T}(\varphi_{3,5}^{(5\, 6)})$ since $I_1=\{(3,3),(3,5)\}$, which are the coefficients of the permutation $(5\, 6)$ applied to $\frac{1}{x_{53}x_{63}}=\grDoubleEdgeSource{5}{3}{6}$ and $\frac{1}{x_{53}x_{65}}=\grDoubleEdge{3}{5}{6}$, expressed as the linear combination of admissible products. The former is unchanged by $(5\, 6)$ and hence, stays admissible, such that
\begin{align*}
\boldsymbol{\adm}^{T}(\varphi_{3,3}^{(5\, 6)})&=\begin{pmatrix}
1&0&0&0&0&0
\end{pmatrix}\,.
\end{align*}
The latter becomes $\frac{1}{x_{56}x_{63}}=\grDoubleEdge{3}{6}{5}$, which is non-admissible, since an arrow points from a higher vertex to a lower vertex. Following the algorithm in \subsecref{sec:rec:algo} and according to \eqn{sec:algo:admissibleSubbranch}, it can be expressed as the following linear combination of admissible sequences
\begin{equation*}
\grDoubleEdge{3}{6}{5}=-{\tiny\begin{tikzpicture}
	\tikzset{vertex/.style = {shape=circle,draw,minimum size=0.5em}}
	\tikzset{edge/.style = {->,> = latex'}}
	\tikzset{baseline={(0, -0.4em)}}
	\node[vertex] (3) at (0,0.35) {$3$};
	\node[vertex] (6) at (1,0.35) {$6$};
	\node[vertex] (5) at (0.5,-0.35) {$5$};
	\draw[double,edge] (3) to (6);
	\draw[double,edge] (6) to (5);
	\draw[edge] (3) to (5);
	\end{tikzpicture}}={\tiny\begin{tikzpicture}
	\tikzset{vertex/.style = {shape=circle,draw,minimum size=0.5em}}
	\tikzset{edge/.style = {->,> = latex'}}
	\tikzset{baseline={(0, -0.4em)}}
	\node[vertex] (3) at (0,0.35) {$3$};
	\node[vertex] (6) at (1,0.35) {$6$};
	\node[vertex] (5) at (0.5,-0.35) {$5$};
	\draw[edge] (3) to (6);
	\draw[edge] (3) to (5);
	\end{tikzpicture}}-{\tiny\begin{tikzpicture}
	\tikzset{vertex/.style = {shape=circle,draw,minimum size=0.5em}}
	\tikzset{edge/.style = {->,> = latex'}}
	\tikzset{baseline={(0, -0.4em)}}
	\node[vertex] (3) at (0,0.35) {$3$};
	\node[vertex] (6) at (1,0.35) {$6$};
	\node[vertex] (5) at (0.5,-0.35) {$5$};
	\draw[edge] (6) to (5);
	\draw[edge] (3) to (5);
	\end{tikzpicture}}
\end{equation*}
such that
\begin{align*}
\boldsymbol{\adm}^{T}(\varphi_{3,5}^{(5\, 6)})&=\begin{pmatrix}
1&0&-1&0&0&0
\end{pmatrix}\,.
\end{align*}
Therefore, the row $\bB^{(5\, 6)}_1$ is given by 
\begin{align*}
\bB^{(5\, 6)}_1&=\sum_{(i_5,i_6)\in I_{1}}t_{ 5, i^{(5\, 6)}_5}t_{ 6, i^{(5\, 6)}_6} \boldsymbol{\adm}^{T}(\varphi_{i_5,i_6}^{(5\, 6)})\nonumber\\
&=t_{5,3}t_{6,3} \boldsymbol{\adm}^{T}(\varphi_{3,3}^{(5\, 6)})+t_{5,6}t_{6,3} \boldsymbol{\adm}^{T}(\varphi_{3,5}^{(5\, 6)})\nonumber\\
&=\begin{pmatrix}
t_{63}t_{53}+t_{65}t_{63}&0&-t_{65}t_{63}&0&0&0
\end{pmatrix}\,.
\end{align*}
Similar calculations for the remaining two rows of $\bB$ lead to the transformation matrix
\begin{align*}
\bB&=\begin{pmatrix}
t_{63}t_{53}&t_{64}t_{53}&t_{65}t_{53}&t_{63}t_{54}&t_{64}t_{54}&t_{65}t_{54}\\
t_{63}t_{53}+t_{65}t_{63}&t_{64}t_{53}&-t_{65}t_{63}&t_{63}t_{54}&t_{64}t_{54}+t_{64}t_{65}&-t_{65}t_{64}\\
t_{63}t_{53}&t_{64}t_{53}&t_{65}t_{53}&0&0&0\\
t_{63}t_{53}+t_{65}t_{63}&0&-t_{65}t_{63}&t_{63}t_{54}&0&0\\
t_{63}t_{53}&0&t_{65}t_{53}&0&0&0\\
t_{63}t_{53}+t_{65}t_{63}&0&-t_{65}t_{63}&0&0&0
\end{pmatrix}\,,
\end{align*}
such that for $n=6$
\begin{align*}
\langle \boldsymbol{\hat{f}}|&=\bB \langle \boldsymbol{f}^{4,+}|\,.
\end{align*}
Therefore, the matrices $\boldsymbol{e}_0$ and $\boldsymbol{e}_1$ are given by 
\begin{equation*}
\boldsymbol{e}_0=\bB \boldsymbol{\Omega}^4_{42}\bB^{-1}=
\begin{pmatrix}
t_{6542} & 0 & -t_{52}-t_{65} &
-t_{62} & -t_{62} & t_{62} \\
0 & t_{6542} & -t_{52} &
-t_{62}-t_{65} & t_{52} & -t_{52} \\
0 & 0 & t_{642} & 0 & -t_{62} & 0 \\
0 & 0 & 0 & t_{542} & 0 & -t_{52} \\
0 & 0 & 0 & 0 & t_{42} & 0 \\
0 & 0 & 0 & 0 & 0 & t_{4 2} 
\end{pmatrix}
\end{equation*}
and
\begin{equation}\label{app:rec:e1N6}
\boldsymbol{e}_1=\bB \boldsymbol{\Omega}^4_{43}\bB^{-1}=
\begin{pmatrix}
t_{43} & 0 & 0 & 0 & 0 & 0 \\
0 & t_{43} & 0 & 0 & 0 & 0 \\
-t_{53} & 0 & t_{543} & 0 & 0 & 0 \\
0 & -t_{63} & 0 & t_{643} & 0 & 0 \\
-t_{53} & t_{53} & -t_{63}-t_{65} & -t_{53} &
t_{6543} & 0 \\
t_{63} & -t_{63} & -t_{63} & -t_{53}-t_{65} & 0 &
t_{6543}
\end{pmatrix}\,.
\end{equation}
Indeed, in the limit $t_{i4}\rightarrow 0$ the matrices of \rcite{Broedel:2013aza} are recovered, the same behaviour has been checked explicitly for the examples up to $n=9$.

\subsection{Graphical derivation of braid matrices for $n=5$}\label{app:ex:deriv}
As an example of a graphical derivation of the braid matrices in terms of directed graphs presented in \subsecref{sec:derGraph}, let us derive the braid matrices $\boldsymbol{\Omega}^4_{42}$ and $\boldsymbol{\Omega}^4_{43}$ for $n=5$ in \eqn{sec:amp:Omega4}. The two twisted forms which constitute the fibration basis are 
\begin{equation*}
\langle f^{4,+}_{3}|=\left\langle \frac{d x_5}{x_{53}}\right|=\left\langle \grEdge{3}{5}\right|\,,\quad \text{and}\quad \langle f^{4,+}_{4}|=\left\langle \frac{d x_5}{x_{54}}\right|=\left\langle \grEdge{4}{5}\right|\,.
\end{equation*}
Beginning with the former, we find that for $g_3=\grEdge{3}{5}$
\begin{equation}
V_2(g_3)=\{2\}\,,\quad V_3(g_3)=\{3,5\}\,,\quad V_4(g_3)=\{4\}\,.
\end{equation}
According to \eqn{sec:deriv:derivGraphs}, the derivative of $\langle f^{4,+}_{3}|$ with respect to $x_4$ is therefore given by
\begin{align*}
\frac{\partial}{\partial x_4}\langle f^{4,+}_{3}|&=\frac{\partial}{\partial x_4}\left\langle\grEdge{3}{5}\right|\nonumber\\
&=t_{42}\left\langle\grEdge{2}{4}\,\grEdge{3}{5}\right|+t_{43}\left\langle\grEdge{3}{4}\,\grEdge{3}{5}\right|+t_{45}\left\langle\grEdge{5}{4}\,\grEdge{3}{5}\right|\nonumber\\
&=\grEdge{2}{4}\,t_{42}\left\langle\grEdge{3}{5}\right|+\grEdge{3}{4}\,t_{43}\left\langle\grEdge{3}{5}\right|+t_{45}\left\langle\grDoubleEdge{3}{5}{4}\right|\nonumber\\
&=\frac{1}{x_4}t_{42}
\langle f^{4,+}_{3}|+\frac{1}{x_4-1}t_{43}
\langle f^{4,+}_{3}|-t_{45}\left\langle\grTriangleSink{3}{4}{5}\right|\nonumber\\
&=\frac{1}{x_4}t_{42}
\langle f^{4,+}_{3}|+\frac{1}{x_4-1}t_{43}
\langle f^{4,+}_{3}|-t_{45}\left\langle\grTriangleDSink{3}{4}{5}\right|\nonumber\\
&=\frac{1}{x_4}t_{42}
\langle f^{4,+}_{3}|+\frac{1}{x_4-1}t_{43}
\langle f^{4,+}_{3}|-t_{45}\left\langle\grTriangleClock{3}{4}{5}-\grTriangleSource{3}{4}{5}\right|\nonumber\\
&=\frac{1}{x_4}t_{42}
\langle f^{4,+}_{3}|+\frac{1}{x_4-1}t_{43}
\langle f^{4,+}_4)_{3}|+\frac{1}{x_4-1}t_{45}\left(\langle f^{4,+}_4)_{3}|-\langle f^{4,+}_4)_{4}|\right)\nonumber\\
&=\left(\frac{\begin{pmatrix}
	t_{42}&0
	\end{pmatrix}}{x_4}+\frac{\begin{pmatrix}
	t_{43}+t_{45}&-t_{45}
	\end{pmatrix}}{x_4-1}\right) \langle \boldsymbol{f}^{4,+}|\,,
\end{align*}
where the row vectors $\begin{pmatrix}
t_{42}&0
\end{pmatrix}$ and $\begin{pmatrix}
t_{43}+t_{45}&-t_{45}
\end{pmatrix}$ are indeed the first rows of $\boldsymbol{\Omega}^4_{42}$ and $\boldsymbol{\Omega}^4_{43}$, respectively, as given in \eqn{sec:amp:Omega4}. Similarly, we find for $g_4=\grEdge{4}{5}$
\begin{equation}
V_2(g_4)=\{2\}\,,\quad V_3(g_4)=\{3\}\,,\quad V_4(g_4)=\{4,5\}\,,
\end{equation}
such that \eqn{sec:deriv:derivGraphs} implies
\begin{align*}
\frac{\partial}{\partial x_4}\langle f^{4,+}_{4}|&=\frac{\partial}{\partial x_4}\left\langle\grEdge{4}{5}\right|\nonumber\\
&=\grEdge{2}{4}\,t_{42}\left\langle\grEdge{4}{5}\right|+\grEdge{3}{4}\,t_{43}\left\langle\grEdge{4}{5}\right|+t_{52}\left\langle\grTriangleSink{2}{4}{5}\right|+t_{53}\left\langle\grTriangleSink{3}{4}{5}\right|\nonumber\\
&=\frac{1}{x_{4}}t_{42}\langle f^{4,+}_{4}|+\frac{1}{x_{4}-1}t_{43}\langle f^{4,+}_{4}|+t_{52}\left\langle\grTriangleDSink{2}{4}{5}\right|+t_{53}\left\langle\grTriangleDSink{3}{4}{5}\right|\nonumber\\
&=\frac{1}{x_{4}}t_{42}\langle f^{4,+}_{4}|+\frac{1}{x_{4}-1}t_{43}\langle f^{4,+}_{4}|\nonumber\\
&\phantom{=}+t_{52}\left\langle\grTriangleClock{2}{4}{5}-\grTriangleSource{2}{4}{5}\right|+t_{53}\left\langle\grTriangleClock{3}{4}{5}-\grTriangleSource{3}{4}{5}\right|\nonumber\\
&=\frac{1}{x_{4}}t_{42}\langle f^{4,+}_{4}|+\frac{1}{x_{4}-1}\left(-t_{53}\langle f^{4,+}_{3}|+(t_{43}+t_{53})\langle f^{4,+}_{4}|\right)\nonumber\\
&\phantom{=}+t_{52}\left\langle\grTriangleClock{2}{4}{5}+\grEdge{2}{4}\,\left(\frac{t_{53}}{t_{52}}\grEdge{3}{5}+\frac{t_{54}}{t_{52}}\grEdge{4}{5}\right)\right|\nonumber\\
&=\frac{1}{x_{4}}\left(t_{53}\langle f^{4,+}_{3}|+t_{542}\langle f^{4,+}_{4}|\right)+\frac{1}{x_{4}-1}\left(-t_{53}\langle f^{4,+}_{3}|+(t_{43}+t_{53})\langle f^{4,+}_{4}|\right)\nonumber\\
&=\left(\frac{\begin{pmatrix}
	t_{53}&t_{542}
	\end{pmatrix}}{x_4}+\frac{\begin{pmatrix}
	-t_{53}&t_{43}+t_{53}
	\end{pmatrix}}{x_4-1}\right)\langle \boldsymbol{f}^{4,+}|\,,
\end{align*}
where we used that
\begin{align*}
t_{52}\left\langle\grEdge{2}{5}\right|+t_{53}\left\langle\grEdge{3}{5}\right|+t_{54}\left\langle\grEdge{4}{5}\right|&=0\,,
\end{align*}
since the left-hand side is a total derivative, to express $\left\langle\grEdge{2}{5}\right|$ in terms of the two (fibration) basis vectors $\left\langle\grEdge{3}{5}\right|$ and $\left\langle\grEdge{4}{5}\right|$ of the twisted cohomology. And indeed, the row vectors $\begin{pmatrix}
t_{53}&t_{542}
\end{pmatrix}$ and $\begin{pmatrix}
-t_{53}&t_{43}+t_{53}
\end{pmatrix}$ obtained are the second rows of the braid matrices $\boldsymbol{\Omega}^4_{42}$ and $\boldsymbol{\Omega}^4_{43}$, respectively, as \eqn{sec:amp:Omega4} approves.

\section{Validity of the partial-fractioning algorithm}
\label{app:proofAlgo}
In this section, we prove the validity of the algorithm presented in \subsecref{sec:rec:algo}. This is done by showing that for an admissible sequence $(i_5,i_6,\dots,i_n)$, a permutation $\sigma\in S_{n}$ and $h$ the highest vertex with a non-admissible factor $\frac{1}{x_{h,i_h^{\sigma}}}=\grEdge{i_h^{\sigma}}{h}$, i.e.\ $5\leq h<i_h^{\sigma}$ in
\begin{equation*}
	\prod_{k=5}^n\frac{1}{x_{k,i_k^{\sigma}}}=\prod_{k=5}^n\grEdge{i^{\sigma}_k}{k}\,,
\end{equation*}
there exists a positive integer $l$ and vertices
\begin{equation*}
h^l<h<h^{l-1}<h^{l-2}<\dots < h^{1}<i_{h}^{\sigma}
\end{equation*}
such that the graph
\begin{align}\label{sec:algo:branchNonAdmEx}
b_h&={\tiny\begin{tikzpicture}
	\tikzset{vertex/.style = {shape=circle,draw,minimum size=0.5em}}
	\tikzset{edge/.style = {->,> = latex'}}
	\tikzset{baseline={(0, -0.4em)}}
	\node[vertex] (1) at (0,0) {$h^l$};
	\node[vertex] (2) at (1.5,0) {$h^{l-1}$};
	\node[vertex] (3) at (3,0) {$h^{l-2}$};
	\node[] (4) at (4.5,0) {\dots};
	\node[vertex] (5) at (5.5,0) {$h^1$};
	\node[vertex] (6) at (6.5,0) {$i^{\sigma}_h$};
	\node[vertex] (7) at (7.5,0) {h};
	\draw[edge] (1) to (2);
	\draw[edge] (2) to (3);
	\draw[edge] (3) to (4);
	\draw[edge] (4) to (5);
	\draw[edge] (5) to (6);
	\draw[edge] (6) to (7);
	\end{tikzpicture}}
\end{align}
is a subgraph of the branch containing the vertex $h$. The argument is based on the fact that the sequence $(i_5,i_6,\dots,i_n)$ is admissible.

Since $i^{\sigma}_h>h\geq 5$, there exists a vertex $h^1<i^{\sigma}_h$, such that
\begin{equation*}
\grDoubleEdge{h^1}{i^{\sigma}_h}{h}
\end{equation*}
is a subgraph of $\prod_{k=5}^n\grEdge{i^{\sigma}_k}{k}$. The condition $h^1<i^{\sigma}_h$ follows from the fact that $h$ is the highest vertex with non-admissible factor. If $h^1<h$, we are done and $l=1$. Otherwise, $h^1\geq h\geq 5$ and there exists another vertex $h^2<h^1<h$, such that 
\begin{equation*}
{\tiny\begin{tikzpicture}
	\tikzset{vertex/.style = {shape=circle,draw,minimum size=0.5em}}
	\tikzset{edge/.style = {->,> = latex'}}
	\tikzset{baseline={(0, -0.4em)}}
	\node[vertex] (2) at (0,0) {$h^2$};
	\node[vertex] (3) at (1,0) {$h^1$};
	\node[vertex] (7) at (2,0) {$i_h^{\sigma}$};
	\node[vertex] (8) at (3,0) {$h$};
	\draw[edge] (2) to (3);
	\draw[edge] (3) to (7);
	\draw[edge] (7) to (8);
	\end{tikzpicture}}
\end{equation*}
is a subgraph of $\prod_{k=5}^n\grEdge{i^{\sigma}_k}{k}$. Again, if $h^2<h$, we are done and $l=2$. Otherwise, we can iterate this process a finite number of times, say $l$ times, until $h^l\leq h$. Thus we are done, since the case where  $h^l= h$ can not occur because of the admissibility of the sequence $(i_5,i_6,\dots,i_n)$: if $h^l= h$ was true, and we denote $\grEdgeLong{\sigma (i_m)}{\sigma (m)}=\grEdge{i_h^{\sigma}}{h}$, $\grEdgeLong{\sigma \left(i_{m^j}\right)}{\sigma (m^j)}=\grEdge{i_{h^j}^{\sigma}}{h^j}$ for $1\leq j\leq l$, such that by the admissibility of $(i_5,i_6,\dots,i_n)$
\begin{equation*}
m>i_m\,,\quad m^j>i_{m^j}\,.
\end{equation*}
Furthermore, by construction we have $h^{j+1}=i^{\sigma}_{h^{j}}<h^j$, which implies
\begin{equation*}
m^{j+1}=i_{m^{j}}\,.
\end{equation*}
This means that $h=h^l$ would imply
\begin{align*}
m&=m^l
\end{align*}
and, hence, the inequality
\begin{equation*}
m=m^l=i_{m^{l-1}}<m^{l-1}=i_{m^{l-2}}<m^{l-2}<\dots < m^1<i_m<m
\end{equation*}
would hold. This contradiction shows that $h^l<h$.

\bibliography{drinfeldTwistedDeRhamJPA}

\end{document}